\documentclass[journal=jpcc,manuscript=article]{achemso}

\usepackage[version=3]{mhchem} 
\usepackage{subcaption}
\usepackage{float}
\usepackage{gensymb} 
\usepackage{appendix}
\usepackage{chngcntr}
\usepackage{placeins} 

\newcommand{\cmmnt}[1]{}

\author{Chidozie Onwudinanti}
\affiliation[DIFFER]
{Dutch Institute for Fundamental Energy Research, P.O. Box 6336, 5600 HH Eindhoven, The Netherlands}
\alsoaffiliation[MSM]{Materials Simulation and Modelling,
Department of Applied Physics, Eindhoven University of Technology,
5600 MB Eindhoven, The Netherlands
}
\alsoaffiliation[CCER]
{Center for Computational Energy Research,
P.O. Box 6336, 5600 HH Eindhoven, The Netherlands
}
\author{Mike Pols}
\affiliation[MSM]{Materials Simulation and Modelling,
Department of Applied Physics, Eindhoven University of Technology,
5600 MB Eindhoven, The Netherlands
}
\author{Geert Brocks}
\affiliation[MSM]{Materials Simulation and Modelling,
Department of Applied Physics, Eindhoven University of Technology,
5600 MB Eindhoven, The Netherlands
}
\alsoaffiliation[CCER]
{Center for Computational Energy Research,
P.O. Box 6336, 5600 HH Eindhoven, The Netherlands
}
\alsoaffiliation[MESA]{Computational Materials Science, Faculty of Science and
Technology, MESA+ Institute for Nanotechnology,
University of Twente, P.O. Box 217,
7500 AE Enschede, The Netherlands
}
\author{Vianney Koelman}
\affiliation[CCER]
{Center for Computational Energy Research,
P.O. Box 6336, 5600 HH Eindhoven, The Netherlands
}
\alsoaffiliation[DIFFER]
{Dutch Institute for Fundamental Energy Research, P.O. Box 6336, 5600 HH Eindhoven, The Netherlands}
\alsoaffiliation[AP]{Department of Applied Physics, Eindhoven University of Technology,
5600 MB Eindhoven, The Netherlands
}
\author{Adri C.T. van Duin}
\affiliation[Penn State]{Department of Mechanical Engineering, The Pennsylvania State University, University Park, PA 16802, USA}
\author{Thomas Morgan}
\affiliation[DIFFER]
{Dutch Institute for Fundamental Energy Research, P.O. Box 6336, 5600 HH Eindhoven, The Netherlands}

\author{Shuxia Tao}
\affiliation[MSM]{Materials Simulation and Modelling,
Department of Applied Physics, Eindhoven University of Technology,
5600 MB Eindhoven, The Netherlands
}
\alsoaffiliation[CCER]
{Center for Computational Energy Research,
P.O. Box 6336, 5600 HH Eindhoven, The Netherlands
}
\email{s.x.tao@tue.nl}
\title[An \textsf{achemso} demo]
  {A ReaxFF molecular dynamics study of hydrogen diffusion in ruthenium -- the role of grain boundaries}


\begin{document}







\begin{abstract}
Ruthenium thin films can serve as protective caps for multi-layer extreme ultraviolet mirrors exposed to atomic hydrogen. Hydrogen permeation through ruthenium is problematic as it leads to blisters on the mirrors. H has been shown to exhibit low solubility in bulk Ru, and rapidly diffuses in and out of Ru. Therefore, the underlying mechanisms of the blistering effect remains unknown. This work makes use of reactive molecular dynamics simulations to study the influence of imperfections in a Ru film on the behaviour of H. For the Ru/H system, a ReaxFF force field was parametrised which reproduces structures and energies obtained from quantum-mechanical calculations. Molecular dynamics simulations have been performed with the newly-developed force field, to study the effect of tilt and twist grain boundaries on the overall diffusion behaviour of H in Ru. Our simulations show the tilt and twist grain boundaries provide energetically favourable sites for hydrogen atoms and act as sinks and highways for H. They therefore block H transport across their planes, and favour diffusion along their planes. This results in the accumulation of hydrogen at the grain boundaries. The strong effect of the grain boundaries on the hydrogen diffusion suggests tailoring the morphology of ruthenium thin films as a means to curb the rate of hydrogen permeation.
\end{abstract}

\section{Introduction}
Hydrogen inclusion is often detrimental to the mechanical response and associated desirable properties of materials.\cmmnt{In particular, the character of hydrogen (H) interaction with metals and their alloys is often a key factor in the performance and lifetime of numerous technologies.} Such hydrogen-induced damage\cite{Dwivedi2018HydrogenReview} poses problems in many fields, including hydrogen (H) transport and storage\cite{Herlach2000,Nanninga2012}, nuclear fusion\cite{Ueda2014ResearchBeyond}, and extreme ultraviolet (EUV) lithography\cite{Kuznetsov2014}. The latter application employs multi-layer reflective optics which are susceptible to hydrogen-induced damage. Ruthenium (Ru) can serve as a capping layer in these mirrors, therefore the transport of H through the metal is an important factor in determining the operational lifetime of the optical elements.

Although the interaction of H with ruthenium surfaces is well-represented in the literature\cite{Feulner1985TheDesorption,Luppi2006,Bruneau2014}, research on H in the Ru bulk is sparse, and limited to H solubility in Ru and the thermodynamics of hydride formation\cite{McLellan1973a,Driessen1990a}. Our previous study using Density Functional Theory (DFT) gave positive formation energies of $+0.34$/$+0.85$ eV for an H atom occupying one octahedral/tetrahedral site in Ru, indicating low solubility of H in Ru\cite{Onwudinanti2019TinRuthenium}. In contrast to numerous other metals\cite{Turnbull2012HydrogenMetals}, the diffusion of H in Ru has received little attention. This is in part due to the technical challenges in detecting the hydrogen in the Ru bulk because of its low concentration. One indirect measurement of the diffusion rate of H through Ru thin films has been reported\cite{Soroka2020HydrogenFilms}, using optical changes in an yttrium hydride substrate. 
 
H diffusion in the bulk of a metal typically occurs via hopping of H atoms through the interstitial sites in the lattice. However, in real samples, this is altered by the interaction of H with defects in the crystal lattice, such as vacancies, voids, phase boundaries and grain boundaries (GBs)\cite{Turnbull2012HydrogenMetals}. These defects provide micro-structural trap sites for H -- sites at which the energy of inclusion of the solute atom is significantly lower, and residence time longer, than at the usual interstitial sites. The number and nature of these traps is therefore a determining factor in the overall diffusion of the hydrogen in the metal. For example, in nickel GBs are reported to retard diffusion\cite{Yao1991ExperimentalMetals}. In aluminium, GBs have been shown to enhance or suppress H diffusion depending on the size of the grains\cite{Ichimura1991GrainAluminum}, and to block H diffusion across the boundary plane while enhancing diffusion along it\cite{Pedersen2009SimulationsAluminum}. In hydrogen storage, the rate of hydrogenation of magnesium was shown to increase with grain size\cite{Yao2008HydrogenHydrides}. Notably, the hydrogen diffusion model in Ru in Ref. \citenum{Soroka2020HydrogenFilms} assumes GB transport to be dominant. An atomistic view of the role of GBs in H diffusion in Ru is, however, lacking.   

Computer simulations, such as molecular dynamics (MD), are an efficient way to study H diffusion in diverse metals, and is in particular suited to studying the effect of GBs. In nickel, certain grain boundaries have been shown to enhance diffusion along their plane and to hinder diffusion across it, others appear to have no significant effect on diffusivity\cite{Zhou2017MolecularBoundaries}. In \(\alpha\)-iron, GBs have been shown to slow diffusion by trapping H atoms\cite{Teus2014Grain-boundary-iron}. These MD studies illustrate that the GB effects -- enhancement or retardation of diffusion -- depend strongly on the specific grain boundary in consideration; in tungsten, certain GBs provide deep traps which hold H atoms in place\cite{Yu2014MolecularBoundary}, while others provide paths of low resistance\cite{VonToussaint2011MolecularTungsten}.

In earlier work, we calculated the energy barrier to H jumps within the perfect Ru lattice using \emph{ab initio} methods\cite{Onwudinanti2019TinRuthenium}. However, the computational cost of dynamic simulations with quantum-mechanical methods is prohibitive. Moreover, grain boundaries disrupt the periodicity of a crystal lattice, so large unit cells are required to model them accurately. They also come in a large variety of possible configurations, and due to the complexity of the interfaces, the potential energy surface may be quite complicated. Therefore a technique which balances accuracy with reasonable computational load is required for a simulation of these structures. With its dynamic bond breaking and bond formation, the bond-order-based ReaxFF method is well-suited to the study of such features as GBs, and can sample the potential energy surface to an extent which is not accessible to \emph{ab initio} methods.


Here, we develop a ReaxFF force field for the Ru/H system which reproduces the energies and properties obtained with quantum-mechanical methods. The force field is used for molecular dynamics simulations of the diffusion of H through the intact Ru crystal lattice, and through structures with different grain boundaries. We show the effect of these GBs on the rate and pattern of H transport through Ru, and report diffusion coefficients for the simulated structures. We find that GBs have a profound effect on the H diffusion dynamics in Ru at all simulated temperatures. We discuss the implications of these findings for H transport through Ru bulk and thin films.

\section{Computational methods}
The critical component of a molecular dynamics simulation is the force field. This work employs ReaxFF\cite{vanDuin2001ReaxFF:Hydrocarbons,Chenoweth2008DevelopmentCatalysts,Russo2011Atomistic-scaleEngineering}, a bond-order-based force field method which allows the formation and breaking of bonds in a dynamic simulation. We have developed a set of force field parameters for the Ru/H system using a Monte Carlo global optimisation algorithm\cite{Iype2013ParameterizationAlgorithm}, which minimises an objective function of the form:
\begin{equation}\label{eqn:obj_func}
    Error = \sum_{i=1}^{n}\left[\frac{x_{i,ref}-x_{i,ReaxFF}}{\sigma_{i}}\right]^2
\end{equation}
where $x_{i,ref}$ and $x_{i,ReaxFF}$ are the reference value of the property and the value computed with ReaxFF respectively, $\sigma_i$ represents the weighting of the property, and the sum is over all the entries in the training set. 

The force field parameters are optimised against energies and charges obtained from first-principles calculations. The Vienna Ab Initio Simulation Package (VASP)~\cite{Kresse1994,Kresse1996,Joubert1999} is used for all DFT calculations, which are performed with the generalized gradient approach as proposed by Perdew, Burke, and
Ernzerhof (PBE)~\cite{Perdew1996}. The convergence
parameters are as follows: energy cutoff of 400 eV; residual force criterion
of 1 $\times$ 10 $^{-2}$ eV/\r{A}; energy convergence criterion of 1 $\times$ 10 $^{-5}$ eV. Slab calculations
are performed with a $(9\times9\times1)$ $\Gamma$-centred \emph{k}-points
grid, while bulk calculations are done with a $(9\times9\times9)$
grid; all atoms are allowed to relax in the optimization process. The training set includes Ru equations of state for multiple crystal structures, surface formation energies, H adsorption energies on Ru surfaces, hydride formation energies, and bond length scans. The parameters for H are taken from the set developed by Senftle et al.\cite{Senftle2014ASimulations} as a starting point, while the Ru and Ru-H parameters were newly-generated for this study. More information on the force field parameters and training set can be found in the Supporting Information.

The ReaxFF MD calculations are performed in the AMS2020 software package (version 2020.101) under licence from SCM\cite{Ruger2021AMSHttp://www.scm.com.}. All the MD simulations employ periodic boundary conditions in three directions, and are carried out with a velocity Verlet integrator with a timestep of 0.25 fs. The atom locations as a function of time are tracked at intervals of 250 fs, i.e. every 1000 timesteps. Initial velocities of the atoms are set according to a Maxwell-Boltzmann distribution at the target temperature. The system is then brought to equilibrium in a preparatory simulation of at least 0.1 ns duration, in an NpT ensemble with a Berendsen barostat set to 1 atm and a Nos\'{e}--Hoover chain (NHC) thermostat with a chain length of 10 and a damping constant of 25 fs. The main diffusion simulation is performed in an NVT ensemble with the NHC thermostat\cmmnt{, for a duration of 10 to 12 ns}. We extract from the NVT simulation trajectory the mean squared displacement (MSD) of the H atoms, using the MDAnalysis package\cite{Michaud-Agrawal2011MDAnalysis:Simulations,Gowers2016MDAnalysis:Simulations,Maginn2019BestV1.0,Calandrini2011NMoldynFunctions,deBuyl2018Tidynamics:Simulations}. Diffusion coefficients are calculated from the slope of the MSDs according to the Einstein-Smoluchowski relation:
\begin{equation}\label{eqn:einstein}
    D = \frac{\langle |\textbf{r}_i(t+\tau) - \textbf{r}_i(t)|^2 \rangle  }{2d\tau}.
\end{equation} 
where $\textbf{r}$ is the position of the H atom, $\tau$ is the elapsed time, and $d$ is the dimensionality of the system. The average is taken over time steps and all H atoms. The temperature-dependent diffusion coefficients are fitted to the Arrhenius expression $D=D_0\exp(-E_a/k_BT)$, which yields a pre-exponential factor $D_0$ and activation energy $E_a$, where $k_B$ is the Boltzmann constant and $T$ is the temperature. Hydrogen diffusion data are generated from MD simulations and analysed for three (3) representative structures: the pristine Ru crystal and the two (2) different types of GBs described below. The simulated structures are summarised in Table \ref{tbl:sim_structures}.

\begin{figure}
\captionsetup{justification=centering}
\centering
\begin{subfigure}[b]{0.4\textwidth}
\includegraphics[width=\textwidth]{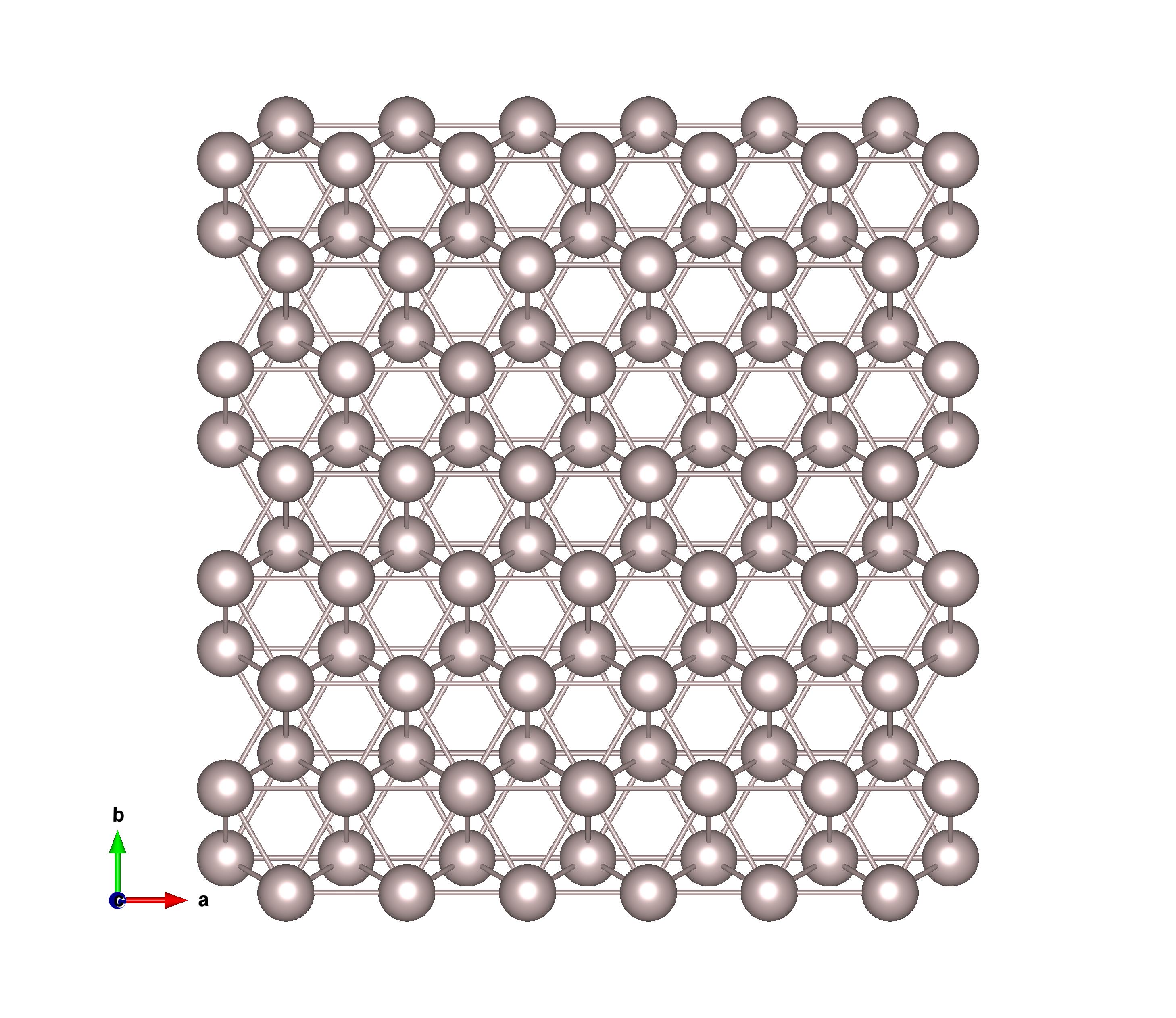}
\caption{\label{fig:pristine}}
\end{subfigure}
\begin{subfigure}[b]{0.4\textwidth}
\includegraphics[width=\textwidth]{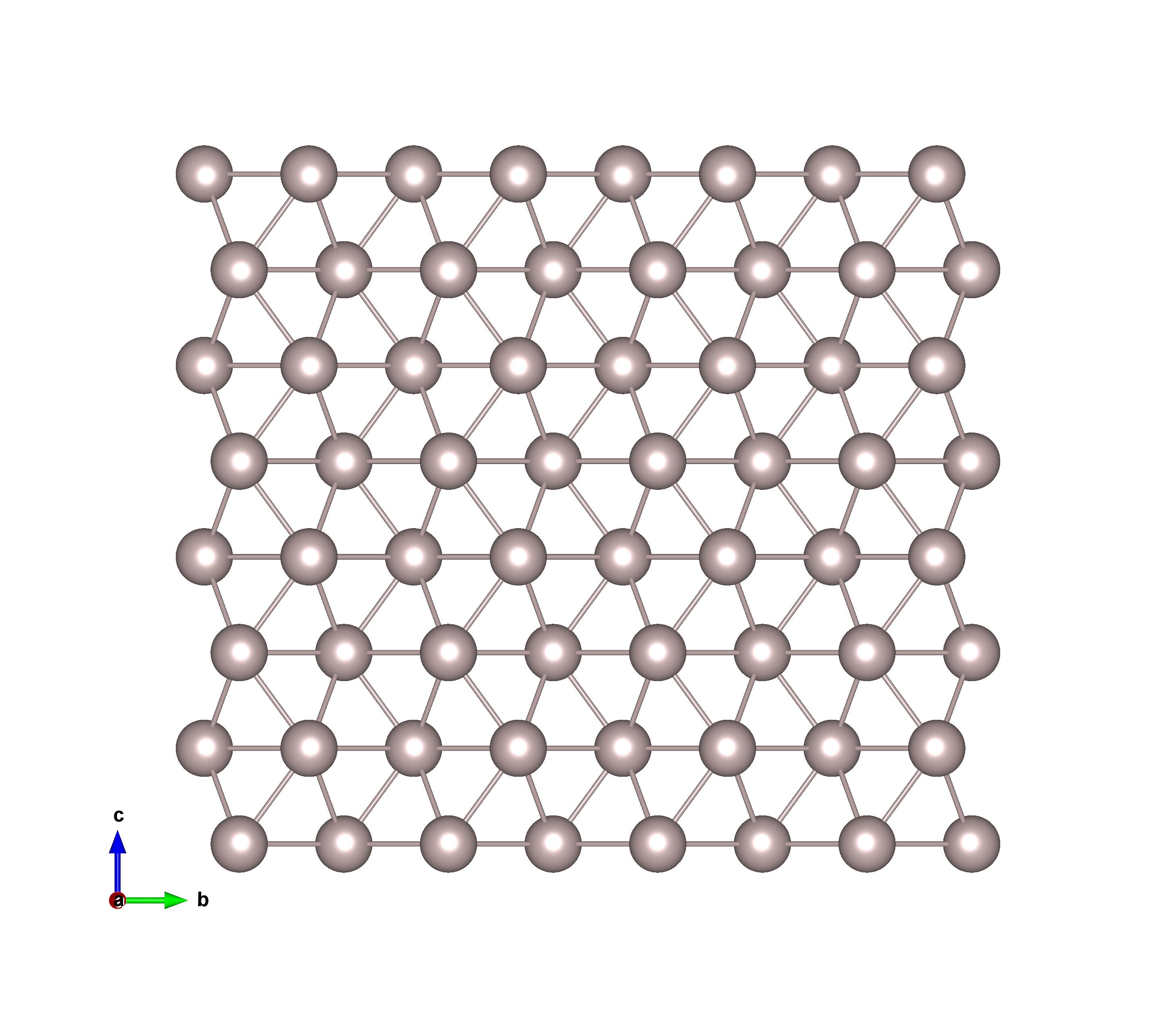}
\caption{\label{fig:pristine_alt}}
\end{subfigure}
\begin{subfigure}[b]{0.4\textwidth}
\includegraphics[width=\textwidth]{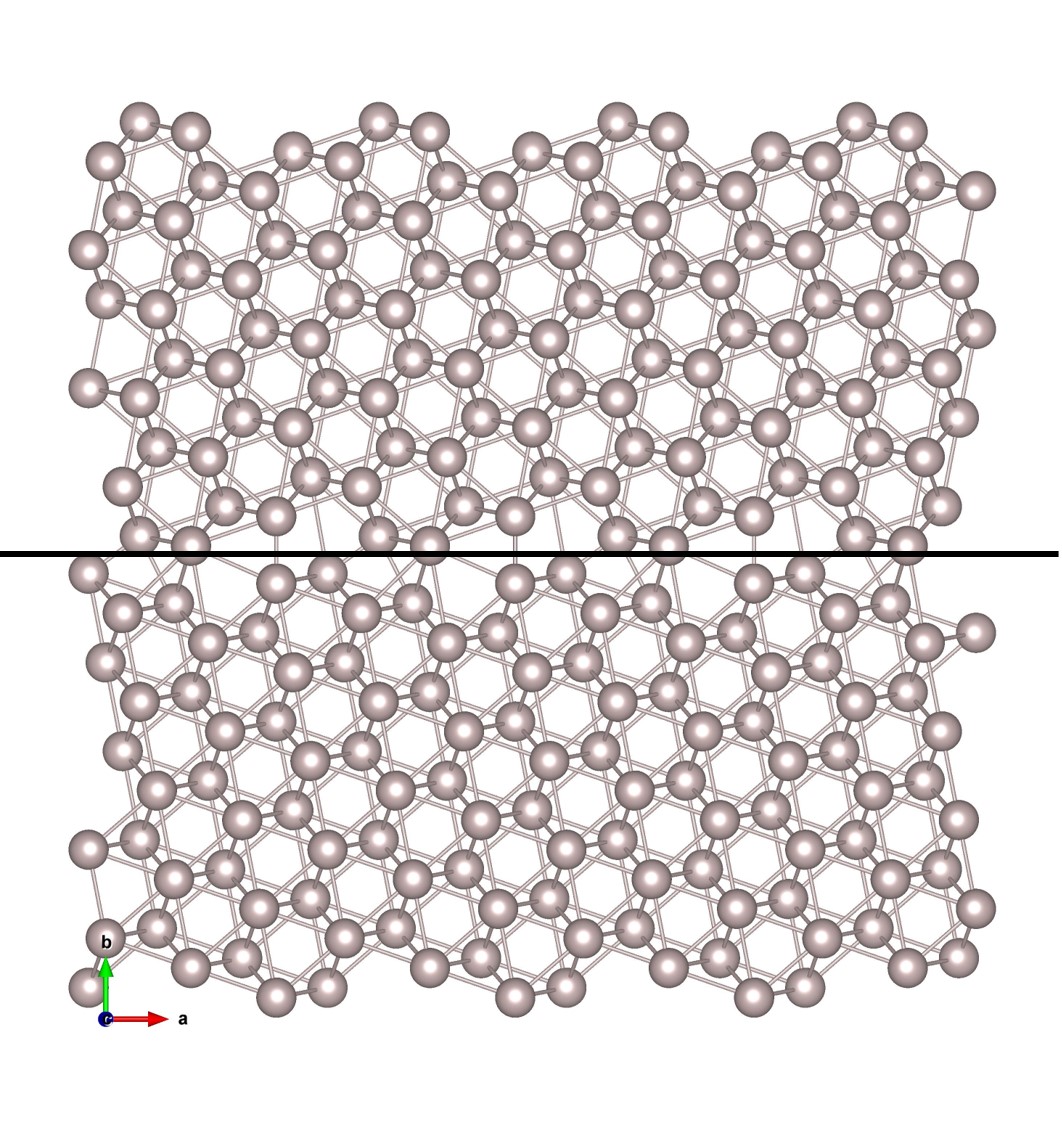}
\caption{\label{fig:tilt_GB_struct}}
\end{subfigure}
\begin{subfigure}[b]{0.4\textwidth}
\includegraphics[width=\textwidth]{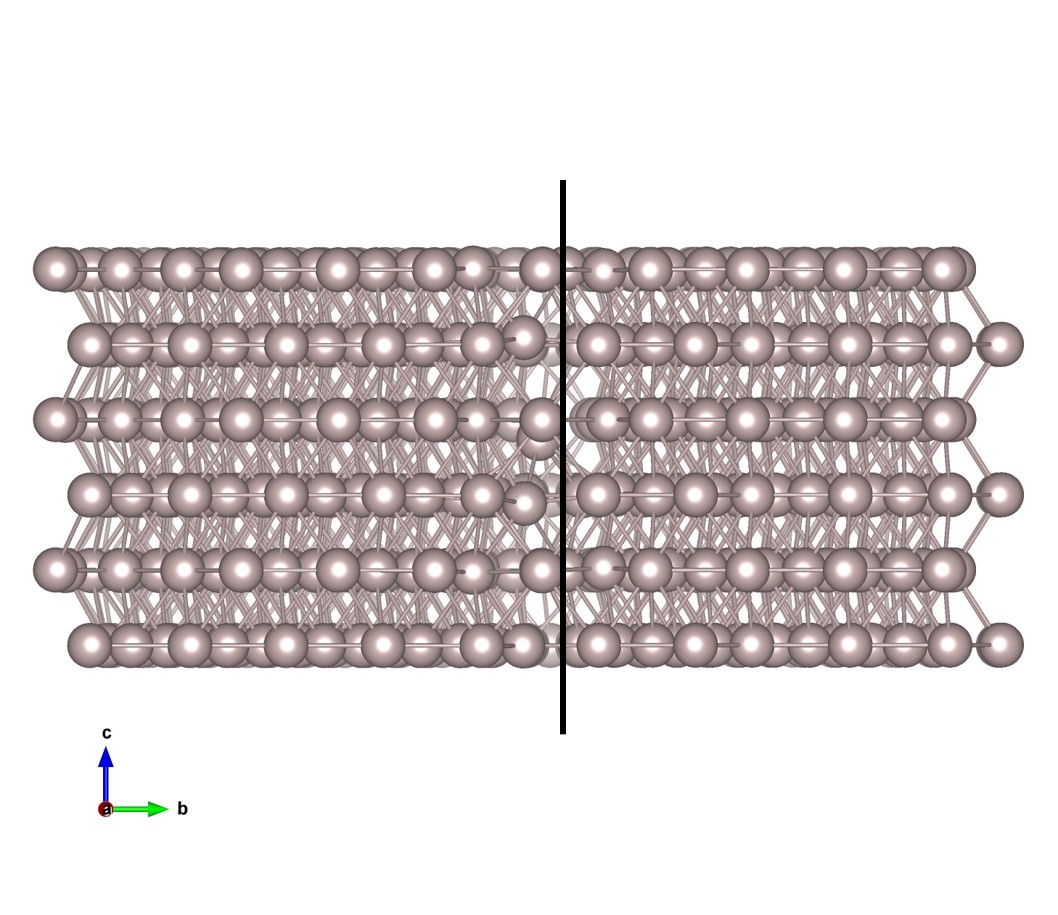}
\caption{\label{fig:tilt_GB_struct_alt}}
\end{subfigure}
\begin{subfigure}[b]{0.4\textwidth}
\includegraphics[width=\textwidth]{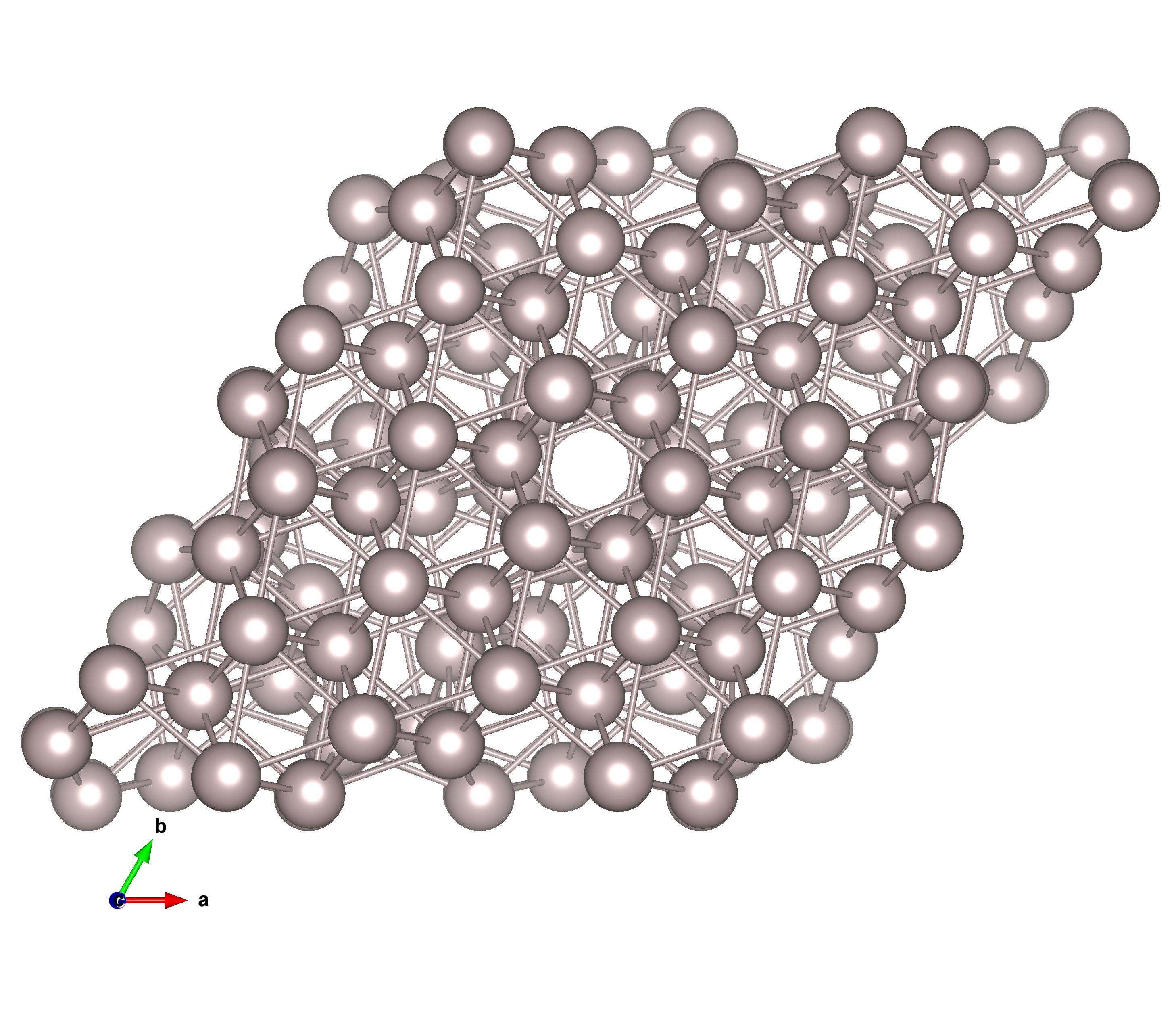}
\caption{\label{fig:twist_GB_struct}}
\end{subfigure}
\begin{subfigure}[b]{0.4\textwidth}
\includegraphics[width=\textwidth]{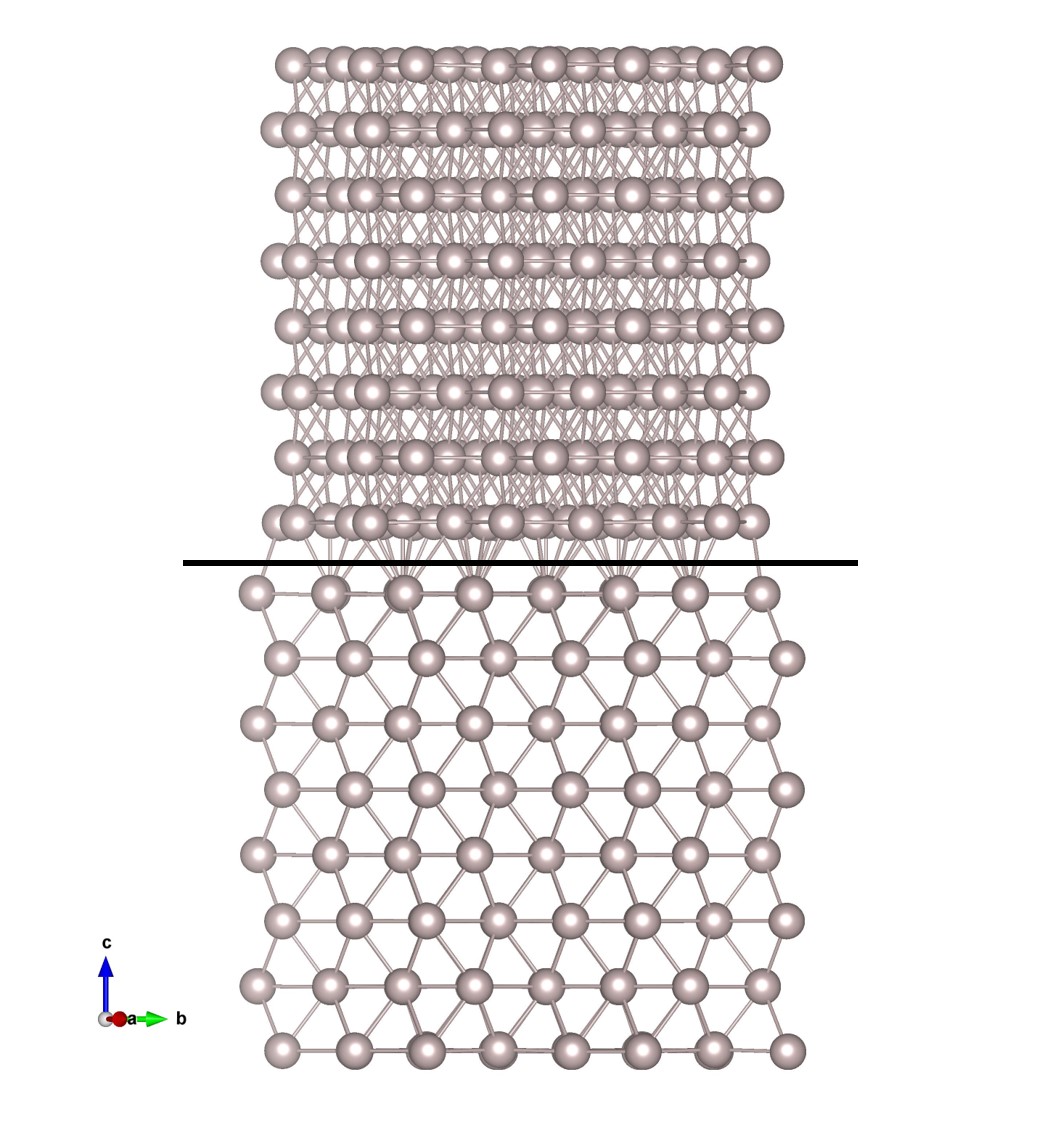}
\caption{\label{fig:twist_GB_struct_alt}}
\end{subfigure}
\caption{(a) Top view and (b) side view of pristine hcp Ru; (c) top view and (d) side view of \(\Sigma\)7 symmetric tilt GB; (e) top view and (f) side view of \(\Sigma\)7 twist GB. The black line marks the GB plane. \label{fig:sim_structures}}
\end{figure}

The crystallography of a GB can be described completely in terms of 5 parameters: 3 to describe the misorientation of the of the two grains, and 2 to describe the inclination of the boundary relative to the axes of either of the crystals. The misorientation is determined by the rotation axis, e.g. [0001] and 38.21$\degree$ for the tilt GB, [0001] and 21.79$\degree$ for the twist GB. The inclination of the boundary is defined by the GB plane, $(01\bar{1}0)$ for the tilt GB, $(0001)$  for the twist GB. The coincidence site lattice (CSL) concept is a convenient way to denote special misorientations, rotation angles at which superposition of two crystals results in a number of lattice points coinciding and forming a sublattice of the two crystal lattices. The CSL is characterised by its \(\Sigma\) value, the ratio of the CSL's unit cell volume to the volume of the generating bulk lattice cell. Because the number of feasible GBs is very large, a thorough exploration of all types is impractical. Therefore, a selection of representative structures is necessary. The first requirement is low formation energy, which implies a high likelihood of occurrence. Another factor is the difference between the selected structures; the more dissimilar the systems simulated, the more information can be extracted. We have selected the \(\Sigma\)7 symmetric tilt GB and the \(\Sigma\)7 twist GB, with rotation about [0001]; they are generated with the free and open-source Atomsk software\cite{Hirel2015Atomsk:Files} and illustrated in Figure \ref{fig:sim_structures}. For a more thorough discussion of the chosen GBs and their properties, see Bruggeman et al.\cite{Bruggeman1972CoincidenceMetals} and Zheng et al.\cite{Zheng2020GrainMetals}. 

\begin{table}
\captionsetup{justification=centering}
\centering
  \caption{Simulated Ru structures}
  \label{tbl:sim_structures}
  \begin{tabular}{lccc}
    \hline
    Structure  & Box dimensions (\r{A}) &  Number of Ru atoms &  Number of H atoms \\
    \hline
    Pristine & 40.6 $\times$ 37.5 $\times$ 34.3 & 3840 & 40\\
    \(\Sigma\)7 tilt GB (tilt) & 27.0 $\times$ 46.8 $\times$ 42.8 & 3690 & 40\\
    \(\Sigma\)7 twist GB (twist) & 36.1 $\times$ 36.1 $\times$ 34.8 & 2800 & 28\\
   
    \hline
  \end{tabular}
\end{table}

\section{Results and discussion}
\subsection{Force field validation}
We have performed a series of calculations to assess the accuracy of the Ru/H force field. The test cases include evaluation of the lattice parameters for the hexagonal close-packed (hcp) Ru crystal and mechanical properties. A comparison of the Ru bulk properties from ReaxFF and DFT is shown in Table \ref{tbl:eos}. To demonstrate the predictive power of the parameters, an additional comparison of the energies of various surfaces (slab models) is also included. 

\begin{table}
\captionsetup{justification=centering}
\centering
  \caption{HCP Ru properties from ReaxFF and DFT.}
  \label{tbl:eos}
  \begin{tabular}{lcccc}
    \hline
    Method  & $a$ (\r{A}) & $c/a$ & $V_0$ (\r{A})\textsuperscript{3} &  $B$ (GPa)  \\
    \hline
    ReaxFF & 2.73 & 1.60 & 14.2 & 332 \\
    DFT & 2.72 & 1.58 & 13.7 & 312 \\
    
    \hline
  \end{tabular}
\end{table}

\begin{figure}[h]
\captionsetup{justification=centering}
\centering
\begin{subfigure}[b]{0.425\textwidth}
\includegraphics[width=\textwidth]{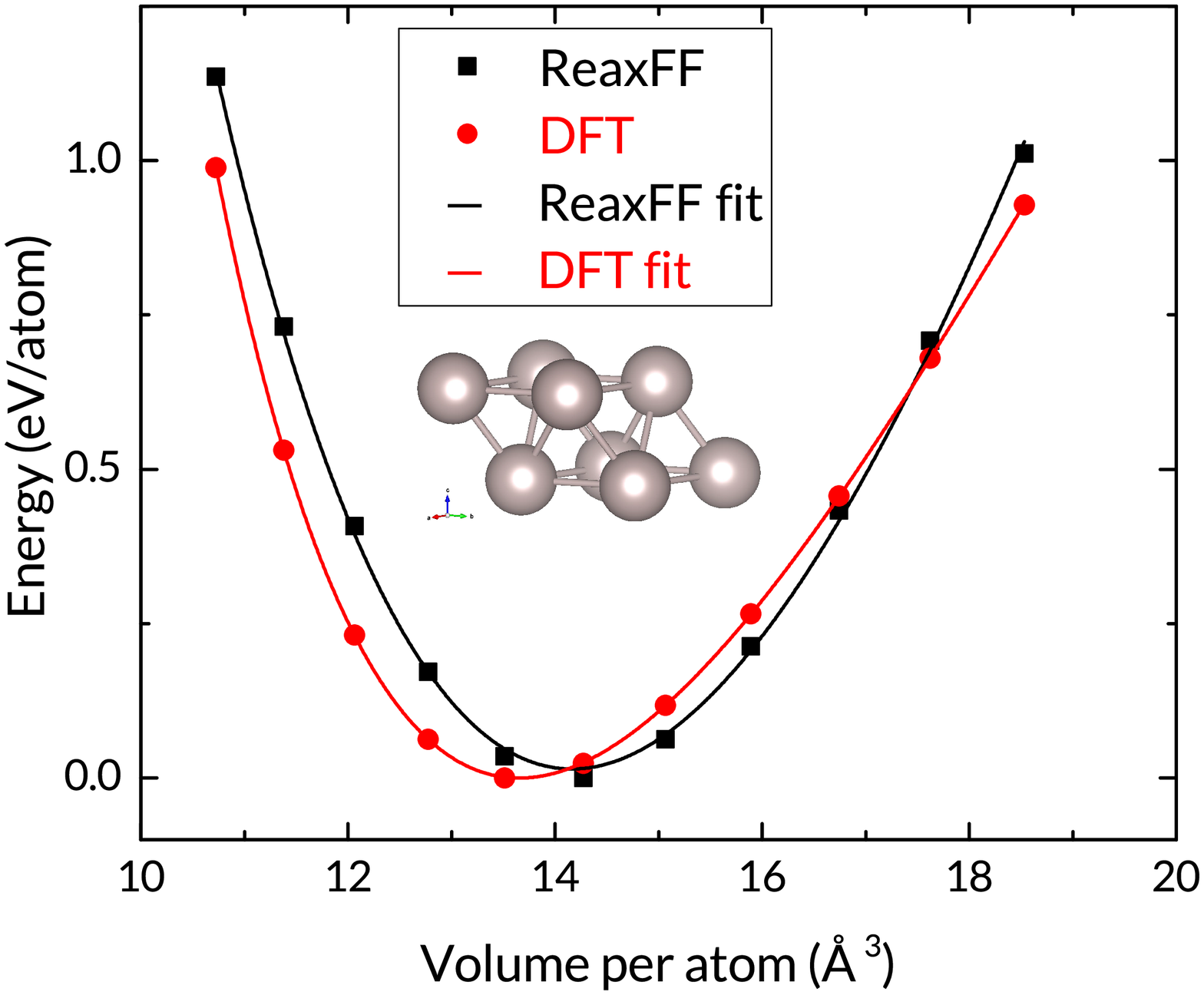}
\caption{\label{fig:hcp_eos}}
\end{subfigure}
\begin{subfigure}[b]{0.525\textwidth}
\includegraphics[width=\textwidth]{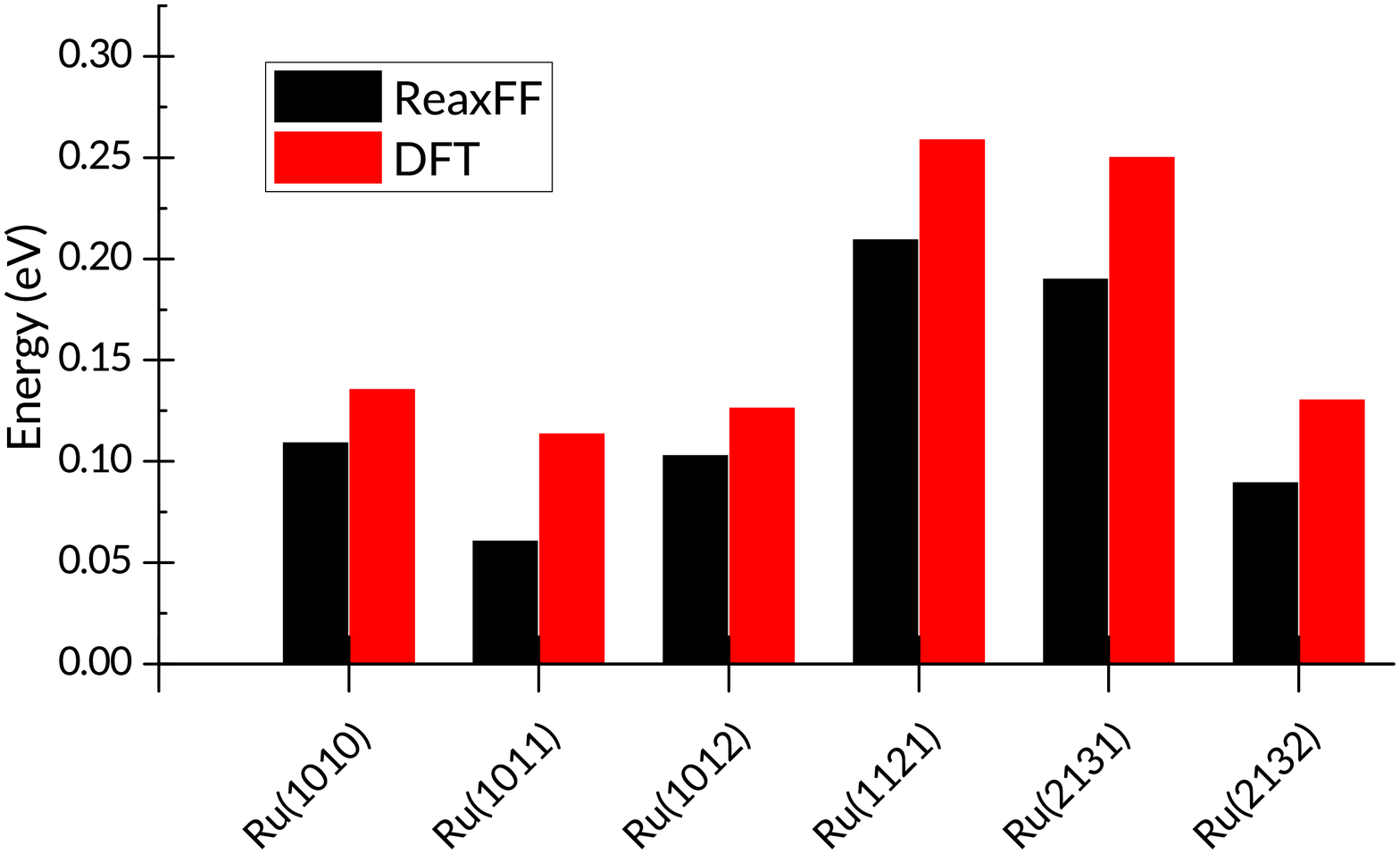}
\caption{\label{fig:slabs}}
\end{subfigure}
\caption{(a) Equations of state for hcp Ru from ReaxFF force field and DFT; and (b) energy per atom of various Ru surfaces relative to Ru(0001)\label{fig:ru_valid}}
\end{figure}

The ReaxFF-computed lattice parameters $a$ (2.73 \r{A}) and $c/a$ (1.60) for hexagonal close-packed (hcp) ruthenium are in good agreement with the DFT-computed values, and with the experimental data, 2.71 \r{A} and 1.58, respectively~\cite{King2012}. The equilibrium volume per atom $V_0$ is overestimated slightly, by 4\%. Mechanical properties from the reference Density Functional Theory (DFT) calculations are also reproduced to a good standard as shown in Figure \ref{fig:hcp_eos}. Bulk moduli calculated from Birch-Murnaghan equations of state are 313 and 333 GPa for ReaxFF and DFT respectively. The force field also shows a good qualitative reproduction of relative energies for a number of Ru slabs with different exposed facets. As shown in Figure \ref{fig:slabs}, although the absolute values are almost uniformly underestimated, the trend is quite well-matched. The largest mismatch does not exceed $0.06$ eV. It should be noted that these slabs were not included in the training set for the parametrisation, so they serve as a test of the force field's performance outside the training space.

\begin{figure}[!h]
\captionsetup{justification=centering}
\centering
\begin{subfigure}[b]{0.25\textwidth}
\includegraphics[width=\textwidth]{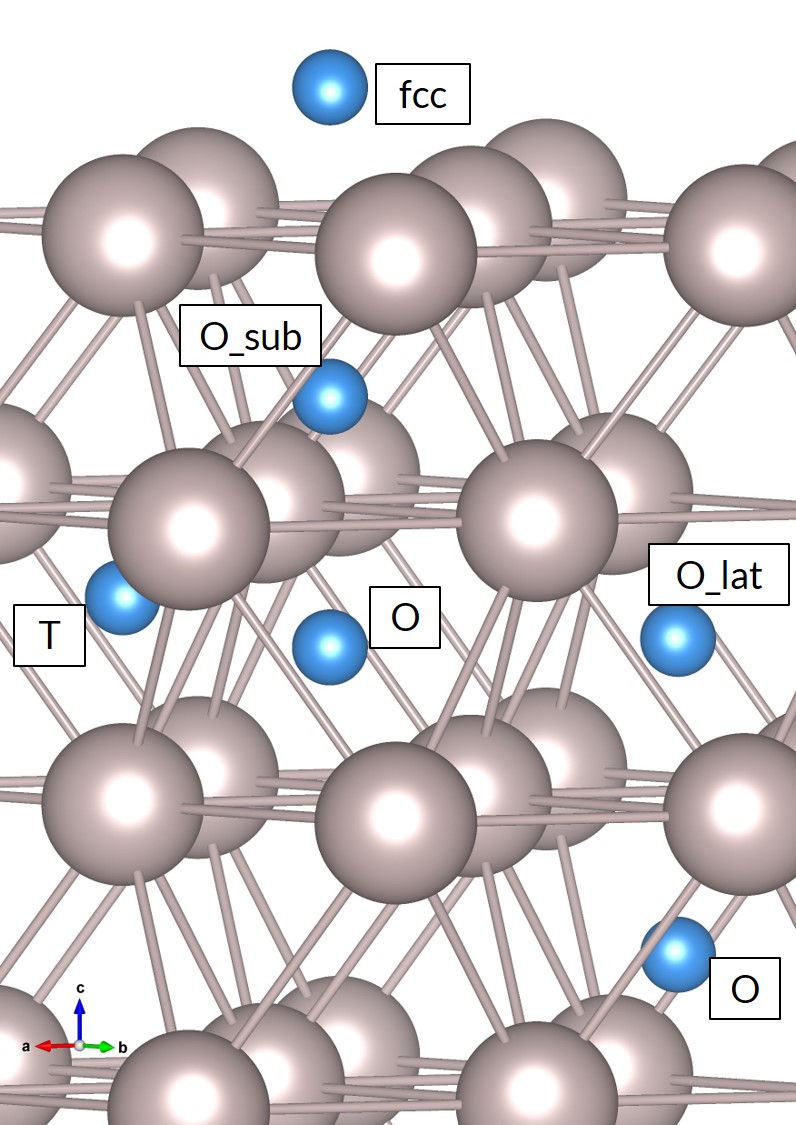}
\caption{\label{fig:H_in_Ru}}
\end{subfigure}
\begin{subfigure}[b]{0.65\textwidth}
\includegraphics[width=\textwidth]{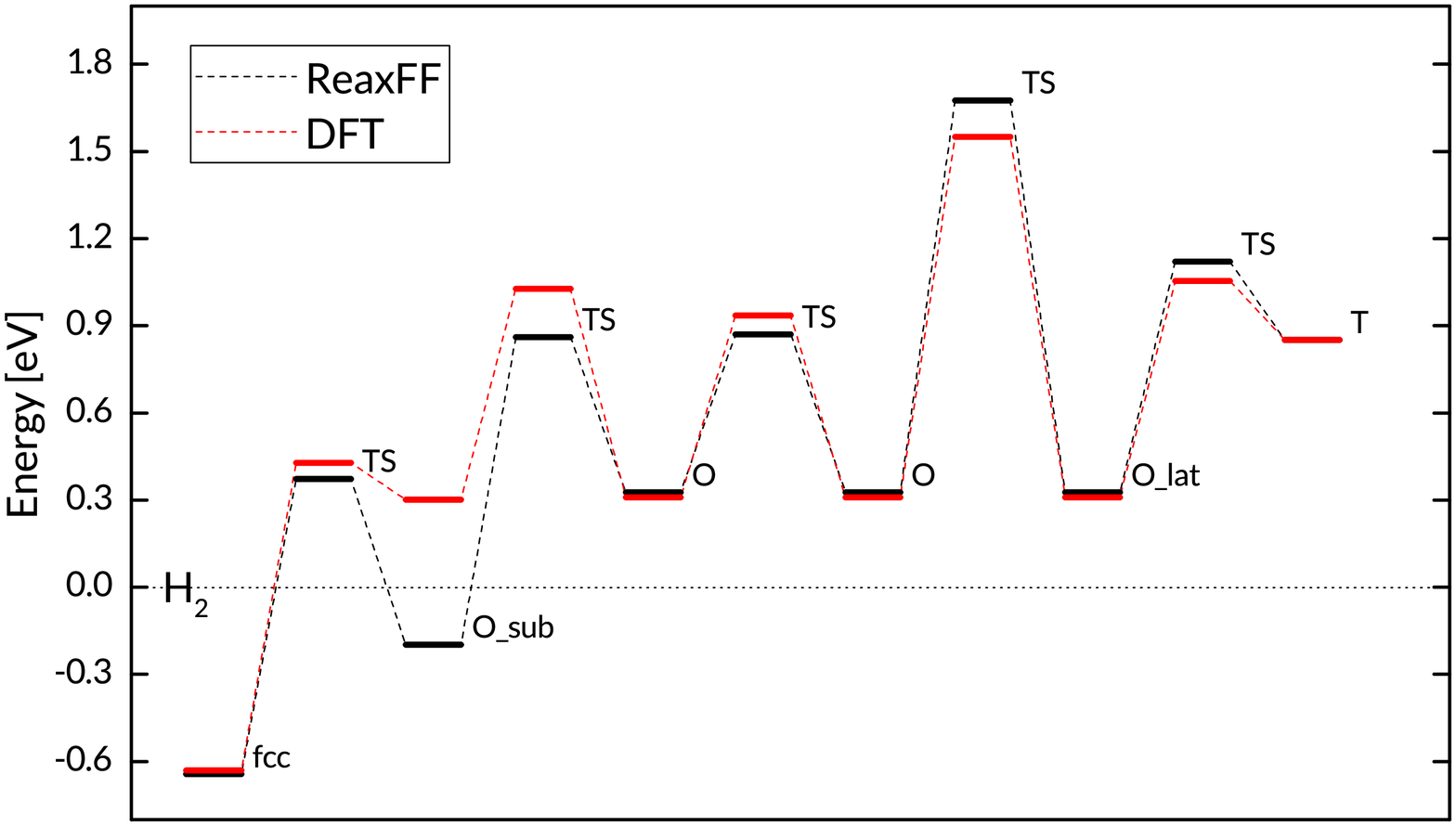}
\caption{\label{fig:energy_profile}}
\end{subfigure}
\caption{A comparison of adsorption energies and hydride formation energies (hydrogen at interstitial sites) on and in Ru, and energies of the transition states along the diffusion paths, obtained with the ReaxFF force field and the DFT reference\cite{Onwudinanti2019TinRuthenium}.\label{fig:reaxff_vs_dft}}
\end{figure}

\begin{figure}[!h]
\captionsetup{justification=centering}
\centering
\begin{subfigure}[b]{0.275\textwidth}
\includegraphics[width=\textwidth]{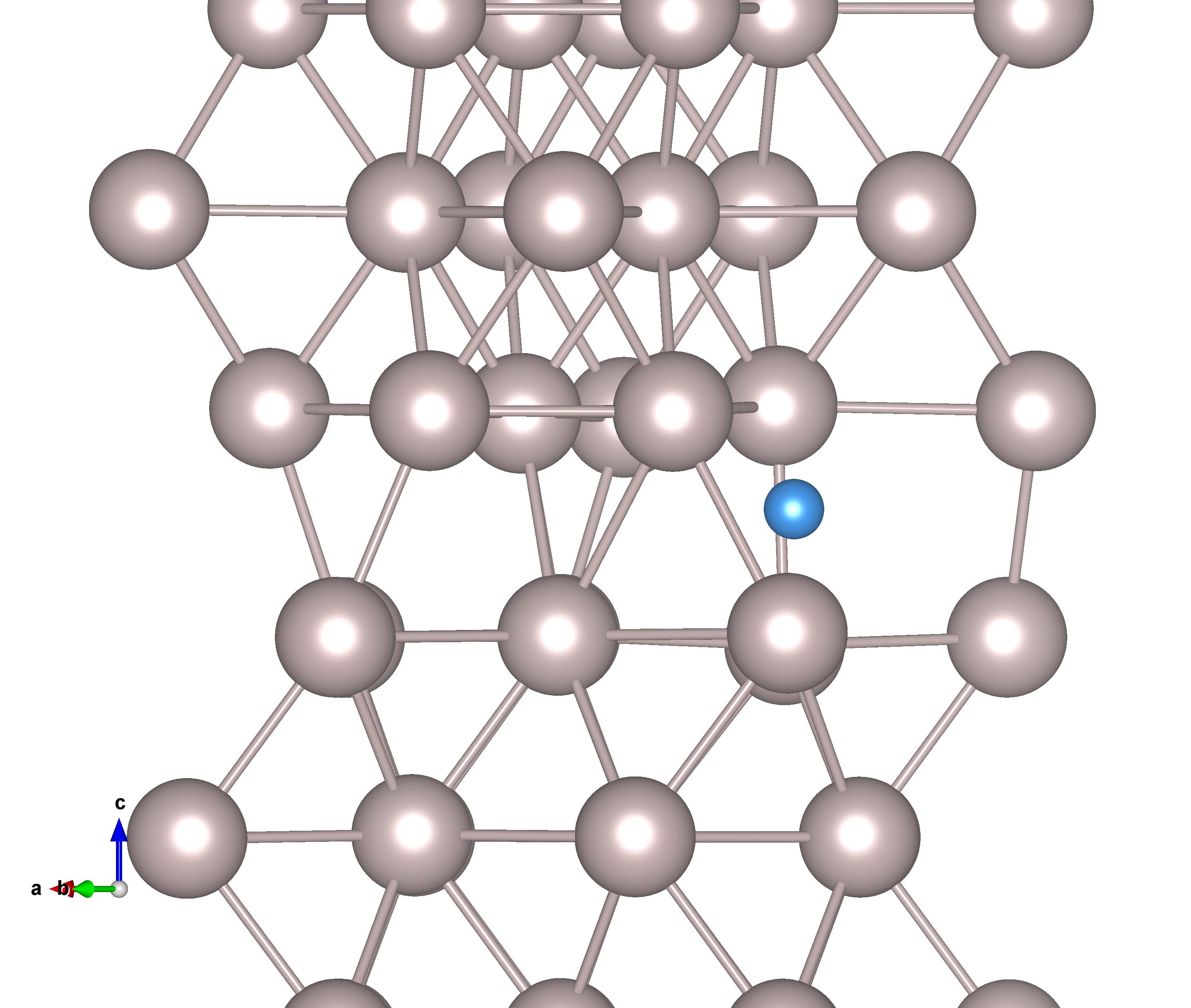}
\caption{\label{fig:H_in_twistGB}}
\end{subfigure}
\quad
\begin{subfigure}[b]{0.275\textwidth}
\includegraphics[width=\textwidth]{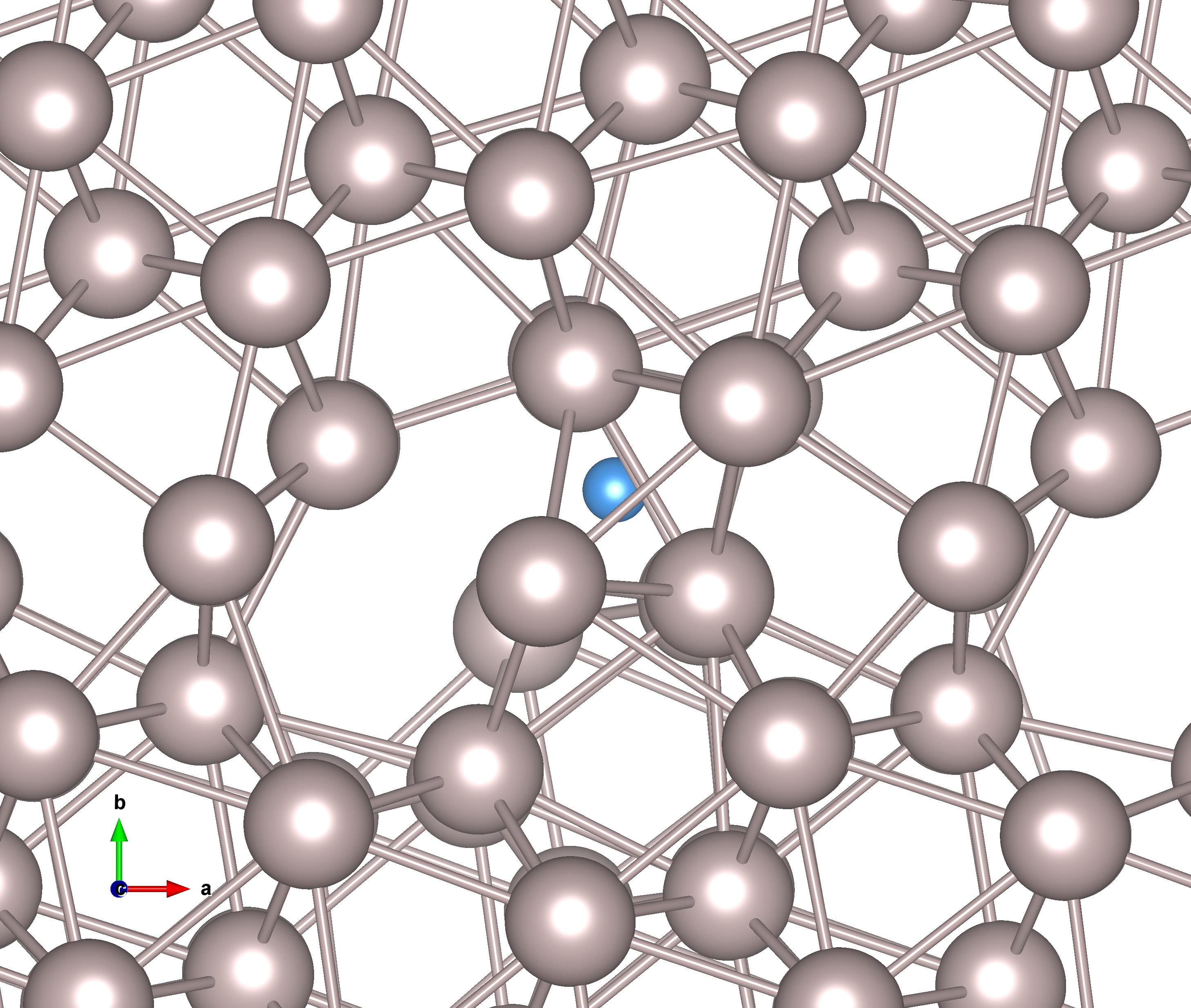}
\caption{\label{fig:H_in_tiltGB}}
\end{subfigure}
\quad
\begin{subfigure}[b]{0.275\textwidth}
\includegraphics[width=\textwidth]{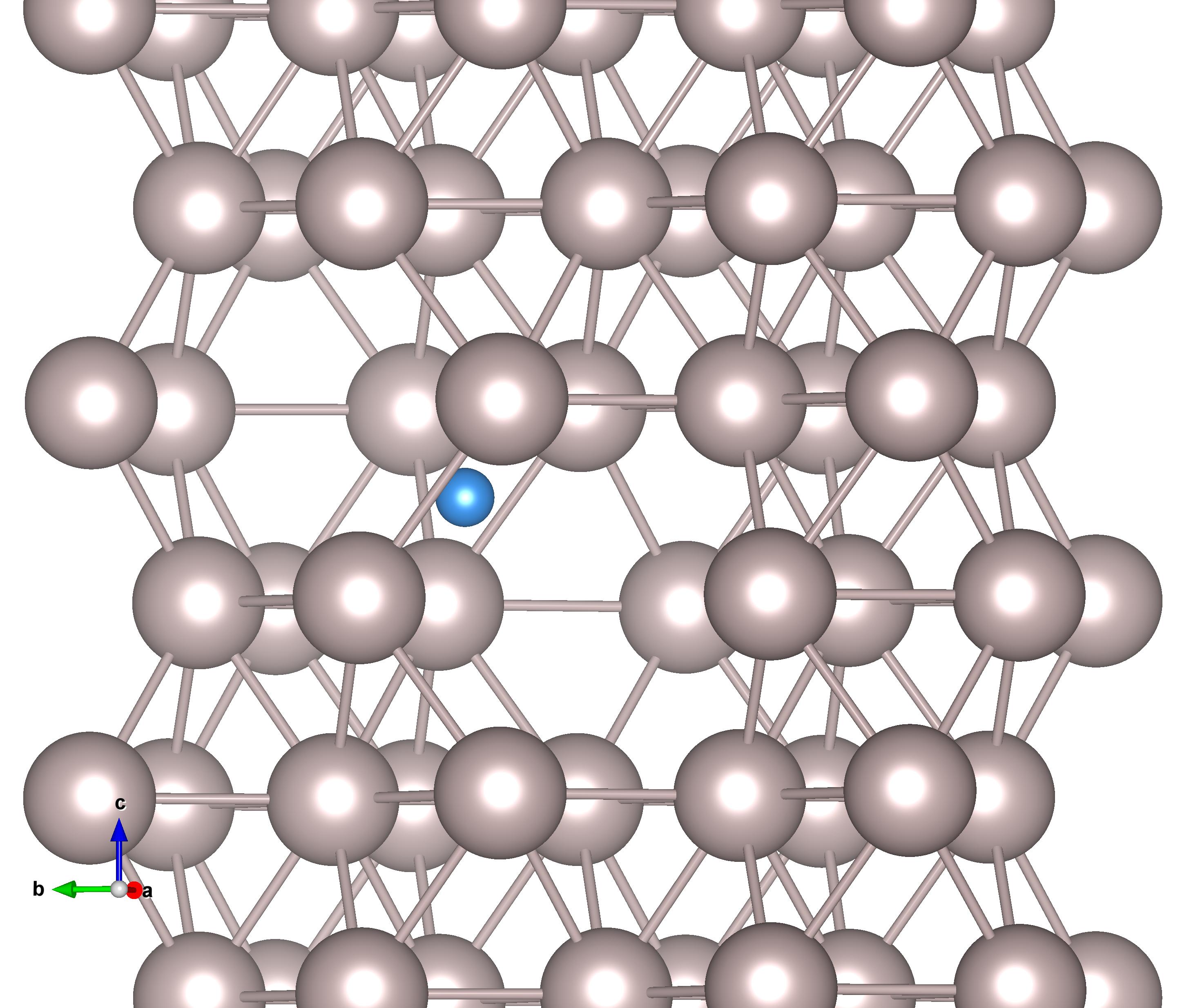}
\caption{\label{fig:H_at_vac_cluster}}
\end{subfigure}

\begin{subfigure}[b]{0.6\textwidth}
\includegraphics[width=\textwidth]{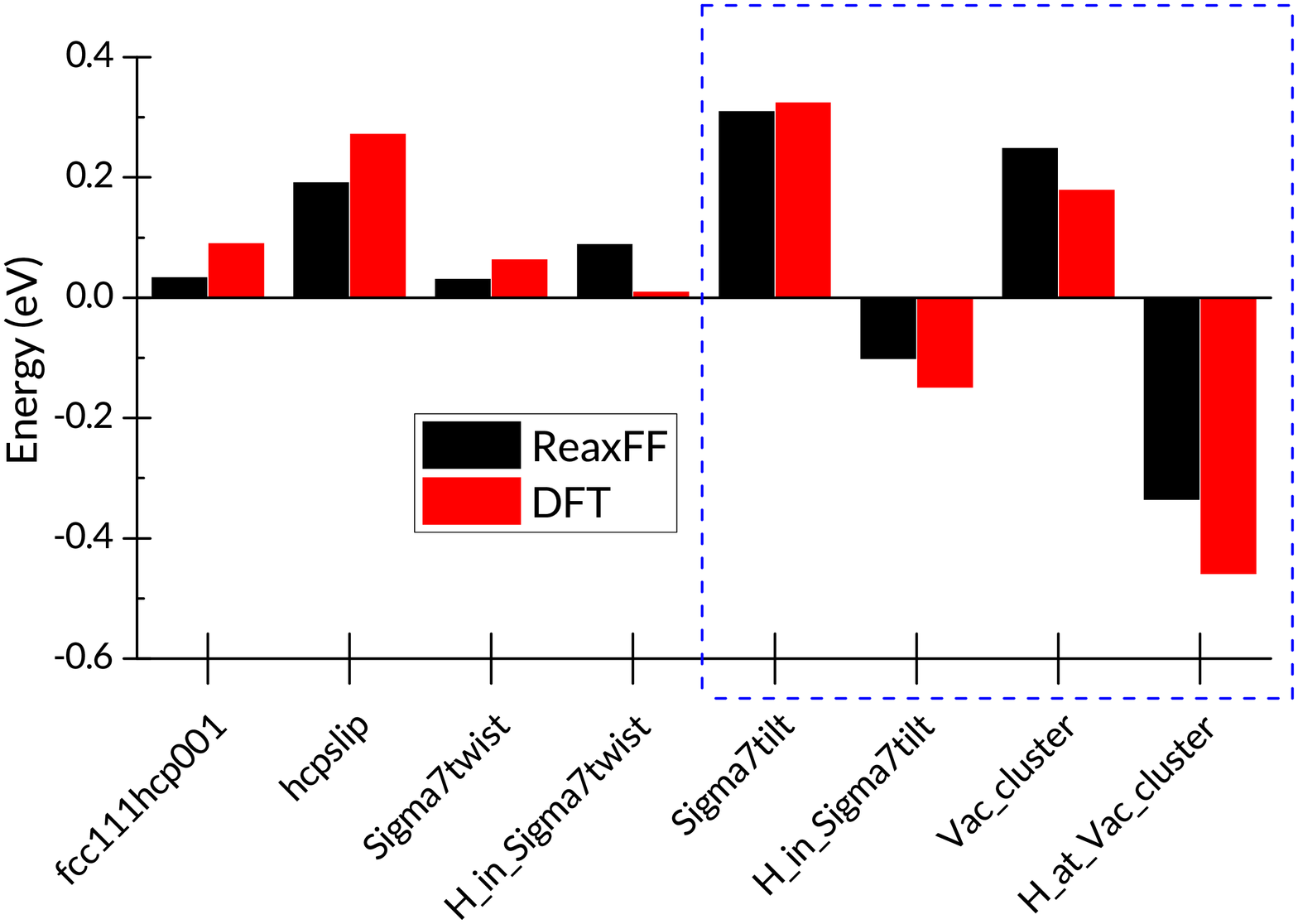}
\caption{\label{fig:GB_form}}
\end{subfigure}
\caption{(a) H in twist GB, (b) H in tilt GB, (c) H at vacancy cluster, and (d) energy per atom for Ru stacking faults, GBs, vacancy clusters, and H sites at defects. \label{fig:defect_valid}}
\end{figure}
For the diffusion simulations, it is especially important that the force field reproduce the energies of H in the interstitial states within the Ru bulk. Figure \ref{fig:reaxff_vs_dft} shows a comparison of ReaxFF and DFT energies for key sites of H in Ru. The interstitial hydride formation energies are calculated as
\begin{equation}
    \Delta E_{H}=\left(E_{M,H_{y}} - E_{M}-\frac{y}{2}E_{H_{2}}\right) ,
\end{equation}
where $y$ is the number of H atoms, while $E_{M_{x}H_{y}}$, $E_{M}$, and $E_{H_{2}}$
stand respectively for the total energy of the metal hydride, the energy of the host metal structure, and the energy of a H\textsubscript{2} molecule.

The agreement is good, within 0.15 eV for all but the octahedral site in the near-surface region, which is lower in energy than the H\textsubscript{2} reference. The main deviation is found near the Ru(0001) surface. However, the discrepancy is unlikely to have a large impact on the simulated bulk diffusion. This conclusion is further strengthened by the force field's reproduction of GBs and defect formation energies. Figure \ref{fig:GB_form} shows the energies of GBs and stacking faults, as well as the energies of H sites at the two GBs and at a vacancy cluster. The structures and energies within the blue rectangle are outside the training set. The reproduction of DFT energies is largely successful, with deviations not exceeding 0.1 eV. 

\subsection{H diffusion in pristine Ru}

\begin{figure}[!h]
\captionsetup{justification=centering}
\centering
\begin{subfigure}[b]{0.45\textwidth}
\includegraphics[width=\textwidth]{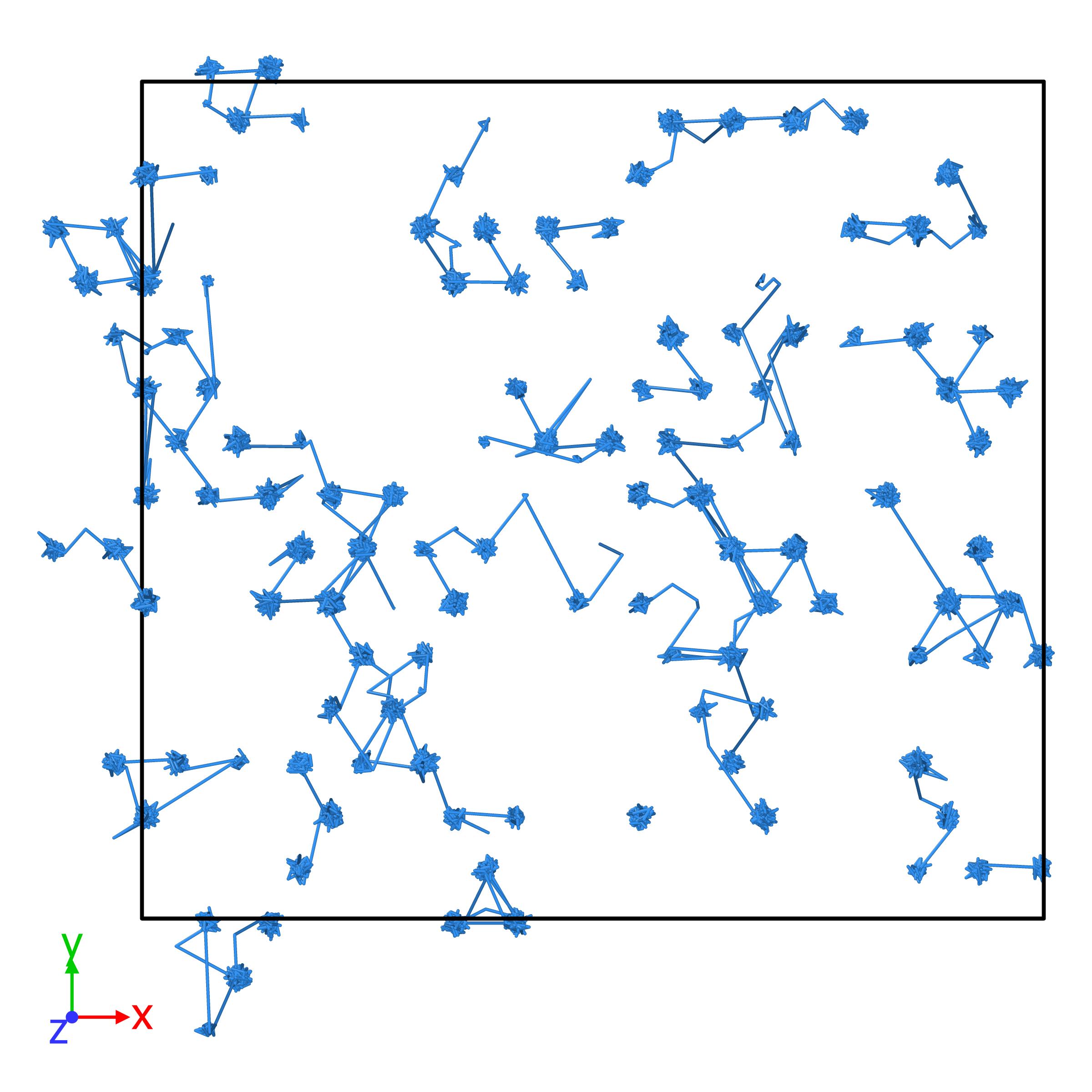}
\caption{\label{fig:pristine_xy}}
\end{subfigure}
\quad
\begin{subfigure}[b]{0.45\textwidth}
\includegraphics[width=\textwidth]{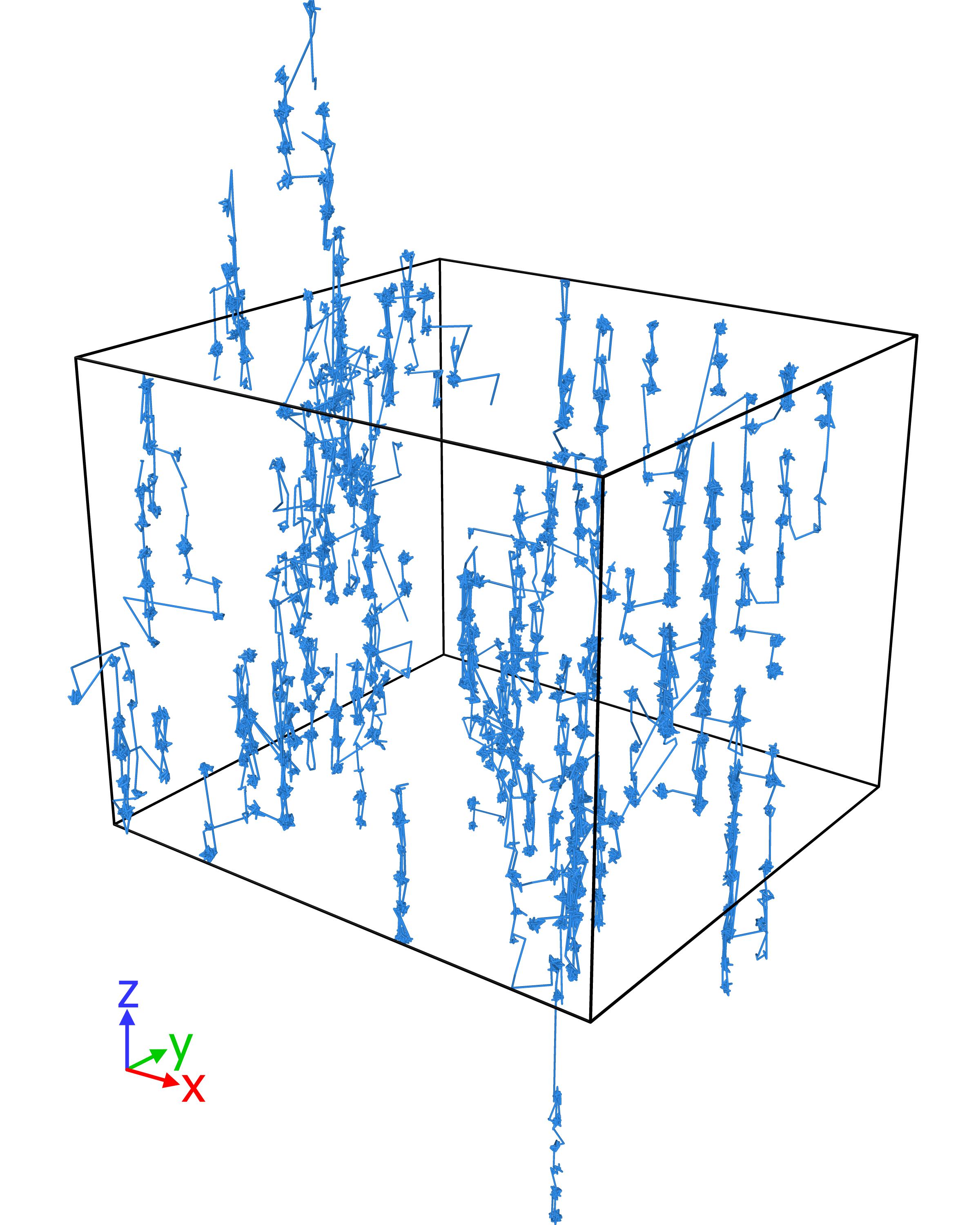}
\caption{\label{fig:pristine_3D}}
\end{subfigure} \\
~\\
\begin{subfigure}[b]{0.45\textwidth}
\includegraphics[width=\textwidth]{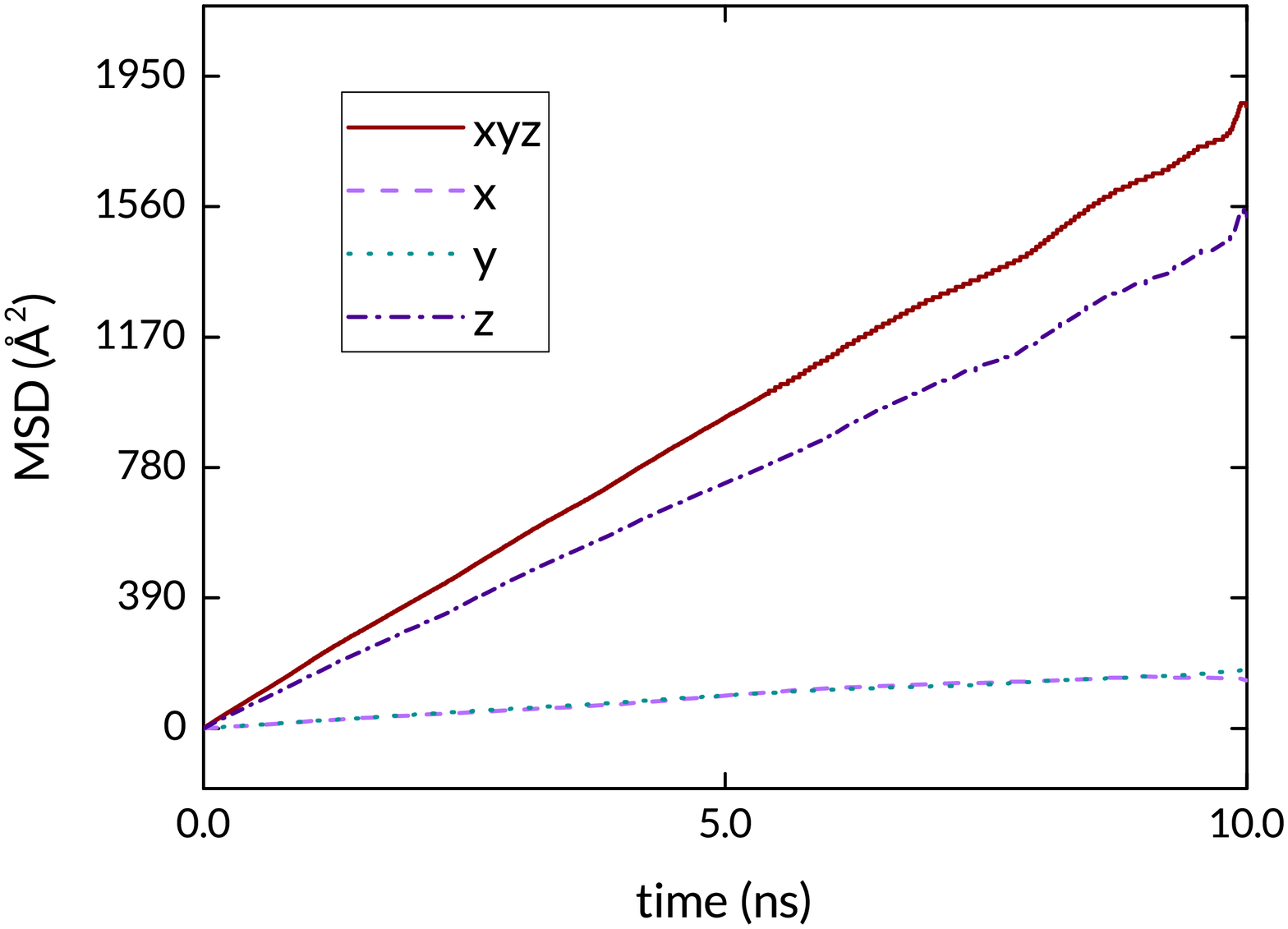}
\caption{\label{fig:pristine_700K}}
\end{subfigure}
\quad
\begin{subfigure}[b]{0.45\textwidth}
\includegraphics[width=\textwidth]{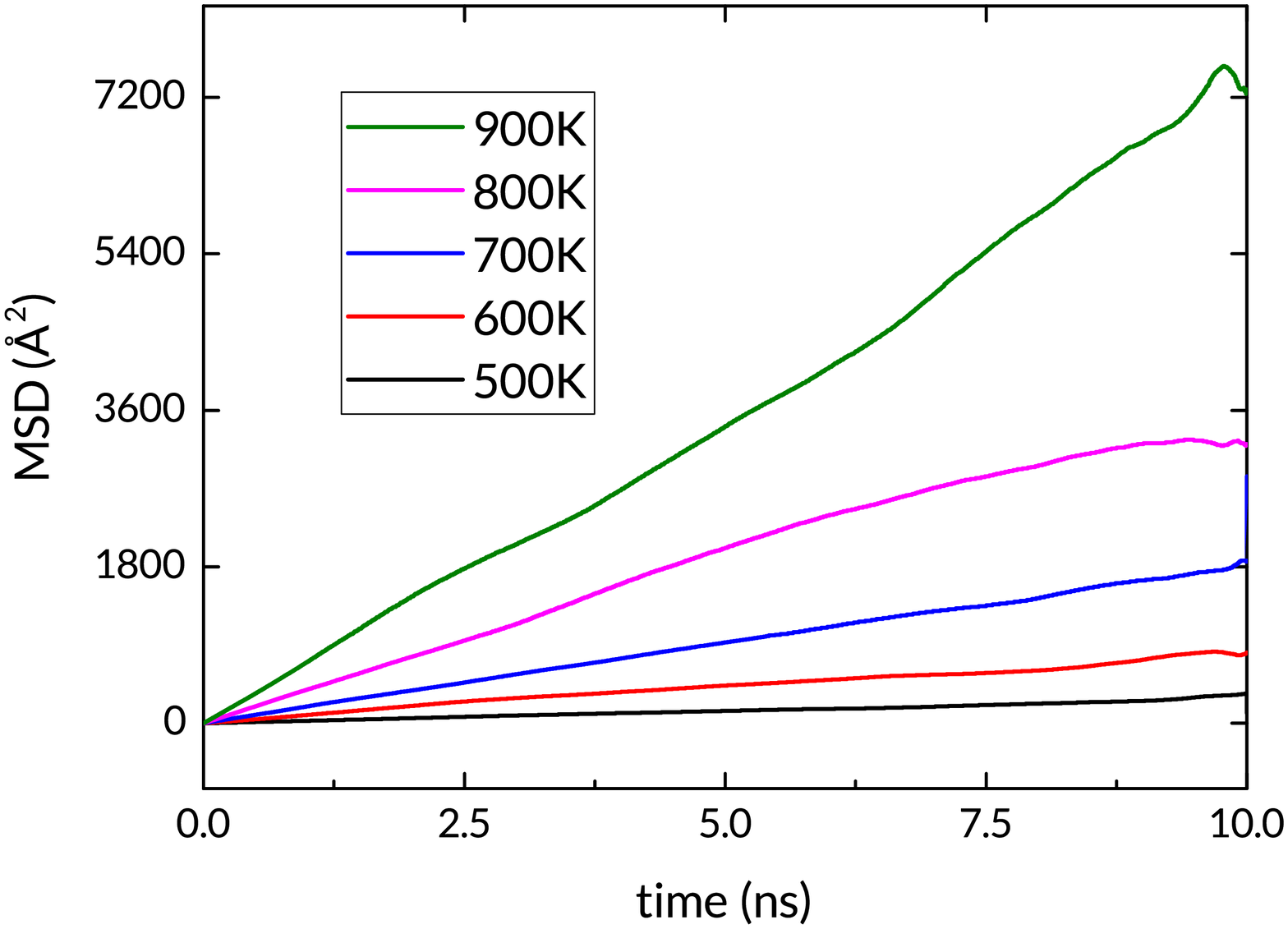}
\caption{\label{fig:pristine_MSD}}
\end{subfigure}
\caption{(a) \& (b) H trajectories in pristine hcp Ru simulation at 700 K. (c) MSD at 700K; (d) MSD for all simulated temperatures.\label{fig:pristine_traj}}
\end{figure}
H transport in Ru proceeds via a series of jumps between interstitial sites, over the energy barriers shown in Figure \ref{fig:reaxff_vs_dft}. The barriers indicate a preference for octahedral-to-octahedral jumps in the \emph{c} direction of the hcp lattice, which is always the Z direction in our coordinate system. This can be seen from the trajectories and the MSD plot in Figure \ref{fig:pristine_traj}. First we observe that the H paths include nodes situated at the octahedral sites, confirming that the solute atoms indeed spend many timesteps at these sites between successful hops. It is also apparent that more successful jumps occur in the Z direction. The trajectories are unwrapped from the periodic translation into the simulation box, so the extent of the map in Figure \ref{fig:pristine_3D} shows the disparity in vertical and horizontal displacement. Figure \ref{fig:pristine_700K} shows the MSD for the 700K NVT simulation. In the diffusive regime, a three-dimensional random walk through the interstitial sites, the MSD has a linear dependence on the elapsed time. The diffusive regime is reached quickly, as the H atoms quickly reach an equilibrium distribution in the intact lattice. The fluctuation at the long time scales is due to the progressively smaller number of data points with the large time lags. The MSD contribution of each of the spatial dimensions is also plotted; the full MSD is the sum of the MSDs in each spatial dimension. In keeping with the observed difference in trajectories, the MSD contribution of the X and Y directions is much smaller than that of the Z component. As expected, there is a monotonic increase in MSD as the temperature is increased, as shown in Figure \ref{fig:pristine_MSD}.
 
\subsection{H diffusion in Ru with tilt GB}
\begin{figure}[!h]
\captionsetup{justification=centering}
\centering
\begin{subfigure}[b]{0.4\textwidth}
\includegraphics[width=\textwidth]{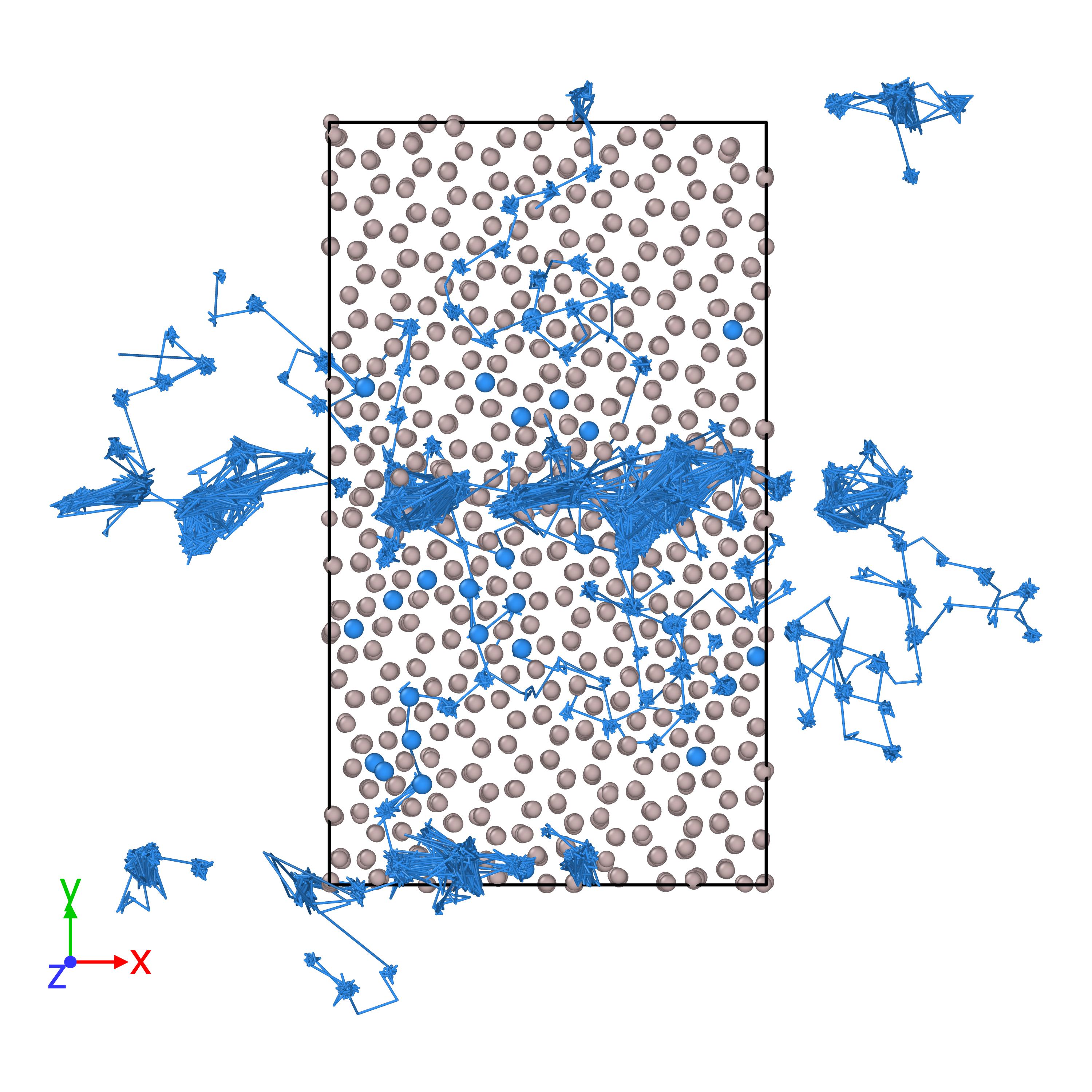}
\caption{\label{fig:tilt_xy}}
\end{subfigure}
\quad
\begin{subfigure}[b]{0.45\textwidth}
\includegraphics[width=\textwidth]{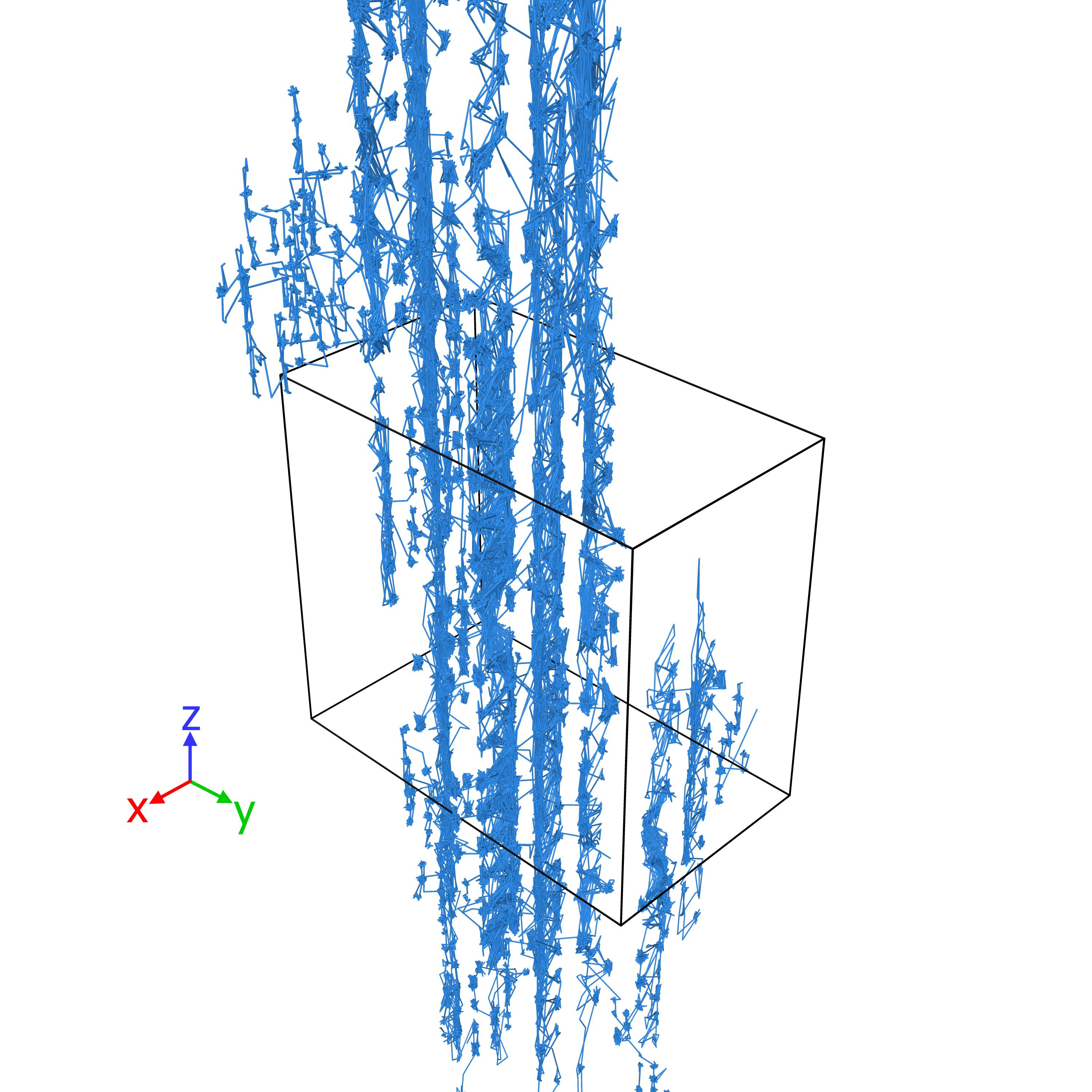}
\caption{\label{fig:tilt_3D}}
\end{subfigure} \\
\begin{subfigure}[b]{0.45\textwidth}
\includegraphics[width=\textwidth]{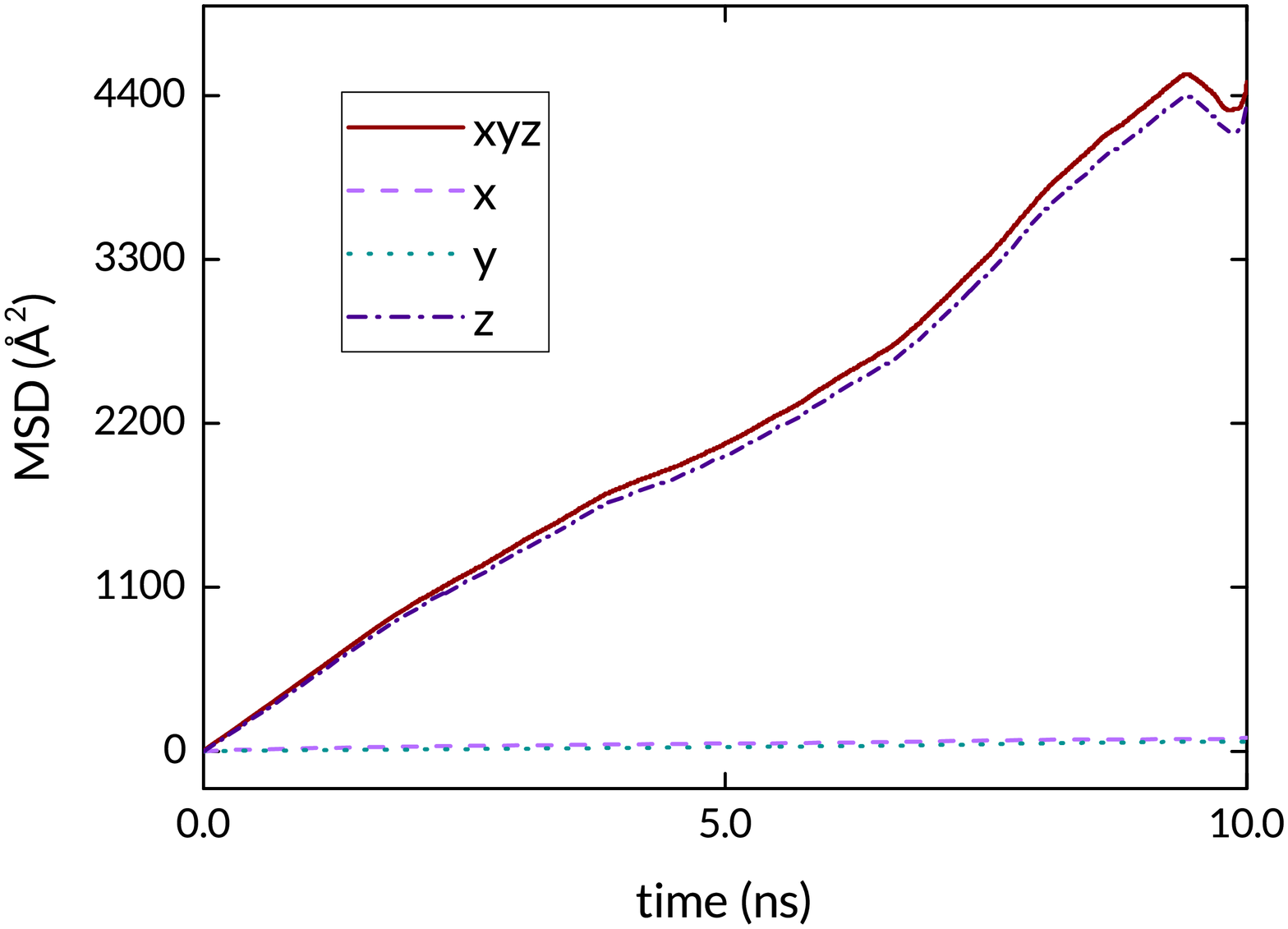}
\caption{\label{fig:tilt_700K}}
\end{subfigure}\quad
\begin{subfigure}[b]{0.45\textwidth}
\includegraphics[width=\textwidth]{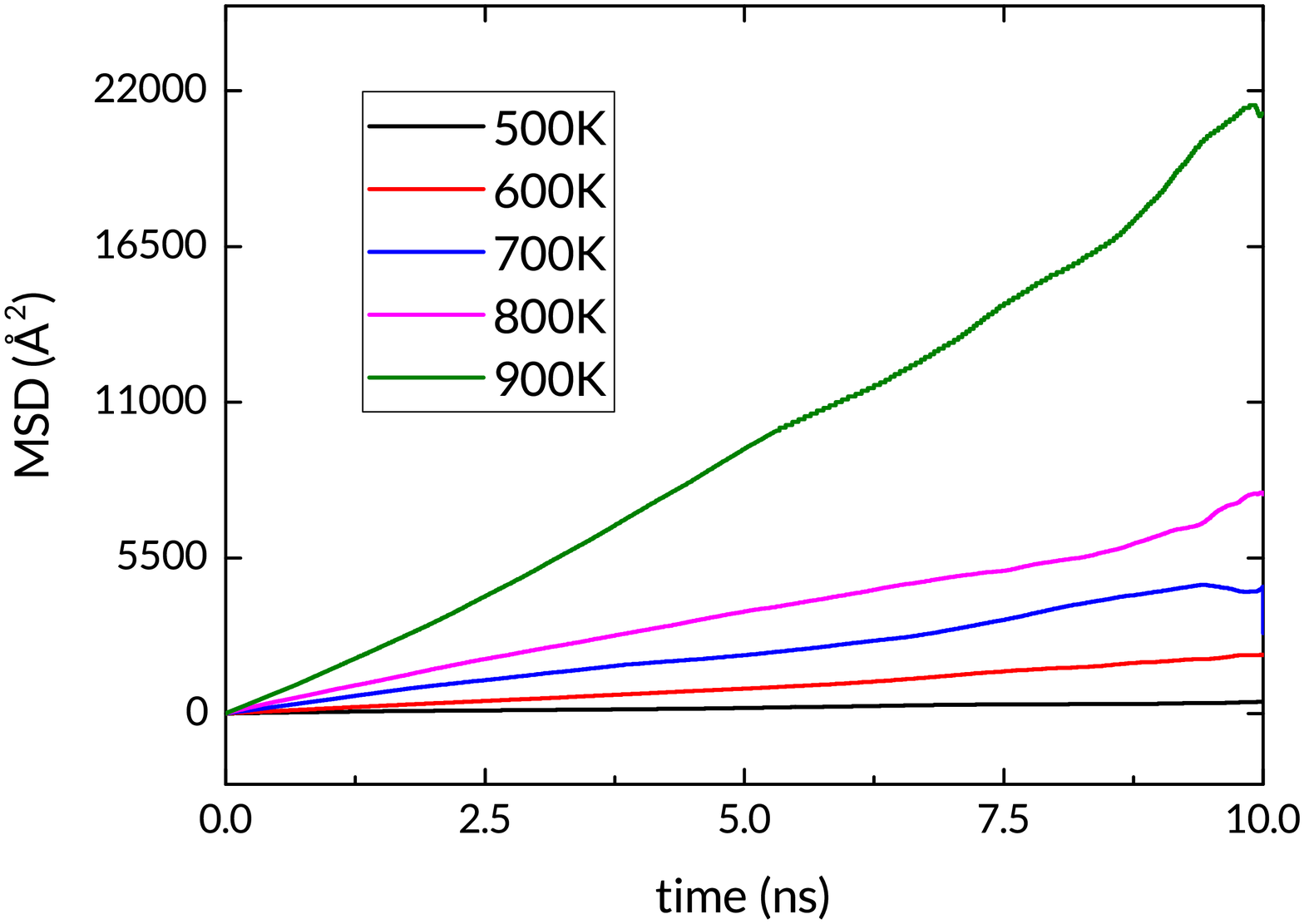}
\caption{\label{fig:tilt_MSD}}
\end{subfigure}\\
\begin{subfigure}[b]{0.225\textwidth}
\includegraphics[width=\textwidth]{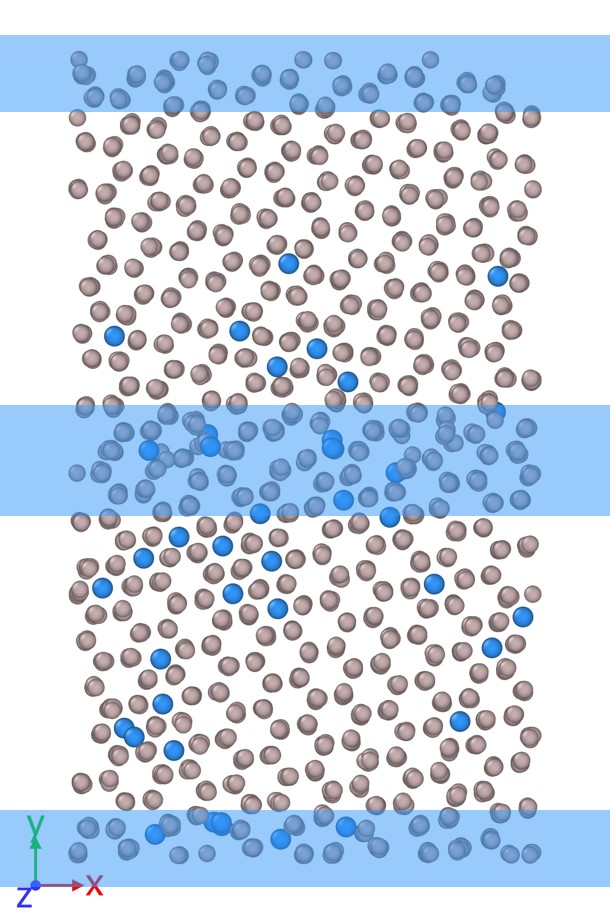}
\caption{\label{fig:tilt_early}}
\end{subfigure}
\begin{subfigure}[b]{0.225\textwidth}
\includegraphics[width=\textwidth]{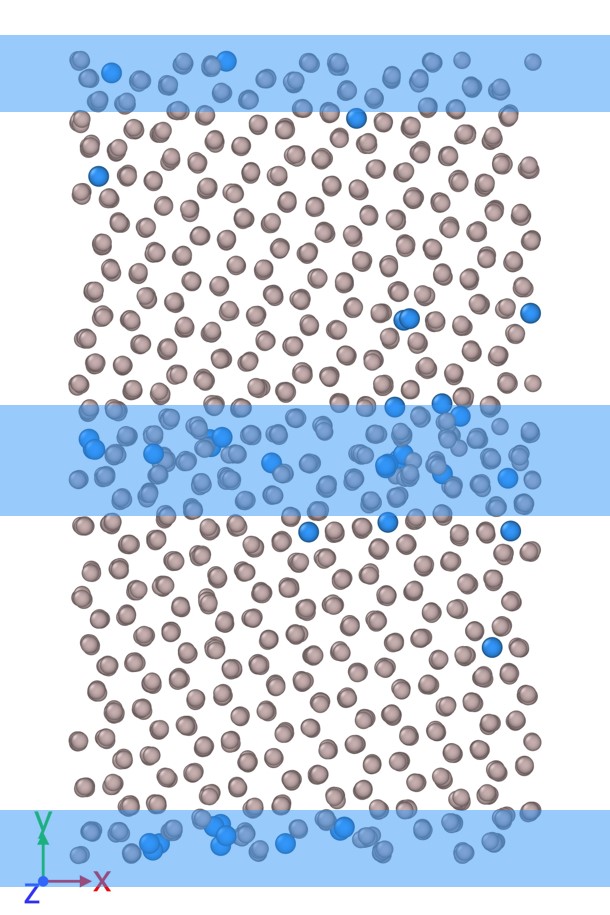}
\caption{\label{fig:tilt_late}}
\end{subfigure}\quad
\begin{subfigure}[b]{0.45\textwidth}
\includegraphics[width=\textwidth]{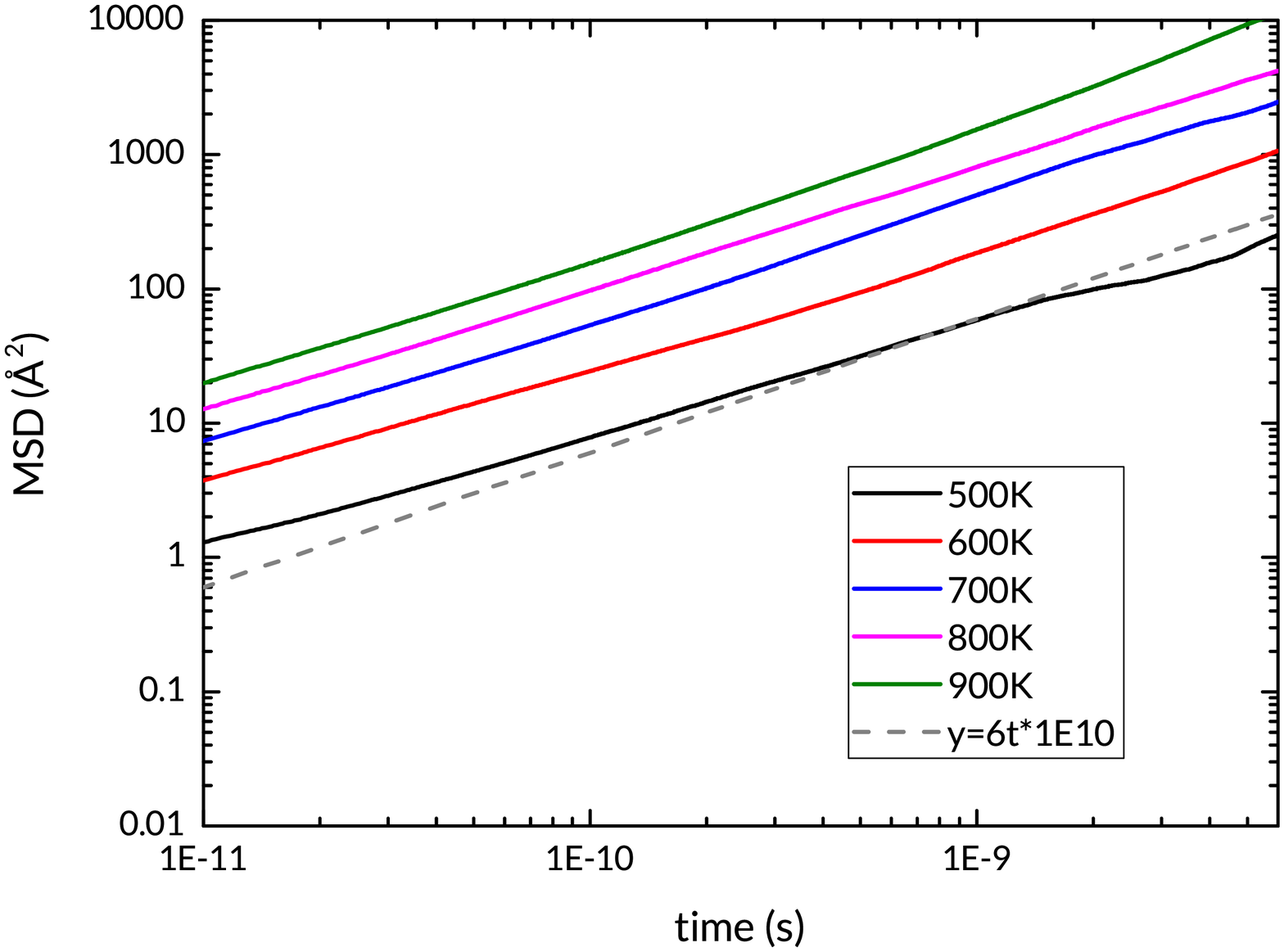}
\caption{\label{fig:tilt_diff_regime}}
\end{subfigure}
\caption{(a) \& (b) H trajectories in tilt GB simulation at 700 K; (c) MSD at 700K; (d) MSD for all simulated temperatures; snapshots of (e) start and (f) end of tilt GB NVT simulation at 700 K, blue bands show GBs; (g) log-log MSD plot for all simulated temperatures of tilt GB. The dashed line has a slope of 1, which indicates a diffusive regime.\label{fig:tilt_GB_traj}}
\end{figure}

The introduction of a tilt grain boundary has a marked effect on the rate and direction of H transport in Ru. As the trajectories in Figure \ref{fig:tilt_GB_traj} show, here too the extent of the unwrapped trajectories in the XY plane is much smaller than in the Z direction. In each of the grains, the predominance of jumps along the Z direction remains, and is overall enhanced within the GB. Here the atoms end up in channels, within which they remain, travelling mainly along the Z direction. The MSD over the same duration as the pristine structure is doubled. Furthermore, it can be seen that there is a much smaller, essentially negligible contribution from diffusion in the \emph{xy} plane.

\begin{figure}[htbp]
\captionsetup{justification=centering}
\centering
\begin{subfigure}[b]{0.45\textwidth}
\includegraphics[width=\textwidth]{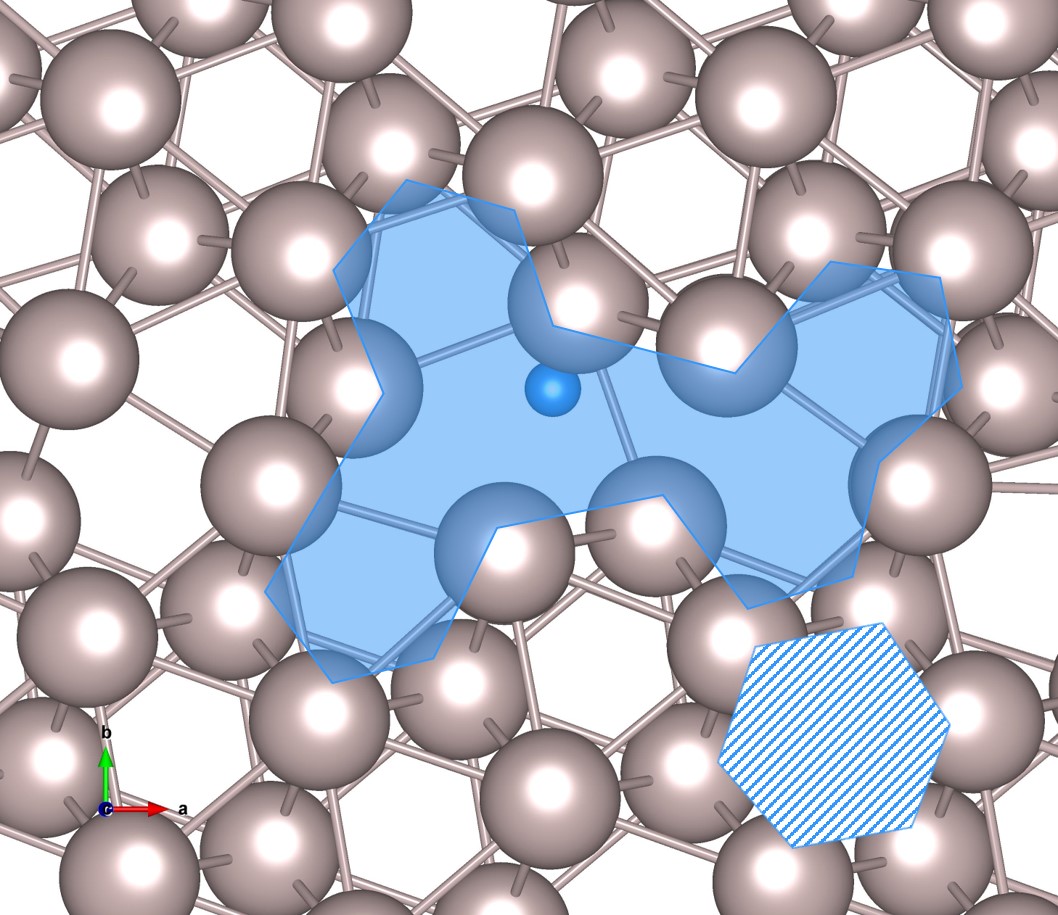}
\caption{\label{fig:tilt_GB_channel}}
\end{subfigure}
\quad
\begin{subfigure}[b]{0.45\textwidth}
\includegraphics[width=\textwidth]{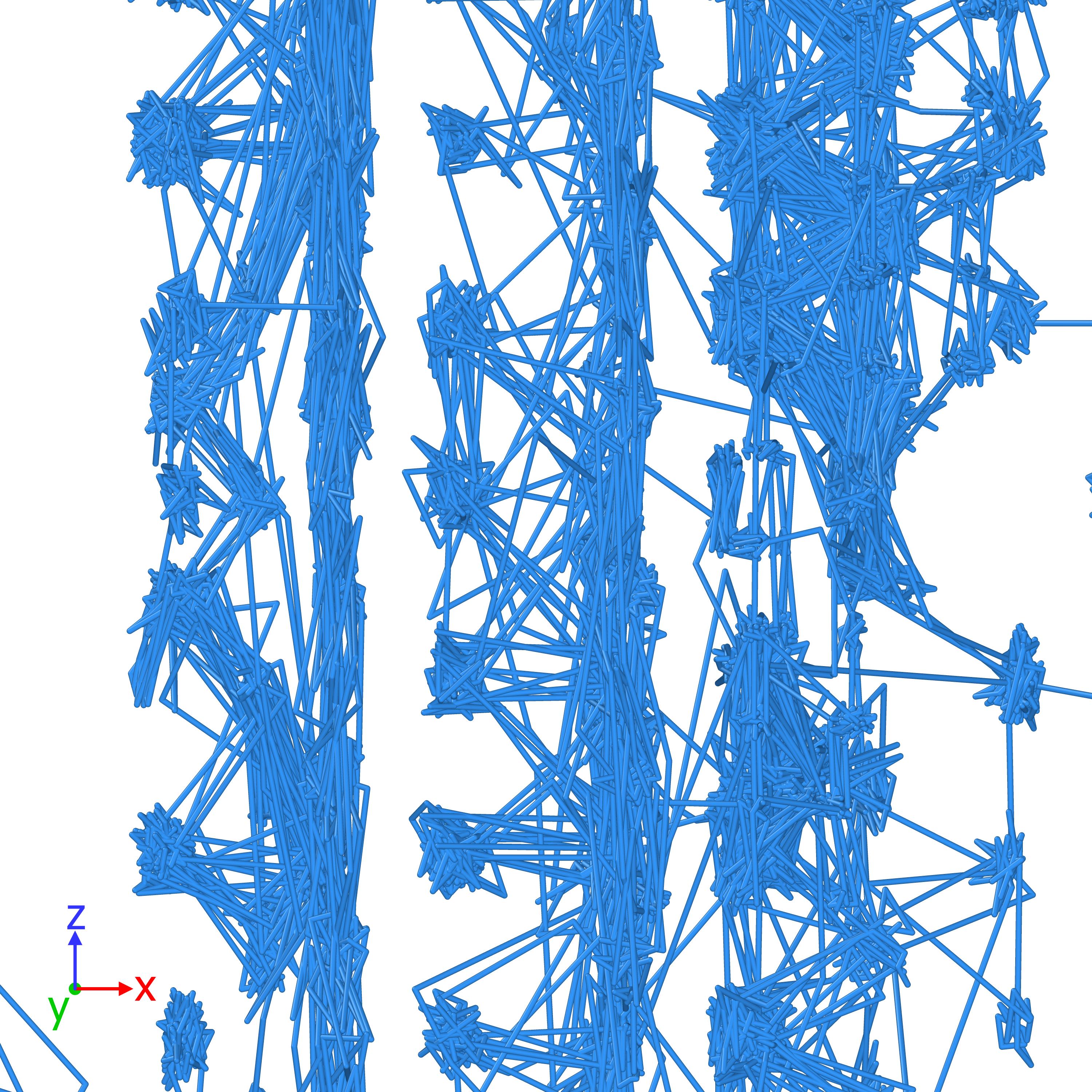}
\caption{\label{fig:tilt_xz}}
\end{subfigure}
\caption{(a) Top view of H channel at tilt GB plane: the shaded hexagon shows the span of the octahedral site, while the semi-transparent irregular polygon shows the Ru-depleted GB channel; and (b) enlarged side view of tilt GB trajectory. \label{fig:tilt_GB_sites}}
\end{figure}

The low energy of the GB sites (Figure \ref{fig:defect_valid}) suggests that the H atoms will tend to be trapped at these sites, and this is reflected in the trajectory maps of Figure \ref{fig:tilt_GB_traj}. There is little transport across the plane of the grain boundary, from one grain to the other. It follows that the equilibrium diffusive regime in this structure is reached only when the population of these GB sites stabilises. Figures \ref{fig:tilt_early} and \ref{fig:tilt_late} show the distribution of H atoms in the structure at the beginning and end of the NVT simulation. It can be seen that the interstitial sites in the grains are depleted, with the H population at the boundary rising accordingly. Figure \ref{fig:tilt_diff_regime} is a log-log plot of the MSD for all the simulated temperatures. It shows that the diffusive regime is reached quickly at the higher temperatures, with all the plots reaching a slope of unity by 500 ps; the exception is the 500K simulation which shows a rough match. This is likely because at such relatively low temperature (a) the H distribution is still out of equilibrium, and/or (b) the number of diffusion events being averaged is small.
A closer view is shown in Figure \ref{fig:tilt_GB_sites} in which the region with the greatest density of trajectory lines can be seen. The tilt GB has channels with larger volume than the native hcp lattice allows, which explains their accommodation of the solute atoms. The image suggests a significant difference in the energy barriers for hops between sites within the channels, and the barrier to exit, with the latter being higher. Figure \ref{fig:tilt_xz} shows a close-up of the side view of the tilt GB, in which 3 separate channels can be demarcated. No hopping between the middle and leftmost channel can be seen in the depicted region.
\FloatBarrier
\subsection{H diffusion in Ru with twist GB}
\begin{figure}[!h]
\captionsetup{justification=centering}
\centering
\begin{subfigure}[b]{0.45\textwidth}
\includegraphics[width=\textwidth]{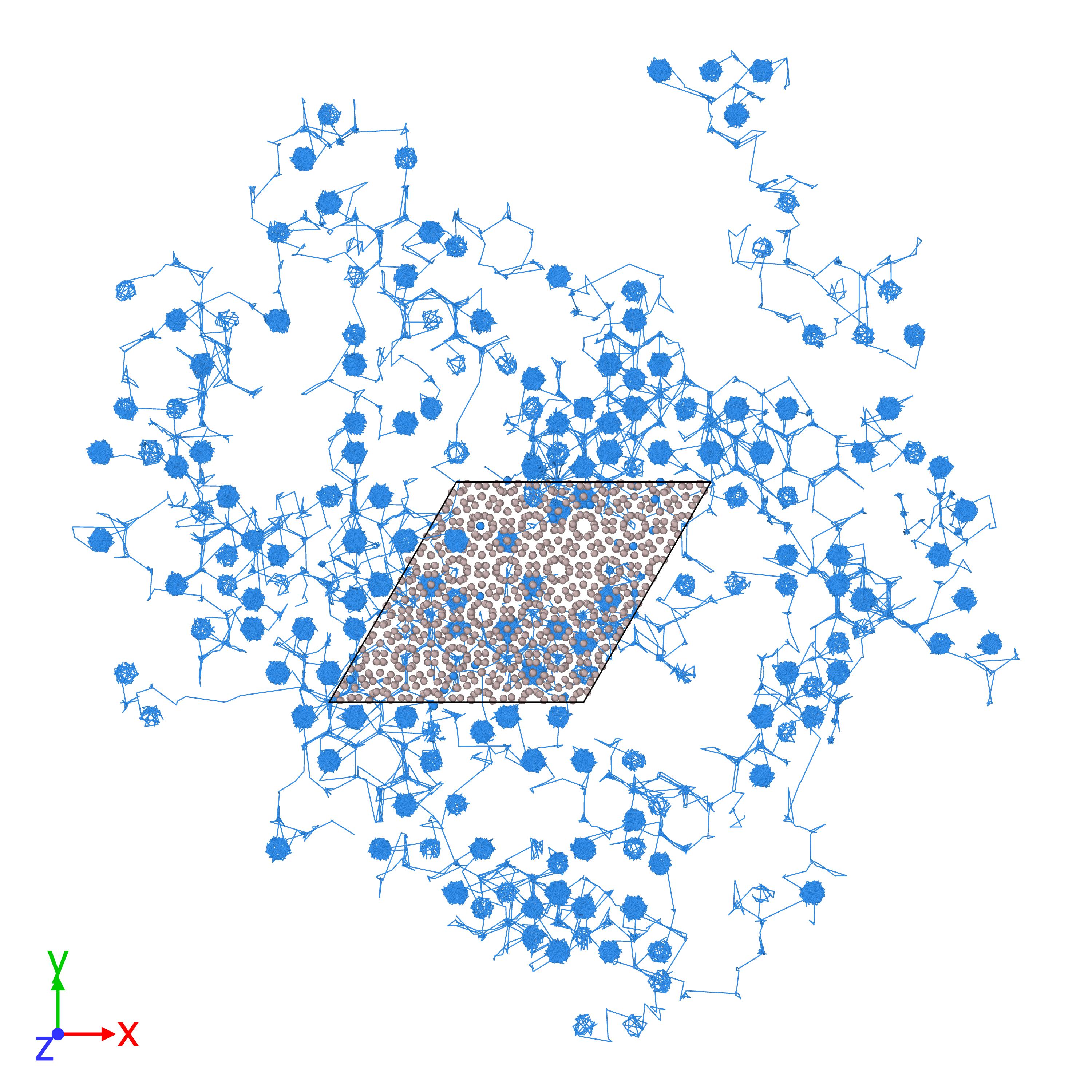}
\caption{\label{fig:twist_xy}}
\end{subfigure}
\quad
\begin{subfigure}[b]{0.45\textwidth}
\includegraphics[width=\textwidth]{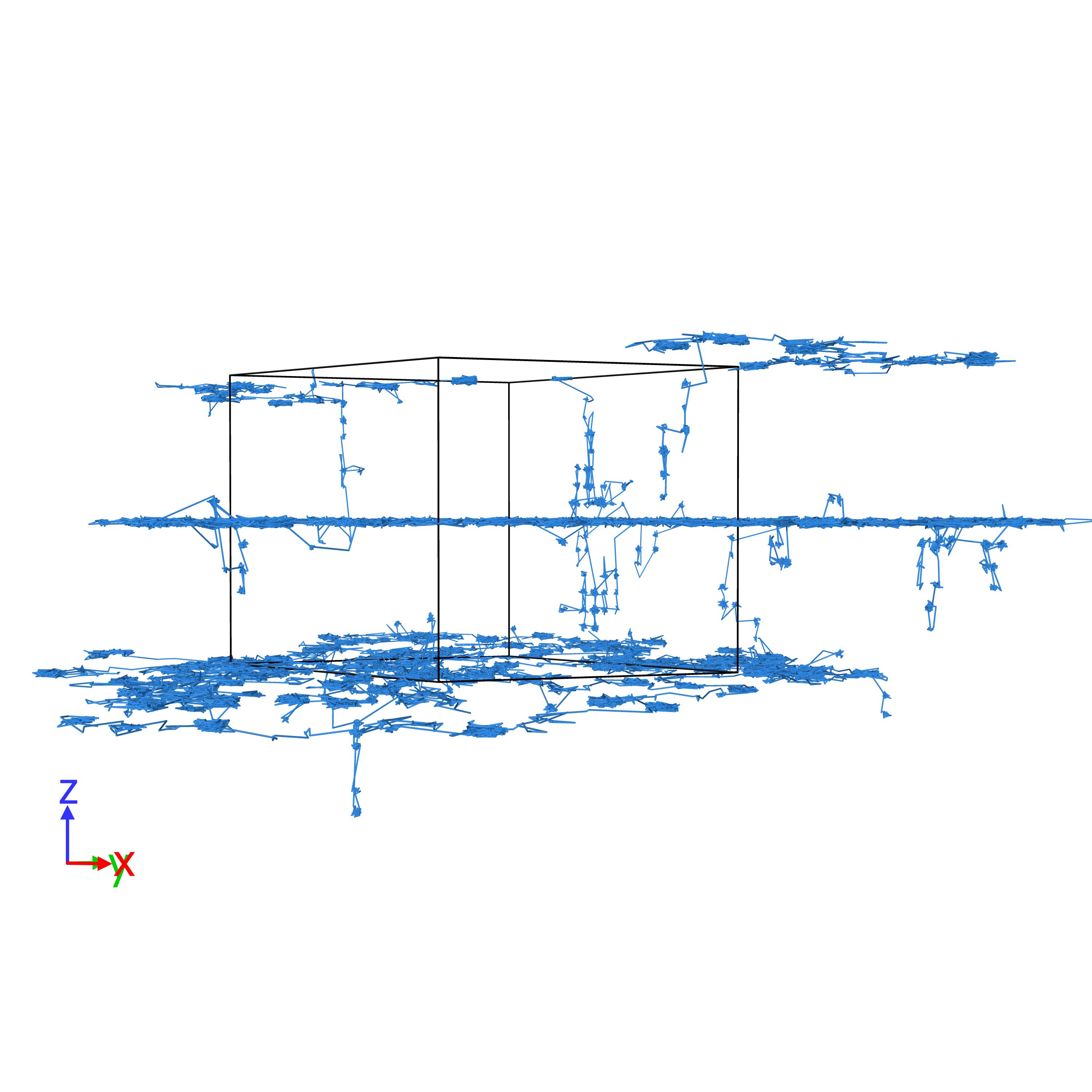}
\caption{\label{fig:twist_3D}}
\end{subfigure} \\
\begin{subfigure}[b]{0.45\textwidth}
\includegraphics[width=\textwidth]{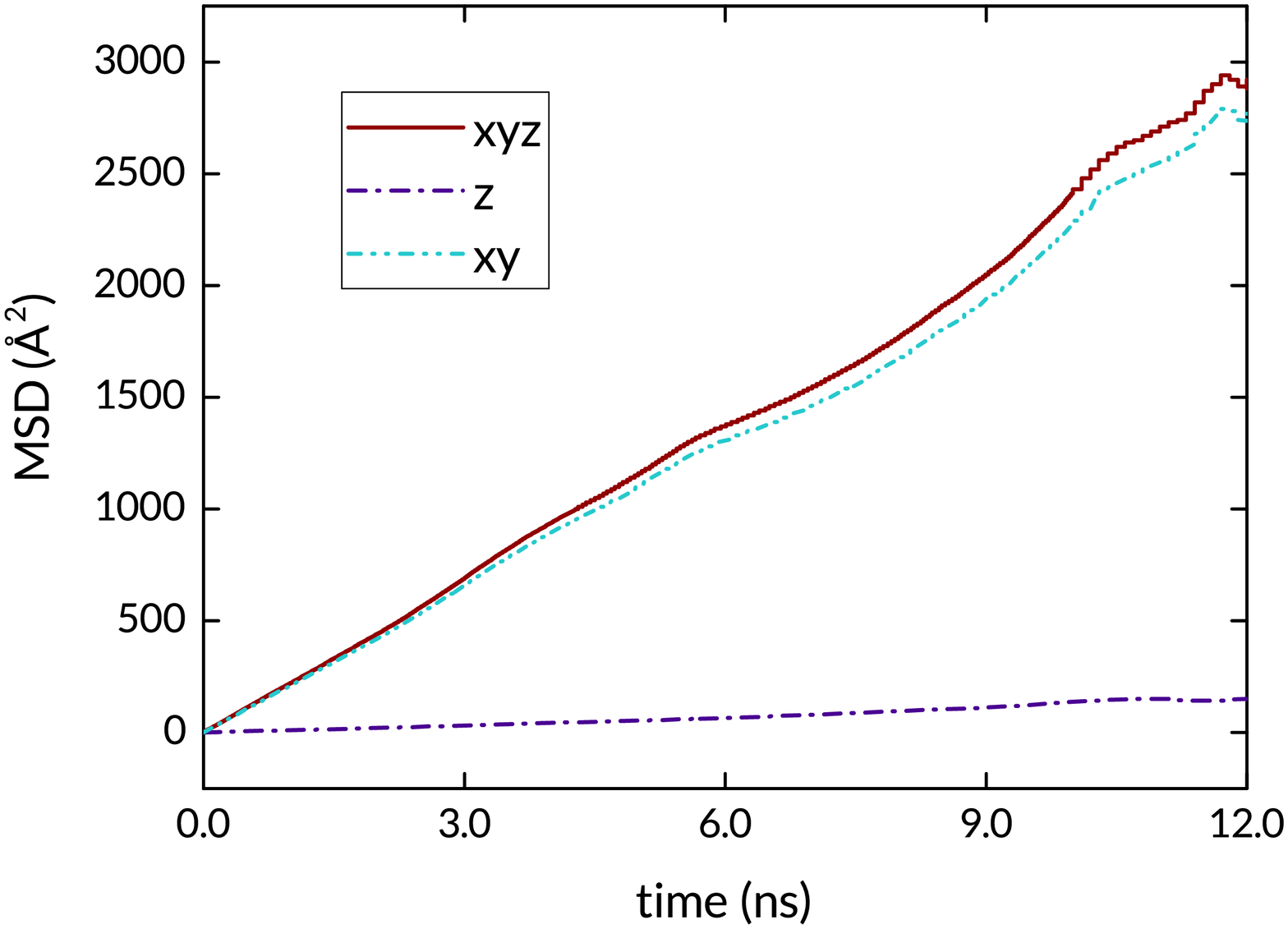}
\caption{\label{fig:twist_700K}}
\end{subfigure}
\quad
\begin{subfigure}[b]{0.45\textwidth}
\includegraphics[width=\textwidth]{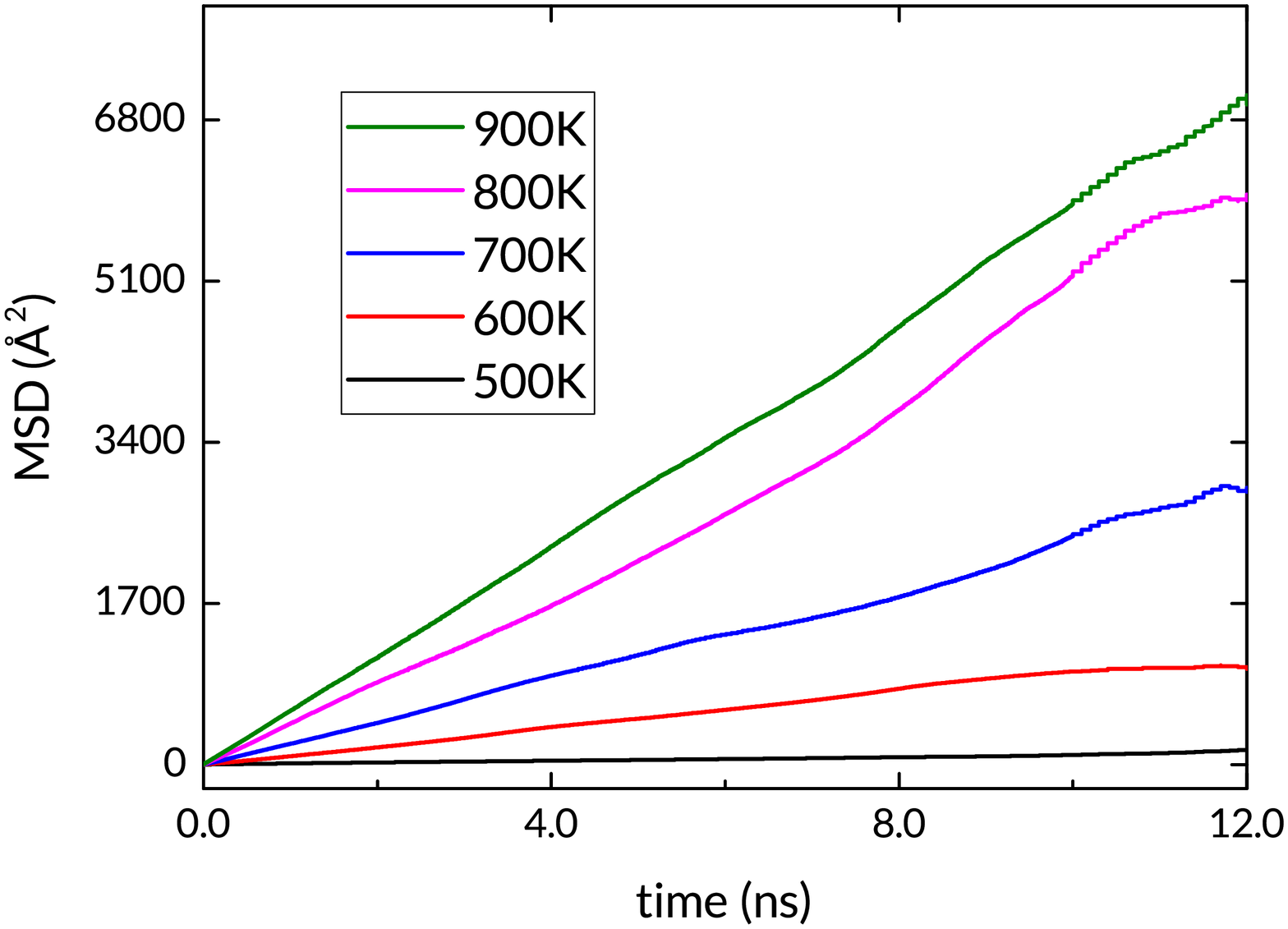}
\caption{\label{fig:twist_MSD}}
\end{subfigure}\\
\begin{subfigure}[b]{0.225\textwidth}
\includegraphics[width=\textwidth]{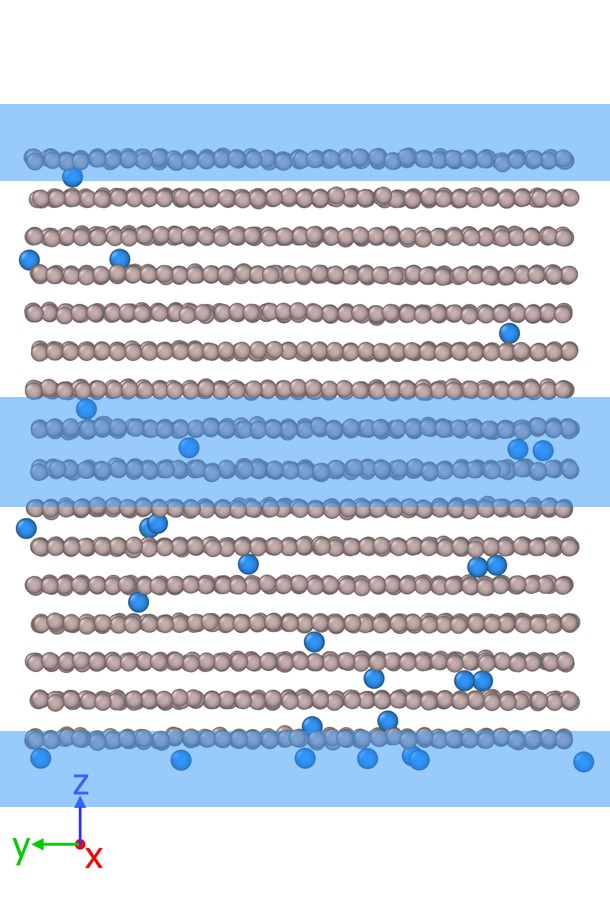}
\caption{\label{fig:twist_early}}
\end{subfigure}
\begin{subfigure}[b]{0.225\textwidth}
\includegraphics[width=\textwidth]{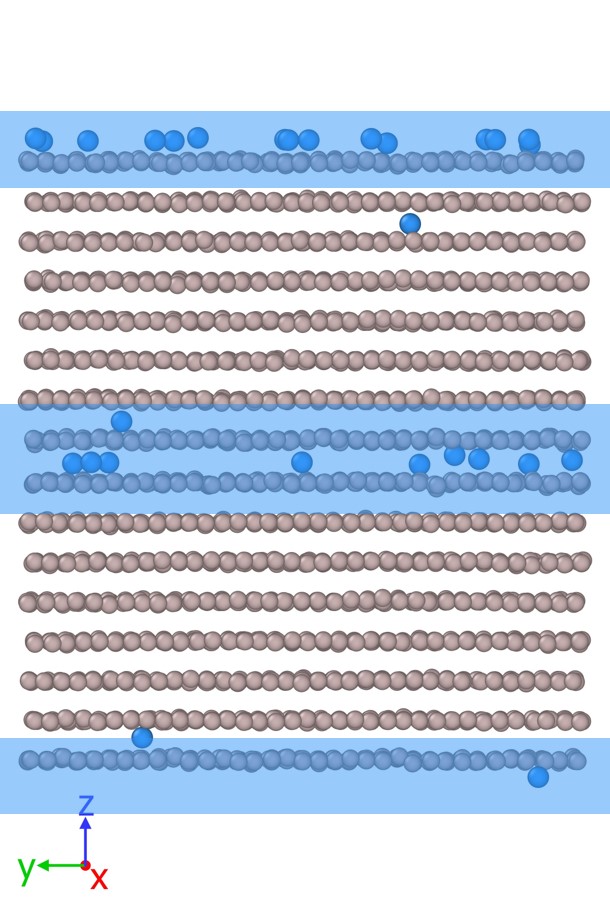}
\caption{\label{fig:twist_late}}
\end{subfigure}\quad
\begin{subfigure}[b]{0.45\textwidth}
\includegraphics[width=\textwidth]{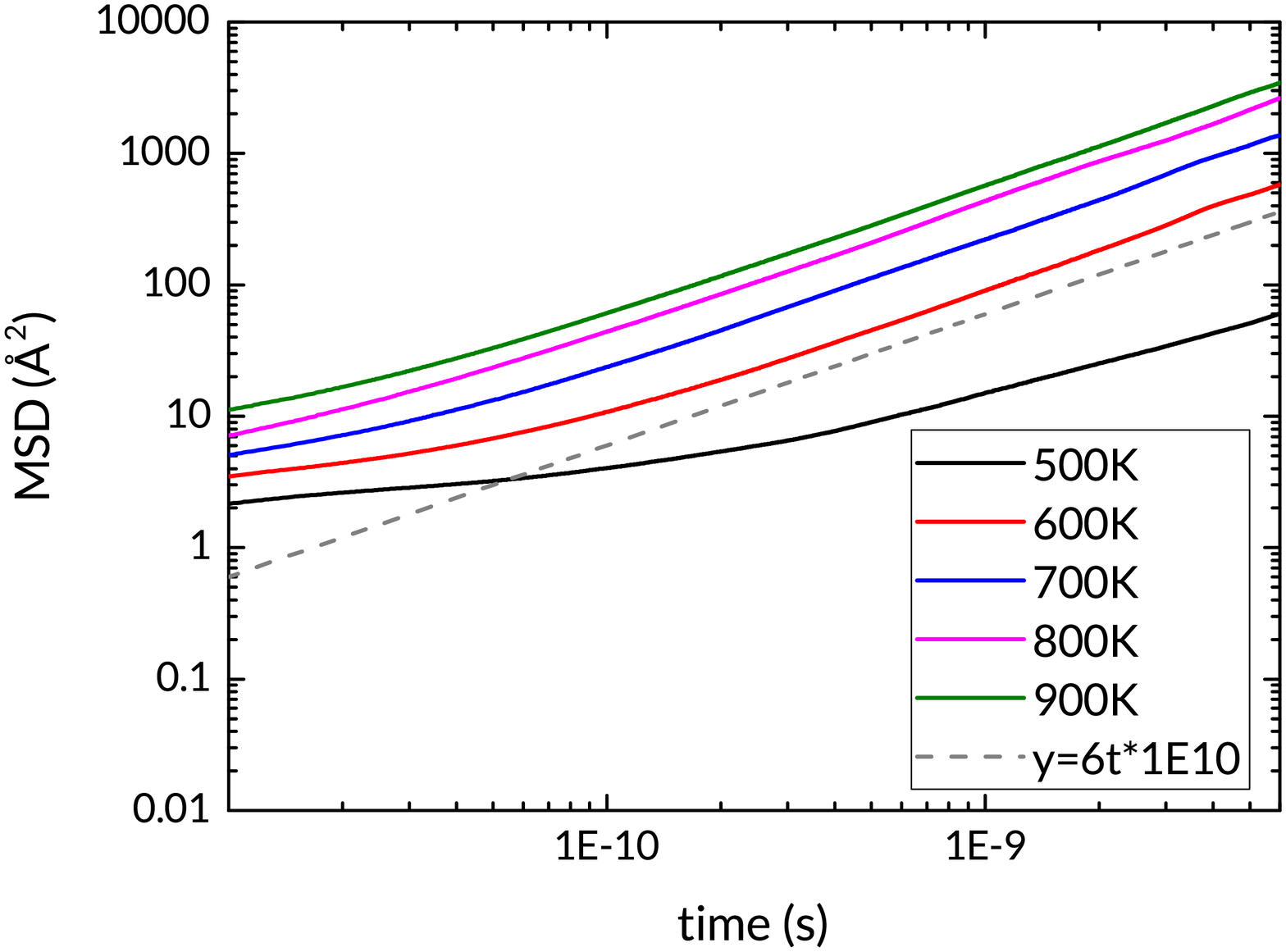}
\caption{\label{fig:twist_diff_regime}}
\end{subfigure}
\caption{(a) \& (b) H trajectories in twist GB simulation at 700 K; (c) MSD at 700K; (d) MSD for all simulated temperatures; snapshots of (e) start and (f) end of tilt GB NVT simulation at 700 K, blue bands show GBs; (g) log-log MSD plot for all simulated temperatures of twist GB. The dashed line has a slope of 1, which indicates a diffusive regime.\label{fig:twist_GB_traj}}
\end{figure}

The second grain boundary structure also influences the diffusion process significantly. The plane of the boundary is horizontal, and the sites in the boundary put H atoms lower in energy than the octahedral site. Figure \ref{fig:twist_GB_traj} shows the H trajectories, which lie predominantly in the plane of the boundary. The movement of H atoms at the twist GB sites contrasts strongly with the tightly-bound vibrations at the octahedral site; the nodes of the latter are quite small, compared to the broadly-smeared loci at the former.
Most importantly for the overall transport, this structure obstructs the otherwise dominant diffusion along the Z direction, with the horizontal reach of the trajectory now greater. The MSD plot reflects this, with the Z contribution practically zero, while the overall MSD magnitude is halfway between the values for the pristine structure and the tilt GB. The diffusion in the grains ultimately leads the solute atoms to accumulate at the boundary (Figure \ref{fig:twist_late}). This proceeds at a temperature-dependent rate. The NVT simulation at 500 K does not reach the slope of a random-walk diffusive regime within the 12 ns duration of the run; all the higher-temperature simulations reach this by the 1 ns mark (Figure \ref{fig:twist_diff_regime}). 
\begin{figure}[htbp]
\captionsetup{justification=centering}
\centering
\begin{subfigure}[b]{0.45\textwidth}
\includegraphics[width=\textwidth]{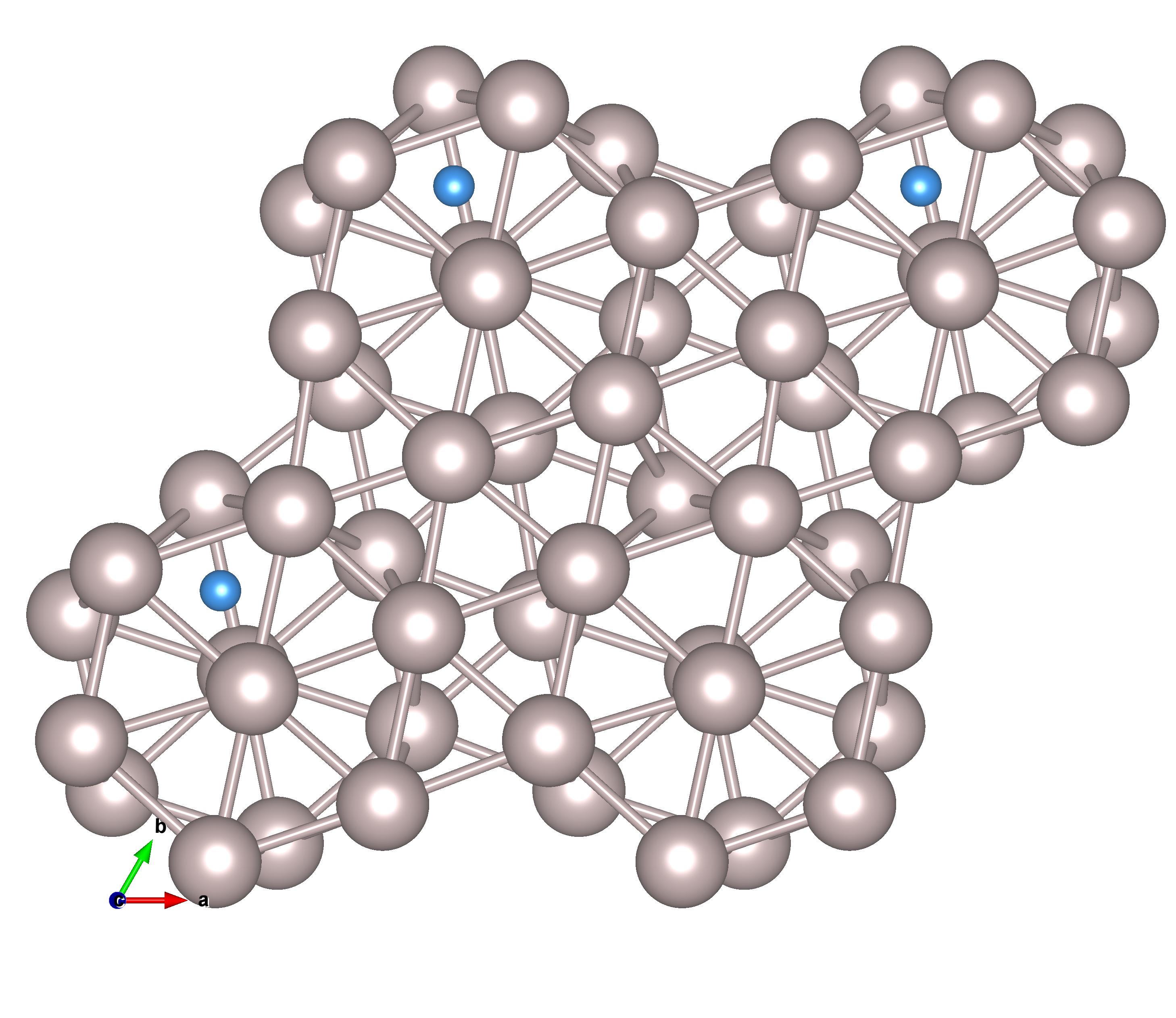}
\caption{\label{fig:twist_GB_plane}}
\end{subfigure}
\quad
\begin{subfigure}[b]{0.45\textwidth}
\includegraphics[width=\textwidth]{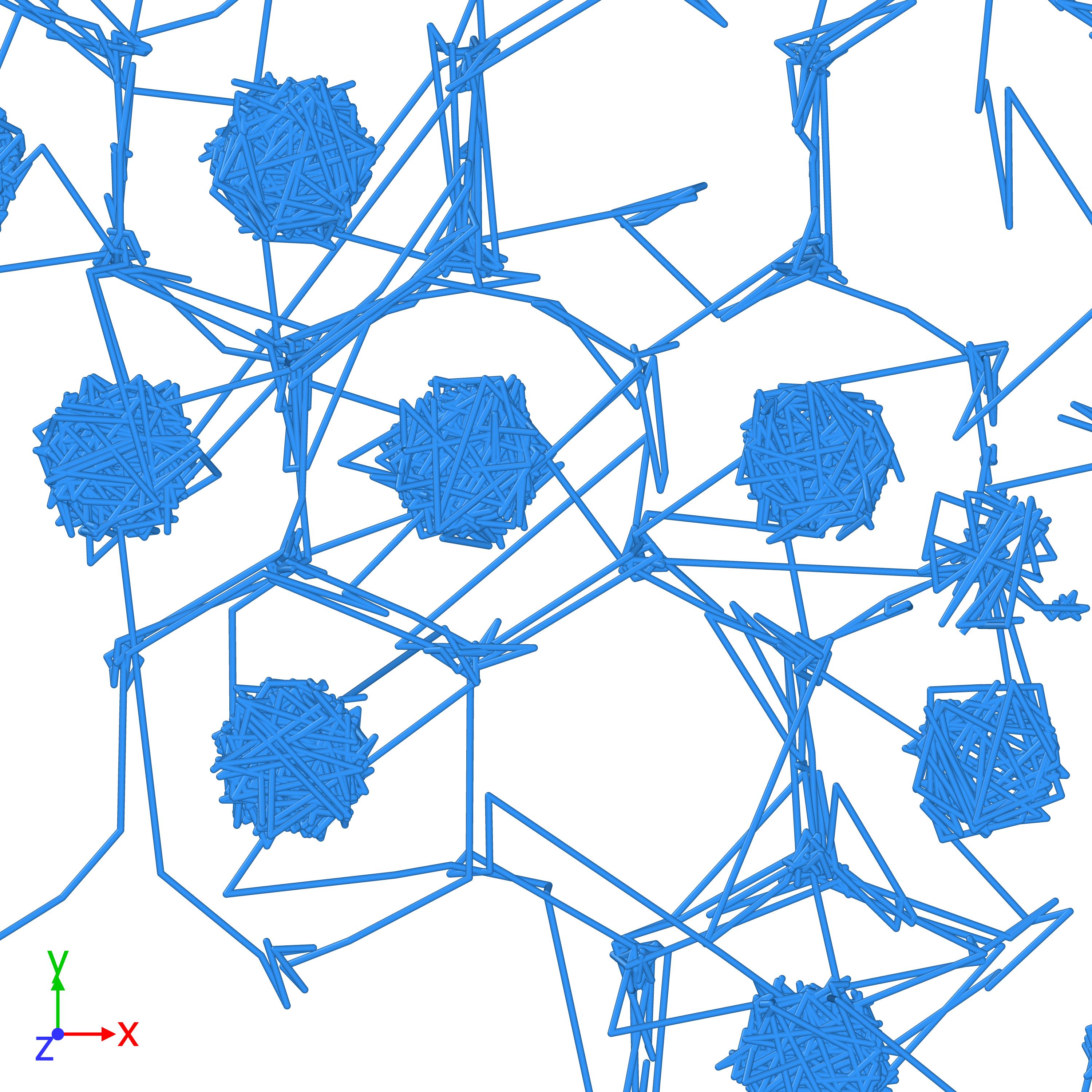}
\caption{\label{fig:twist_wheel}}
\end{subfigure}
\caption{(a) H sites at twist GB plane: the wheels are formed by atoms from the interfacing grains; and (b) enlarged top view of twist GB trajectory. \label{fig:twist_GB_sites}}
\end{figure}
Figure \ref{fig:twist_GB_sites} offers a closer look at the GB plane. The trajectory lines show that H atoms have a strong affinity for the sites in the GB plane. The twist GB has formed hexagonal 'wheels' between the top and bottom grains, within which the H atom moves, jumping intermittently to an identical neighbouring region. These traps show up in the trajectory lines of Figure \ref{fig:twist_xy} as the large nodes. The hexagonal symmetry of the interfacing grains is reflected in the paths around and between these traps.
\FloatBarrier
\subsection{Diffusion coefficients}
\begin{figure}[h]
\captionsetup{justification=centering}
\centering
\includegraphics[width=0.6\textwidth]{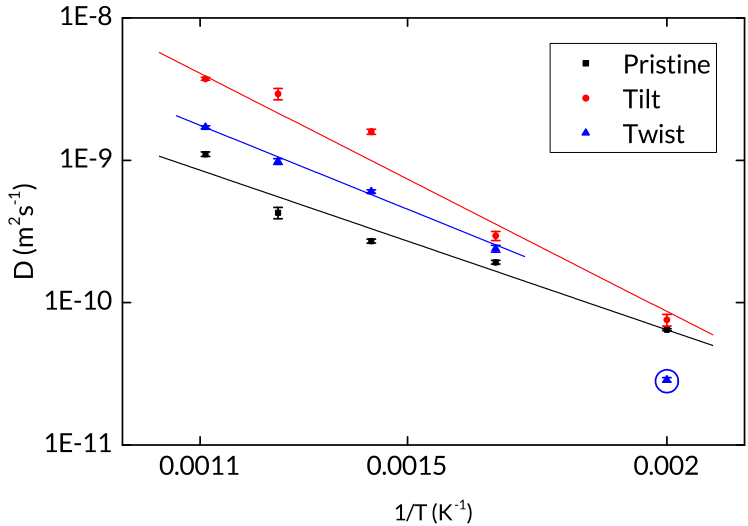}
\caption{Diffusion coefficients for all simulated structures and temperatures; the point circled in blue is excluded from the fit, as it does not represent a diffusive regime.\label{fig:Arr_plot_last}}
\end{figure}

The MSD for each of the diffusion simulations yields the diffusion coefficient at each temperature, according to Equation \ref{eqn:einstein}. In both GBs, the diffusion is mostly restricted to the plane, such that the random-walk dimensionality is reduced to 2. Figure \ref{fig:pristine_MSD} shows a linear dependence of MSD on time for the entire duration of the simulation of the pristine Ru. However, the inhomogeneous GB structures do not reach a diffusive regime as rapidly. They also exhibit greater variation in the slope. Therefore the MSD slope is taken only after 40\% of the simulation time has elapsed, and the slope of the MSD has matched that of a random walk. Furthermore, to account for the noise, the diffusion coefficient is averaged from 10 overlapping intervals between the 40\% mark and the end of the simulation. The diffusion coefficients are plotted in Figure \ref{fig:Arr_plot_last}.
We fit the temperature-dependent diffusion coefficients to the Arrhenius expression. As shown in Figure \ref{fig:twist_diff_regime}, the 500K twist GB simulation does not reach the diffusive regime; the diffusion coefficient is therefore not included in the fit. \cmmnt{The fit is weighted according to the error of each diffusion coefficient. }We obtain a pre-exponential factor $D_0$ and activation energy $E_a$ for temperature-dependent H diffusion in each of the simulated structures. These are, in m\textsuperscript{2}s\textsuperscript{-1} and eV respectively: $2.0\times10^{-8}$ and $0.25$  for the pristine hcp Ru; $4.5\times10^{-7}$ and $0.37$ for the tilt GB structure; and $7.3\times10^{-8}$ and $0.29$ for the twist GB structure. \cmmnt{The activation energy for the pristine hcp Ru is lower than the $0.54$ eV given by a transition state calculation for a single atom hop between octahedral sites. This may indicate that the other diffusion paths are contributing, i.e. tetrahedral-to-tetrahedral and tetrahedral-octahedral-tetrahedral. The H-H interaction may also be contributing, as the concentration in the simulations is far from the dilute limit.}

\section{Conclusion}
We have developed a ReaxFF force field for the Ru/H system which reproduces DFT energies with high accuracy. We have applied the force field to the study of H diffusion in Ru, a topic previously under-represented in literature. We performed simulations of H diffusion though a perfect Ru crystal and through tilt and twist GBs, which have yielded diffusion coefficients for H in hcp Ru crystal and in GBs. While they do not cover all possible GBS and defects which can influence H transport through polycrystalline Ru, the diffusion coefficients and the trajectory maps indicate that the character of H diffusion in Ru depends largely on the number and nature of GBs present.

Both the static calculations and the dynamic simulations show the presence of energetically favourable sites for H atoms in the boundary region. Also important is the fact that diffusion across the GBs is inhibited. These findings are similar to the results obtained for H in Al GBs\cite{Pedersen2009SimulationsAluminum}. However, although Pedersen et al. observed diffusion through H hopping from the GB site out into the grain, we find that the main trajectories lie within the GB. The diffusion coefficients we have extracted imply that at 300K, the diffusion rates in the tilt and twist GBs are slightly lower than that of the perfect crystal. We can surmise that in moving through polycrystalline Ru, H atoms will hop between interstitial sites until they reach a GB, within which they become confined. If the sites at the GB are occupied, an arriving H atom is repelled, as we observed minimal transport across GBs. For thin films, transport through the Ru will depend greatly on the morphology of the film. Since the Ru capping layers are mostly of (0001) orientation\cite{Bajt2008PropertiesIrradiation}, the accumulation of H and preferential transport along the plane of the tilt GB suggests that this type of GB will dominate H diffusion through the films, enabling the so-called short-circuit diffusion through the film. 

These results point to film morphology control as an important tool in preventing the permeation of hydrogen into multilayer mirrors. The results also give insight into the trapping and diffusion of H and other impurities in metal grain boundaries. The developed force field can be applied to the study of other phenomena, including surface interactions, while the results of the diffusion study will be of interest both for research and for technological applications.

\begin{acknowledgement}

This research was carried out under project number T16010a in the framework of the Partnership Program of the Materials innovation institute M2i (www.m2i.nl) and the Technology Foundation TTW (www.stw.nl), which is part of the Netherlands Organization for Scientific Research (www.nwo.nl).

\end{acknowledgement}

\bibliography{mendeley}
\newpage
\begin{suppinfo}

\setcounter{figure}{0}
\renewcommand{\thefigure}{S\arabic{figure}}

\section{ReaxFF training set}
The fitting of parameters for the force field was done with a training set which includes Ru equations of state for multiple crystal structures (hcp, fcc, bcc, sc); Ru clusters; amorphous Ru; surface formation energies (slabs of hcp Ru, fcc Ru, bcc Ru); H adsorption energies on the same slabs; interstitial hydride formation energies (up to 0.25 H/Ru concentration); and bond length scans(Ru-Ru and Ru-H in RuH\textsubscript{4}). Figures \ref{fig:Ru_set} and \ref{fig:Ru_H_set} show samples of the training set. 
\begin{figure}[!h]
\captionsetup{justification=centering}
\centering
\begin{subfigure}[b]{0.225\textwidth}
\includegraphics[width=\textwidth]{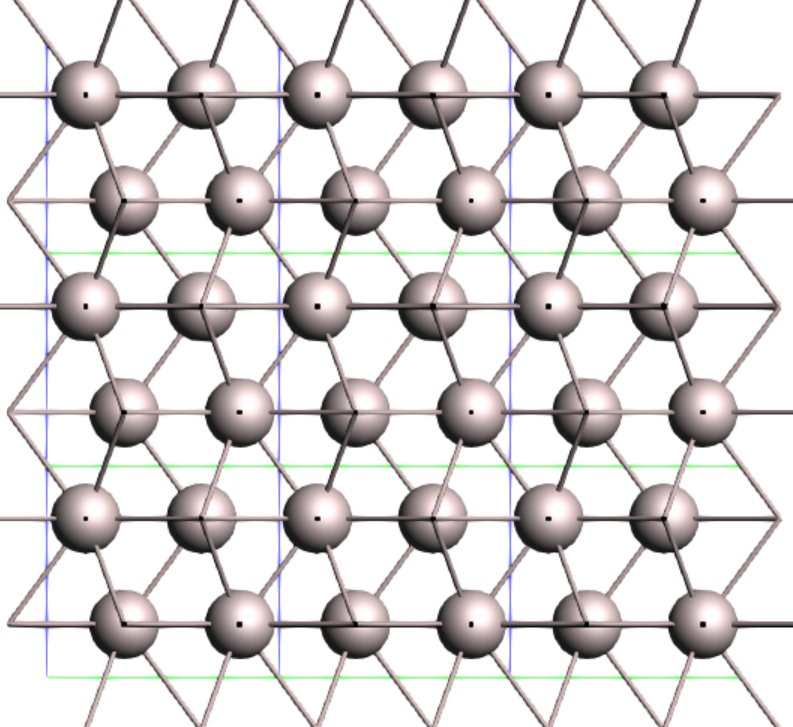}
\caption{HCP}
\end{subfigure}
\begin{subfigure}[b]{0.225\textwidth}
\includegraphics[width=\textwidth]{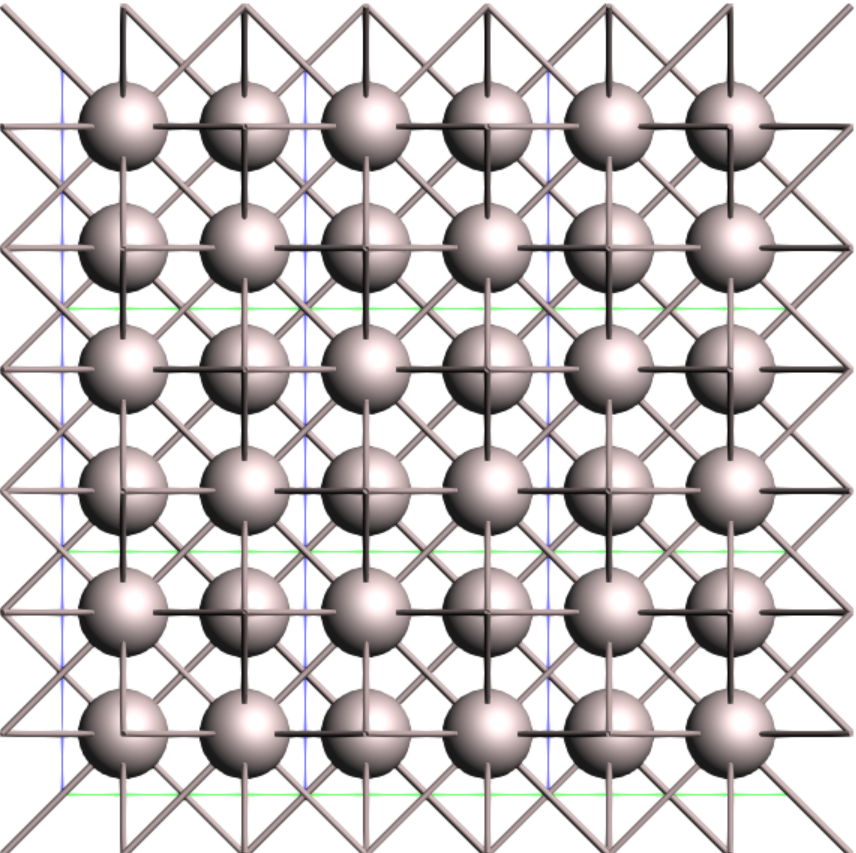}
\caption{FCC}
\end{subfigure}
\begin{subfigure}[b]{0.225\textwidth}
\includegraphics[width=\textwidth]{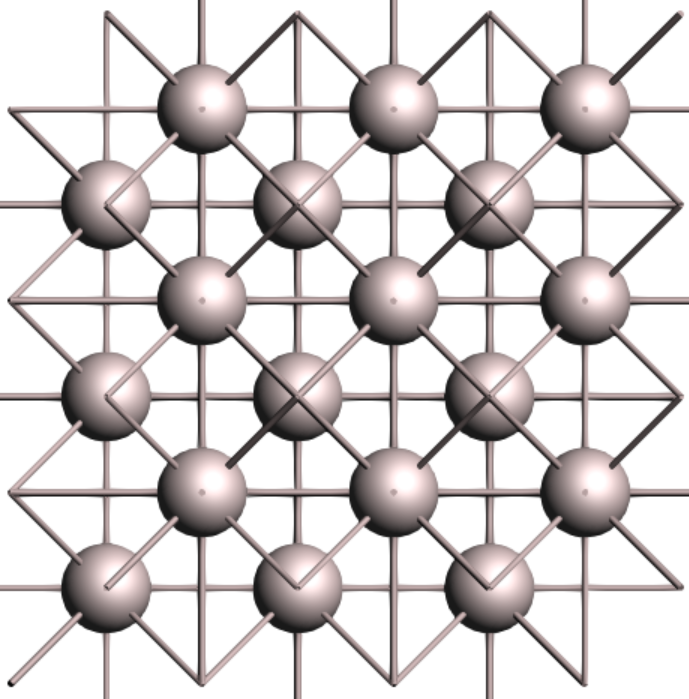}
\caption{BCC}
\end{subfigure}
\begin{subfigure}[b]{0.225\textwidth}
\includegraphics[width=\textwidth]{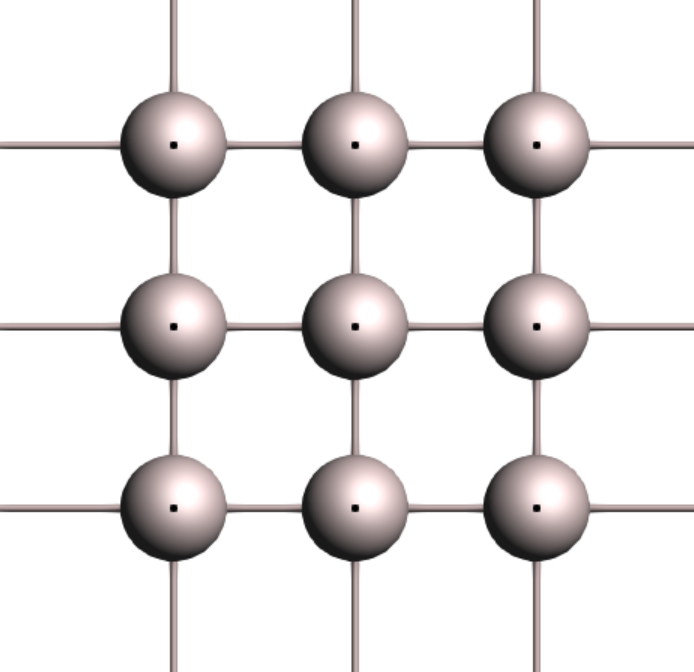}
\caption{SC}
\end{subfigure} \\
~\\
\begin{subfigure}[b]{0.225\textwidth}
\includegraphics[trim={0 0 0 2cm},clip,width=\textwidth]{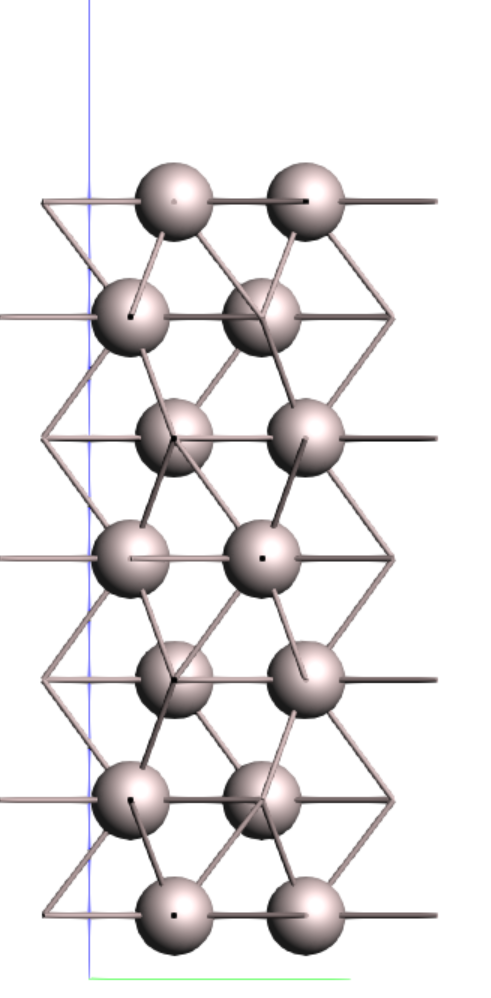}
\caption{Ru(0001) slab}
\end{subfigure}
\begin{subfigure}[b]{0.275\textwidth}
\includegraphics[width=\textwidth]{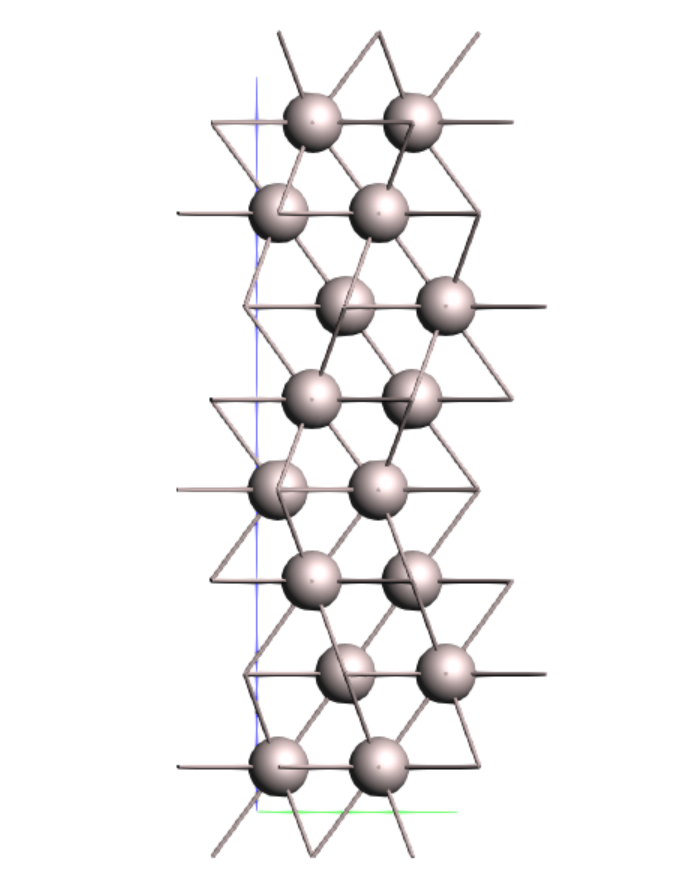}
\caption{fcc111hcp001 stacking fault}
\end{subfigure}
\begin{subfigure}[b]{0.225\textwidth}
\includegraphics[width=\textwidth]{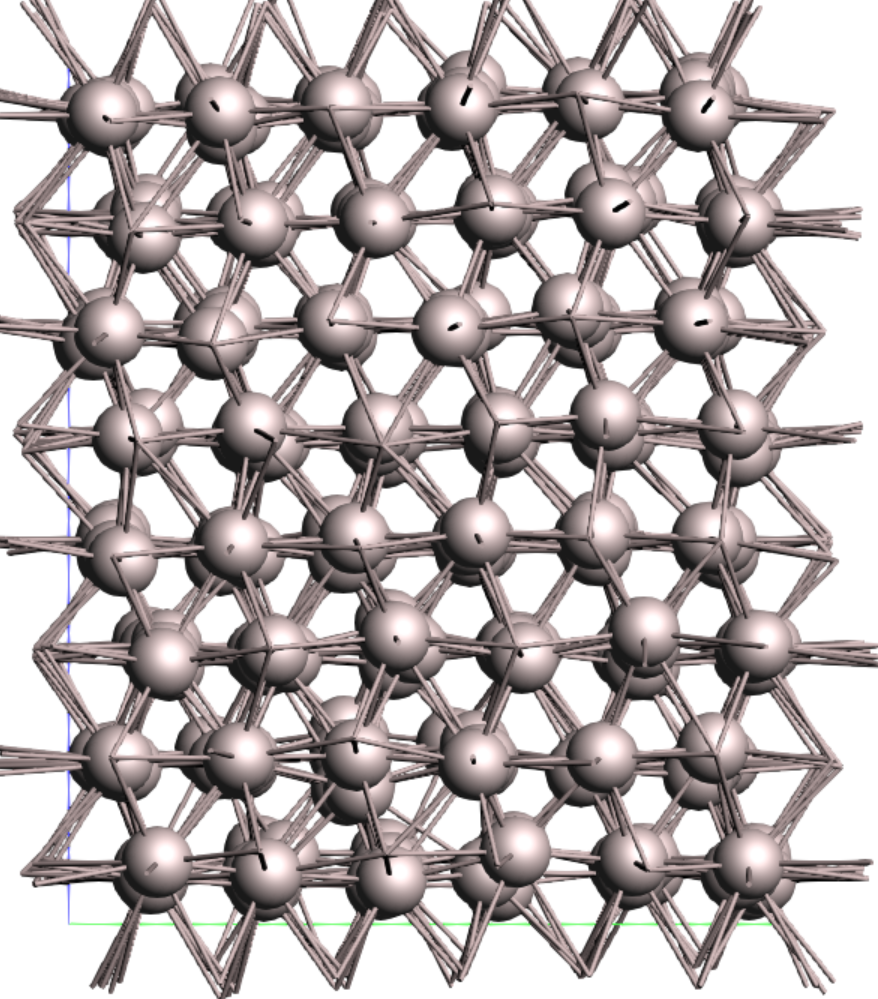}
\caption{High temperature Ru}
\end{subfigure}
\begin{subfigure}[b]{0.225\textwidth}
\includegraphics[width=\textwidth]{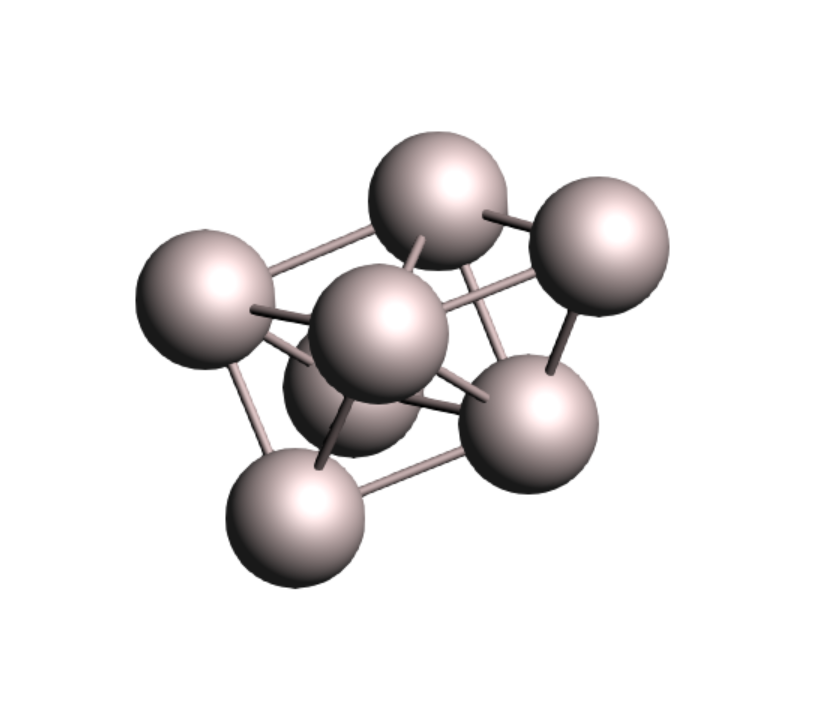}
\caption{Ru cluster}
\end{subfigure}
    \caption{Examples of structures included in the Ru training set.}
    \label{fig:Ru_set}
\end{figure}

\begin{figure}[!h]
\captionsetup{justification=centering}
\centering
\begin{subfigure}[b]{0.225\textwidth}
\includegraphics[width=\textwidth]{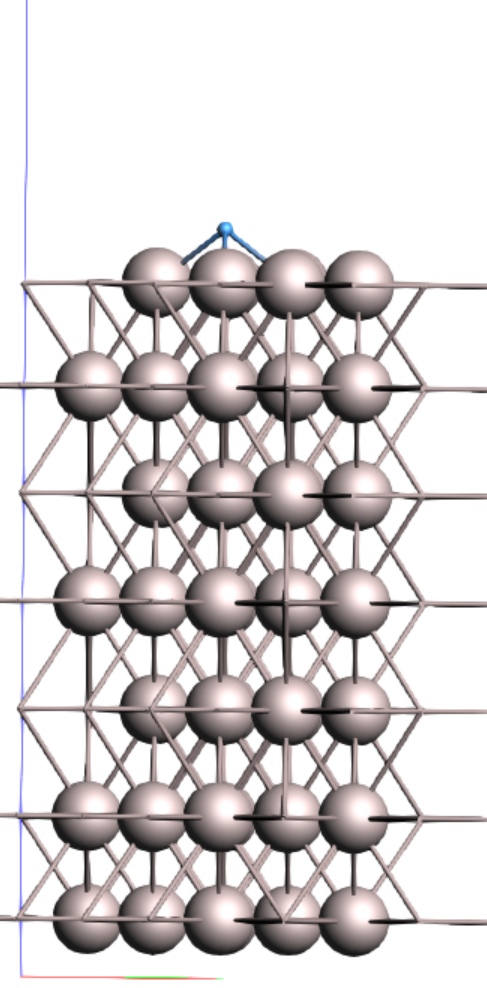}
\caption{H on Ru(0001)}
\end{subfigure}
\begin{subfigure}[b]{0.225\textwidth}
\includegraphics[width=\textwidth]{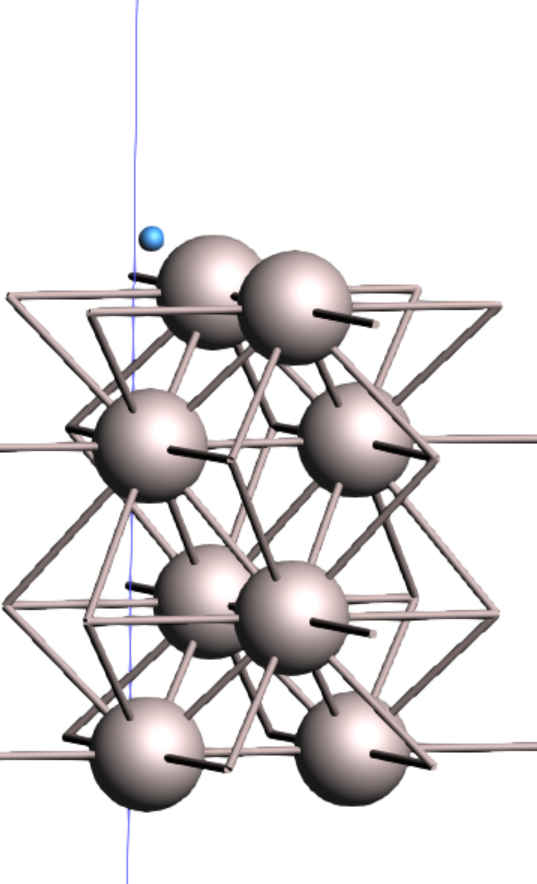}
\caption{H on fcc(100)}
\end{subfigure}
\begin{subfigure}[b]{0.225\textwidth}
\includegraphics[width=\textwidth]{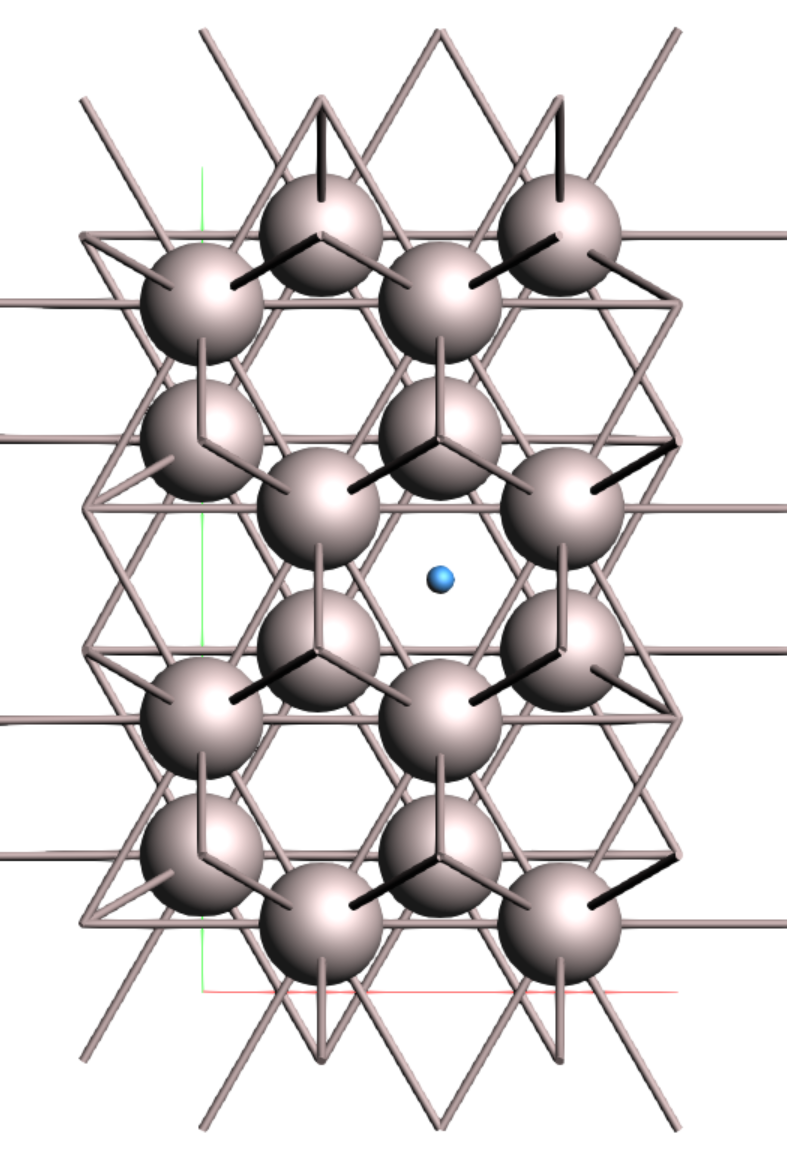}
\caption{H in octahedral site}
\end{subfigure}
\begin{subfigure}[b]{0.225\textwidth}
\includegraphics[width=\textwidth]{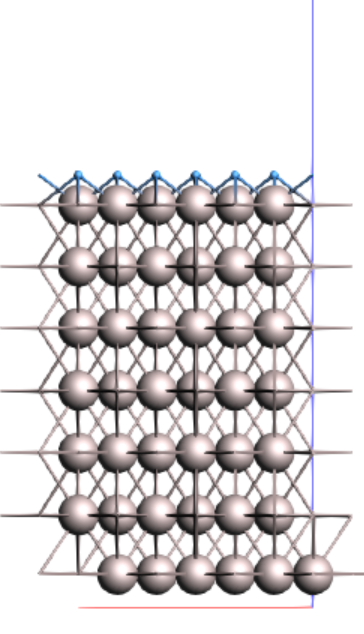}
\caption{1ML H on Ru}
\end{subfigure}\\
~\\
\begin{subfigure}[b]{0.225\textwidth}
\includegraphics[width=\textwidth]{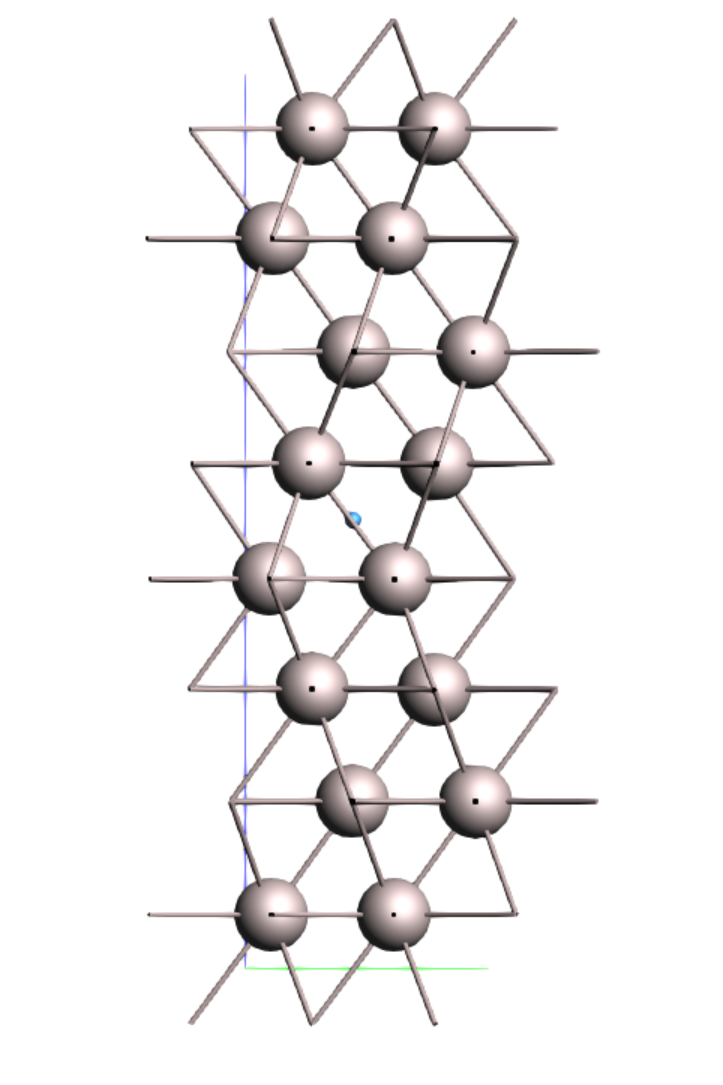}
\caption{H in fcc111hc001}
\end{subfigure}
\begin{subfigure}[b]{0.225\textwidth}
\includegraphics[width=\textwidth]{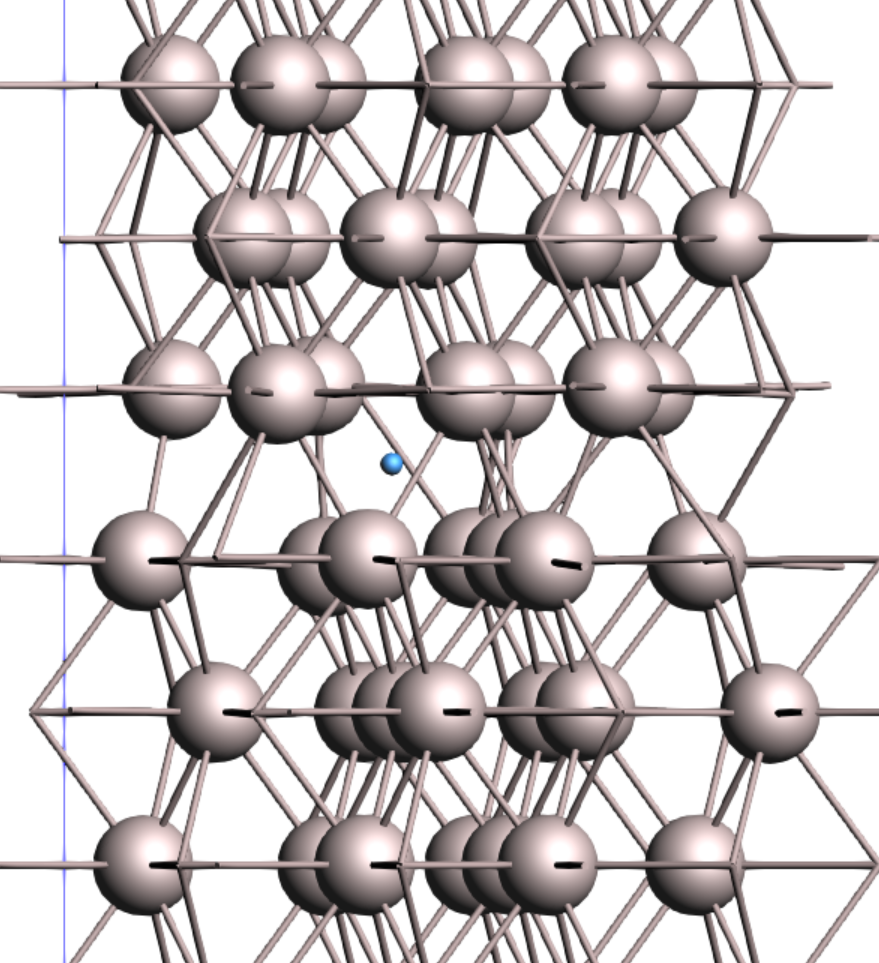}
\caption{H in twist GB}
\end{subfigure}
\begin{subfigure}[b]{0.225\textwidth}
\includegraphics[width=\textwidth]{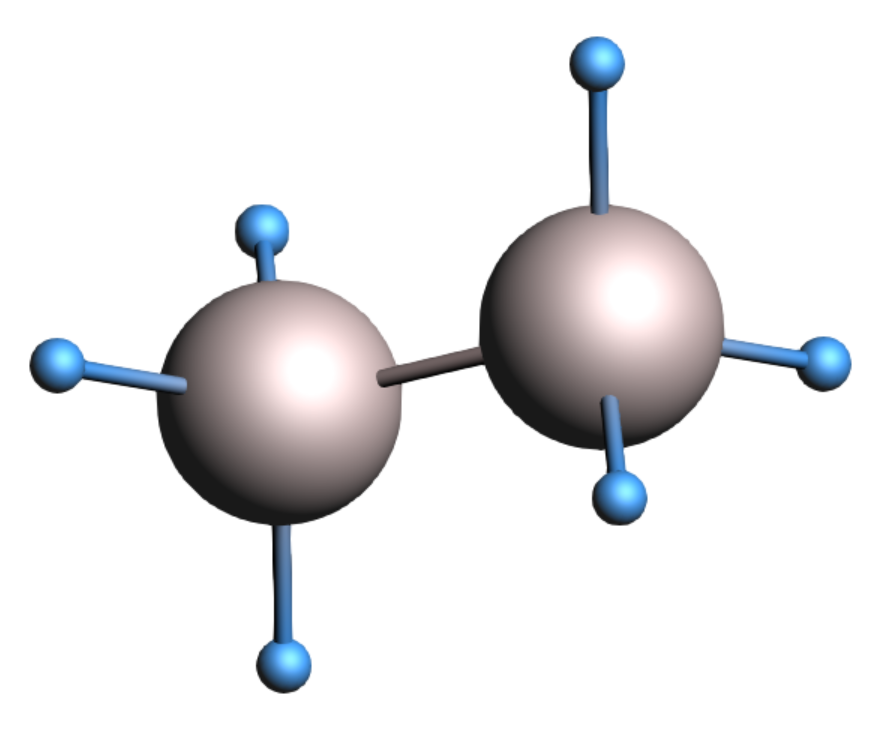}
\caption{Ru\textsubscript{2}H\textsubscript{6}}
\end{subfigure}
\begin{subfigure}[b]{0.225\textwidth}
\includegraphics[width=\textwidth]{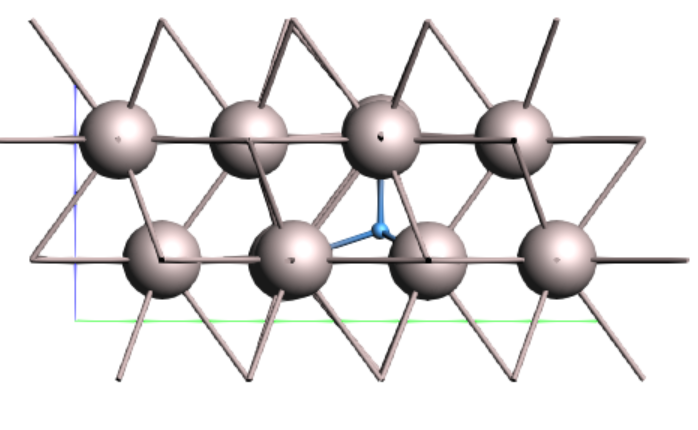}
\caption{H in tetrahedral site}
\end{subfigure} 
    \caption{Examples of structures included in the Ru/H training set.}
    \label{fig:Ru_H_set}
\end{figure}

%
\section{MSD plots}
In this section are MSD plots for each of the simulated structures at temperatures 500, 600, 800, and 900K. Note that the scales on the vertical axis differ significantly.
\begin{figure}[!h]
\captionsetup{justification=centering}
\centering
\begin{subfigure}[b]{0.45\textwidth}
\includegraphics[width=\textwidth]{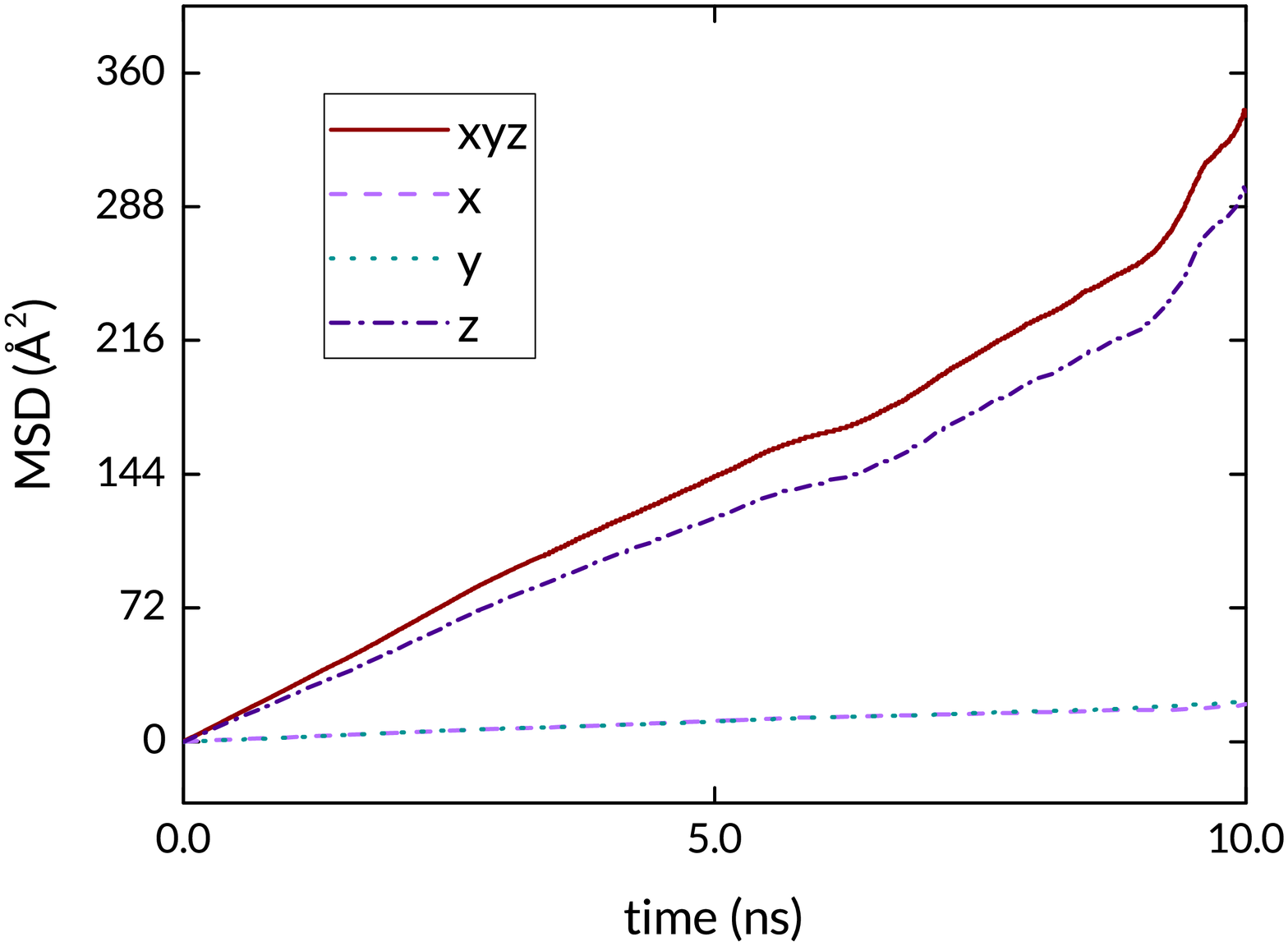}
\caption{500K}
\end{subfigure}
\begin{subfigure}[b]{0.45\textwidth}
\includegraphics[width=\textwidth]{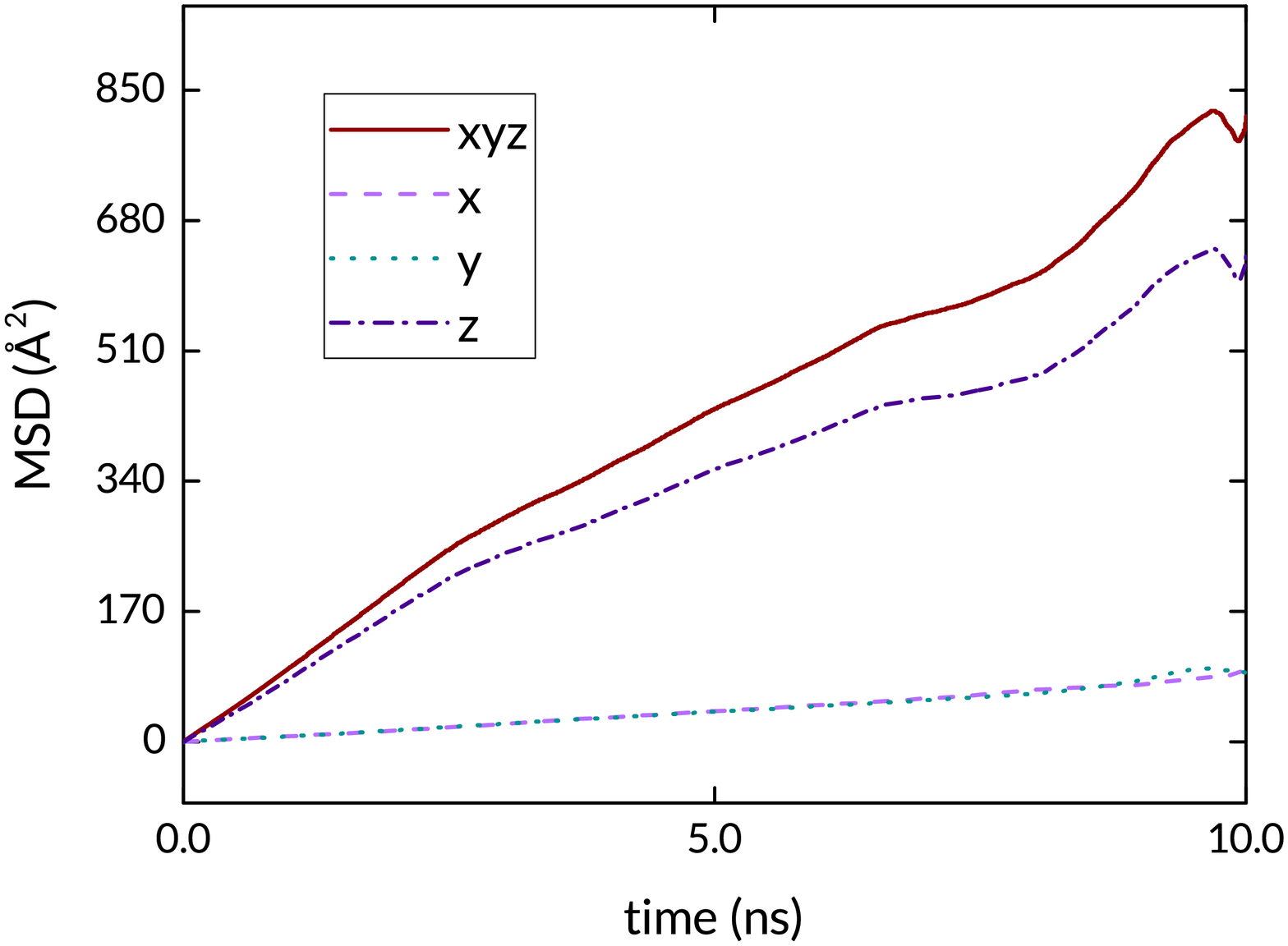}
\caption{600K}
\end{subfigure} \\
~\\
\begin{subfigure}[b]{0.45\textwidth}
\includegraphics[width=\textwidth]{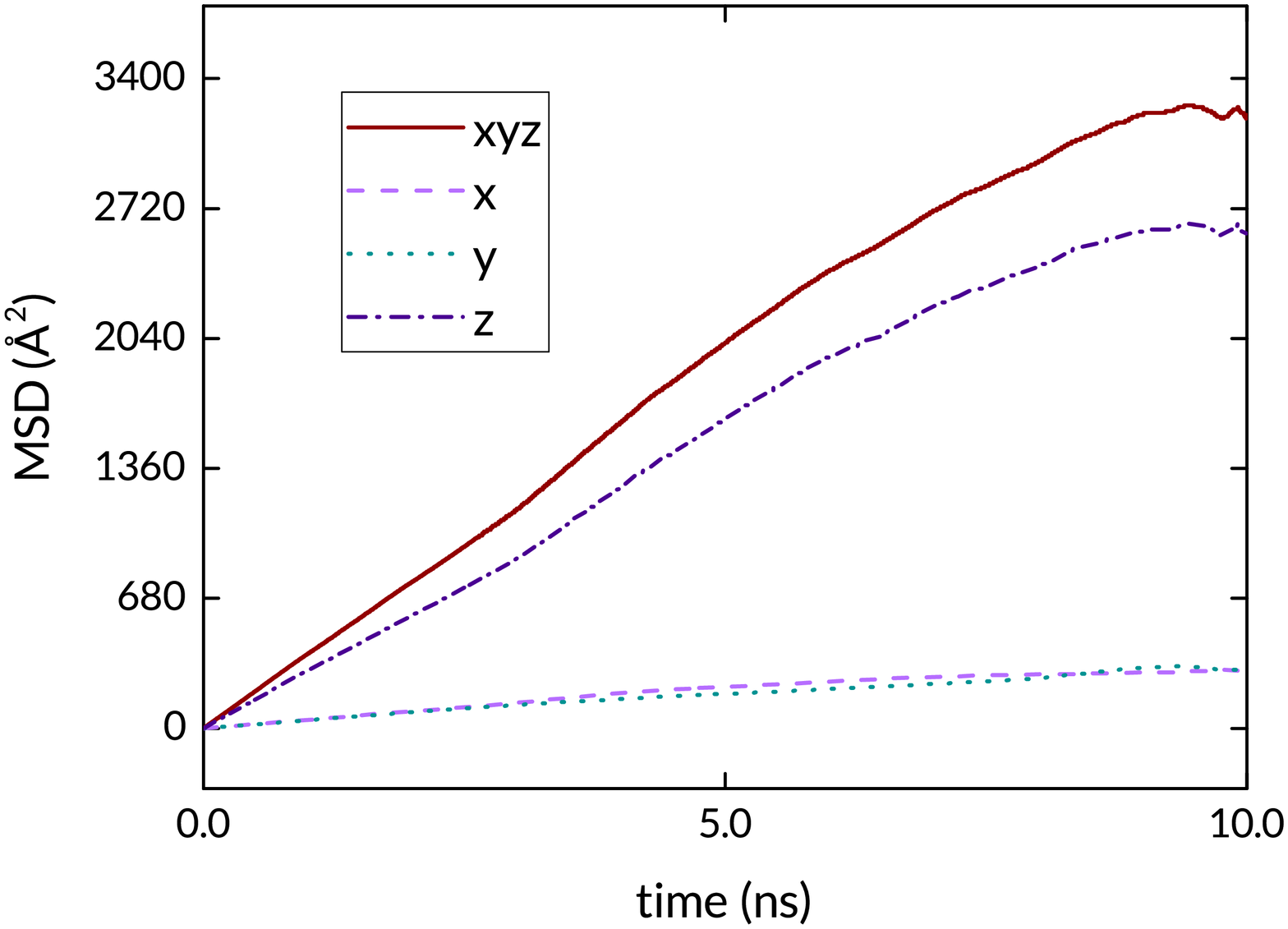}
\caption{800K}
\end{subfigure}
\begin{subfigure}[b]{0.45\textwidth}
\includegraphics[width=\textwidth]{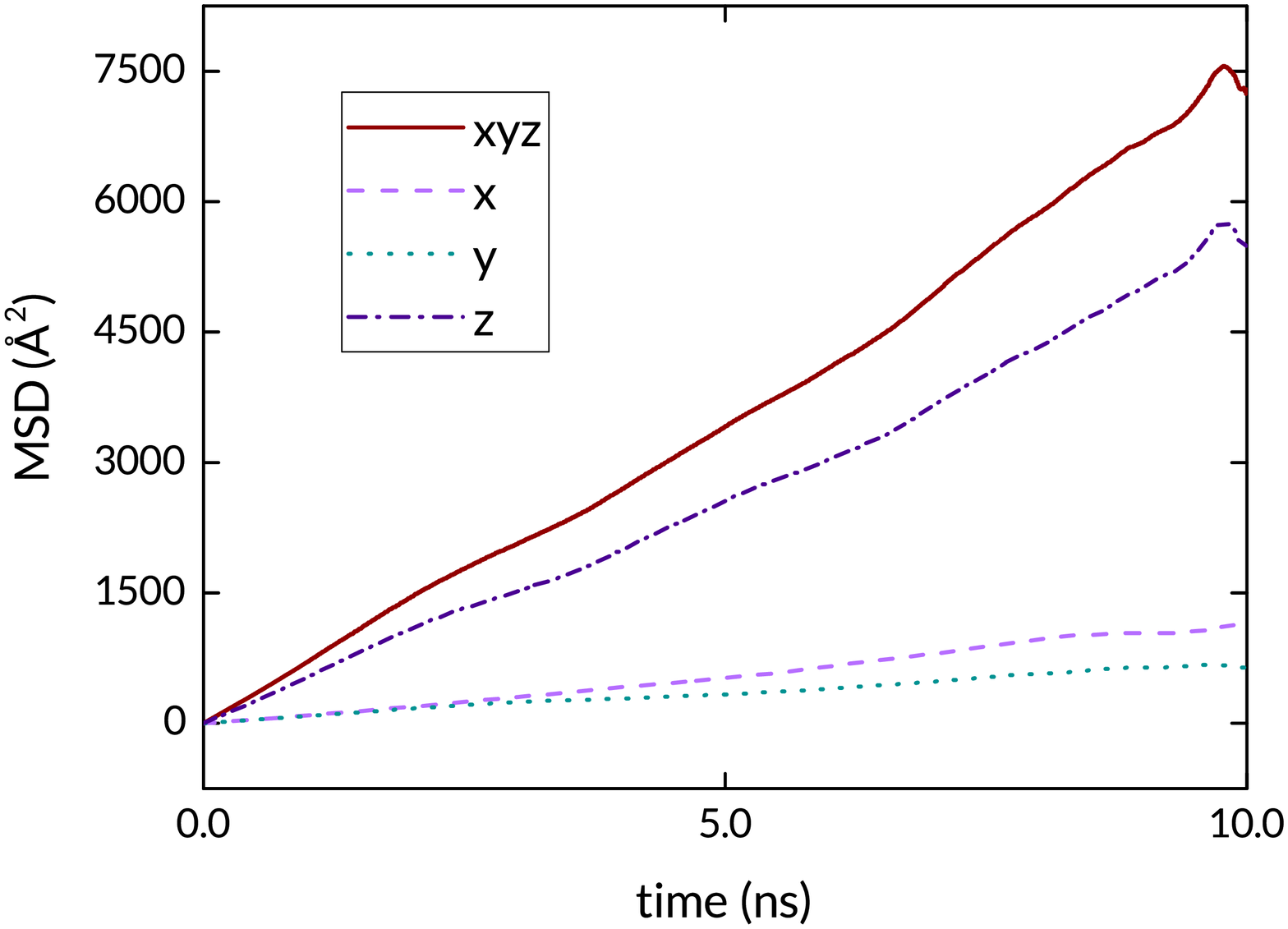}
\caption{900K}
\end{subfigure}
    \caption{MSD plots for NVT simulations of H in pristine Ru.}
    \label{fig:app_pristine}
\end{figure}
\begin{figure}[!htbp]
\captionsetup{justification=centering}
\centering
\begin{subfigure}[b]{0.45\textwidth}
\includegraphics[width=\textwidth]{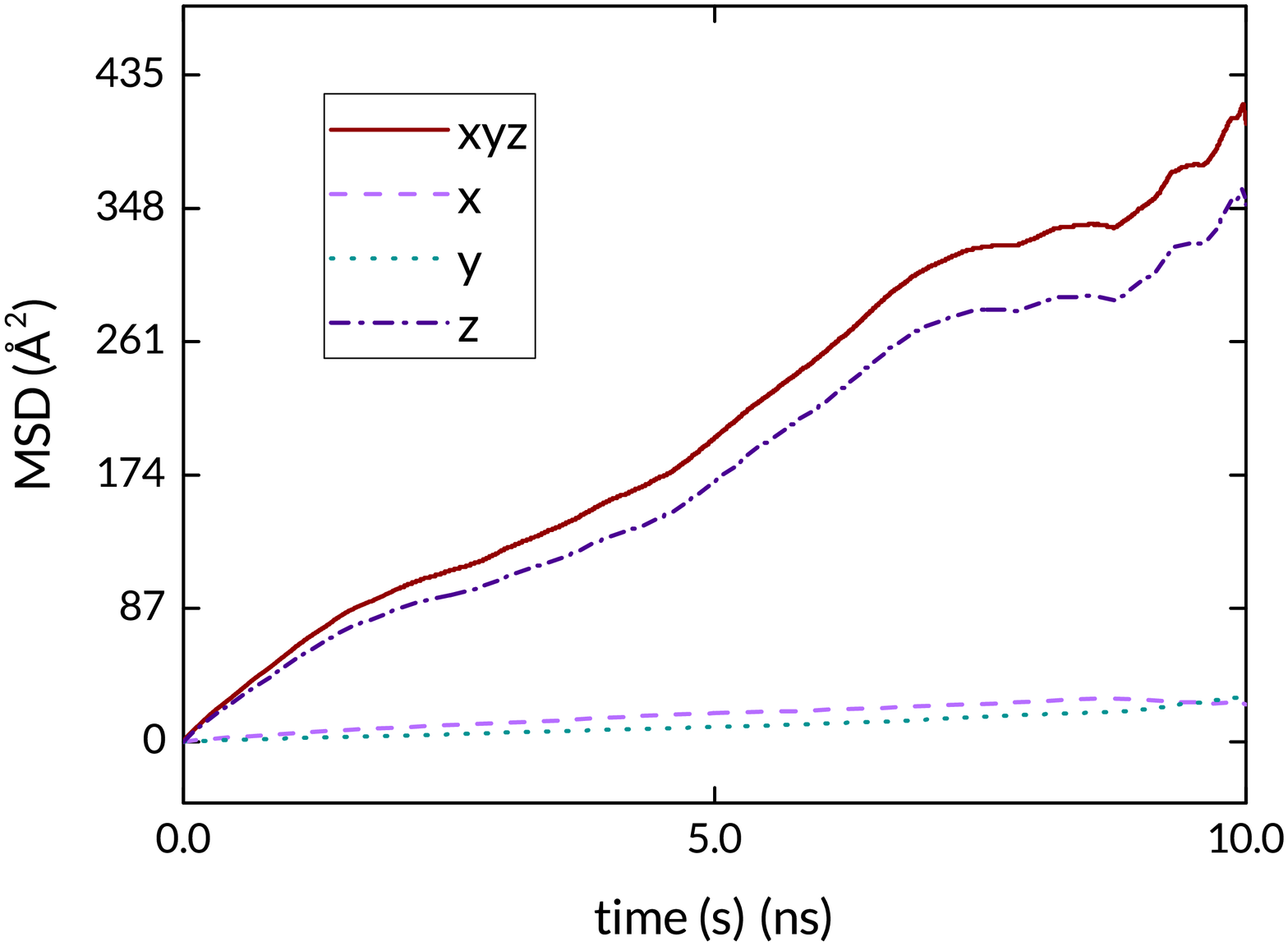}
\caption{500K}
\end{subfigure}
\begin{subfigure}[b]{0.45\textwidth}
\includegraphics[width=\textwidth]{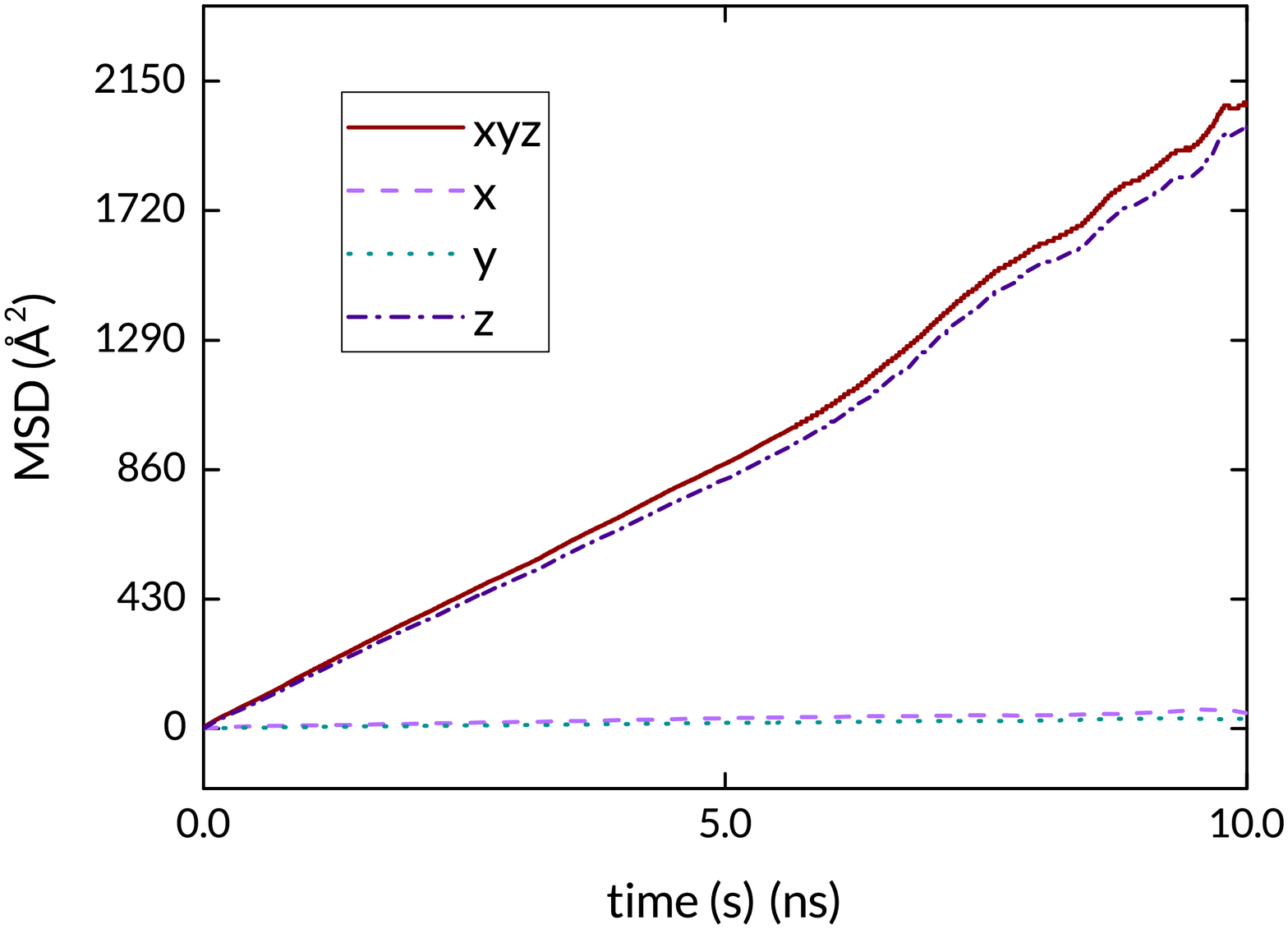}
\caption{600K}
\end{subfigure} \\
~\\
\begin{subfigure}[b]{0.45\textwidth}
\includegraphics[width=\textwidth]{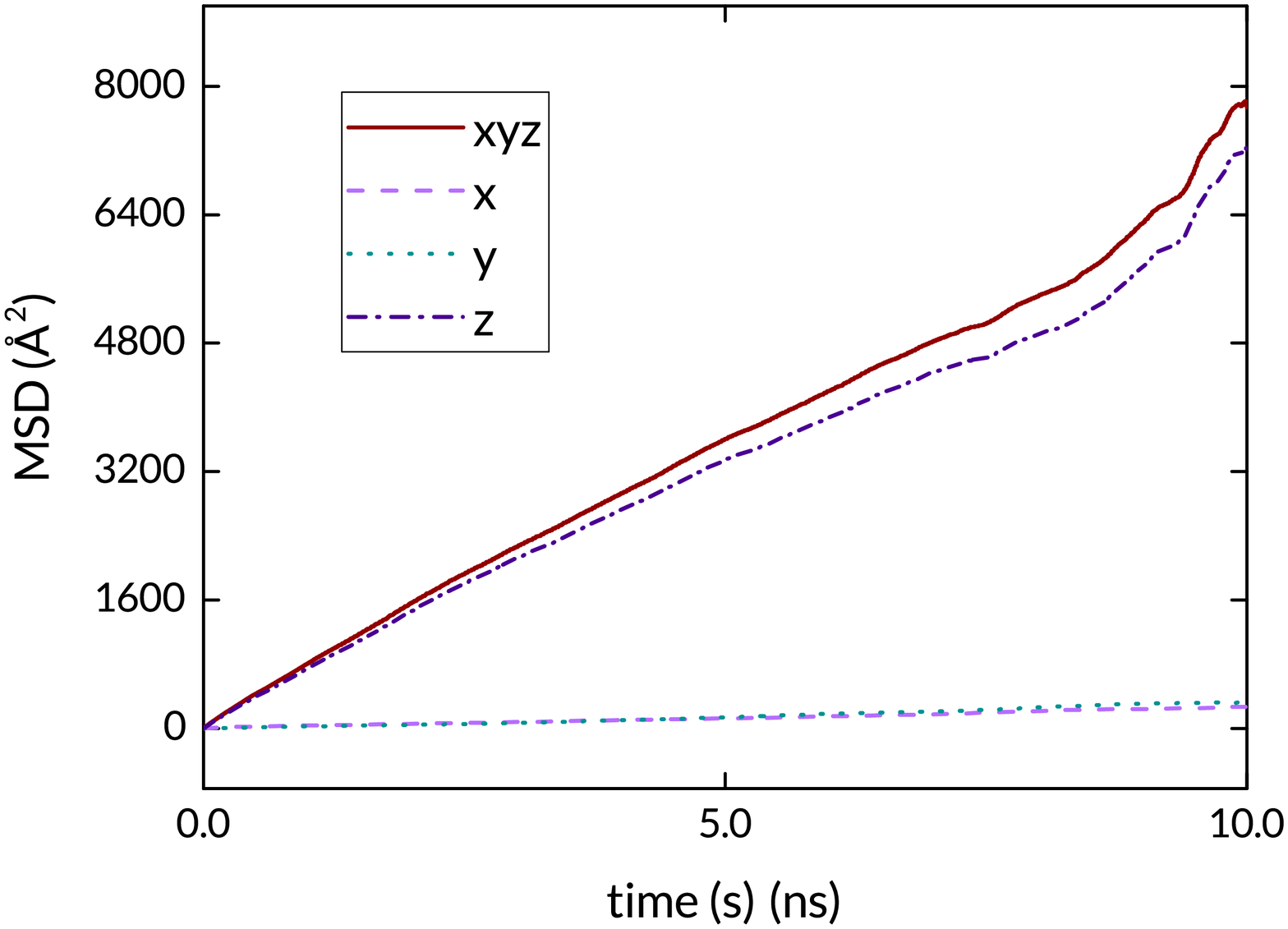}
\caption{800K}
\end{subfigure}
\begin{subfigure}[b]{0.45\textwidth}
\includegraphics[width=\textwidth]{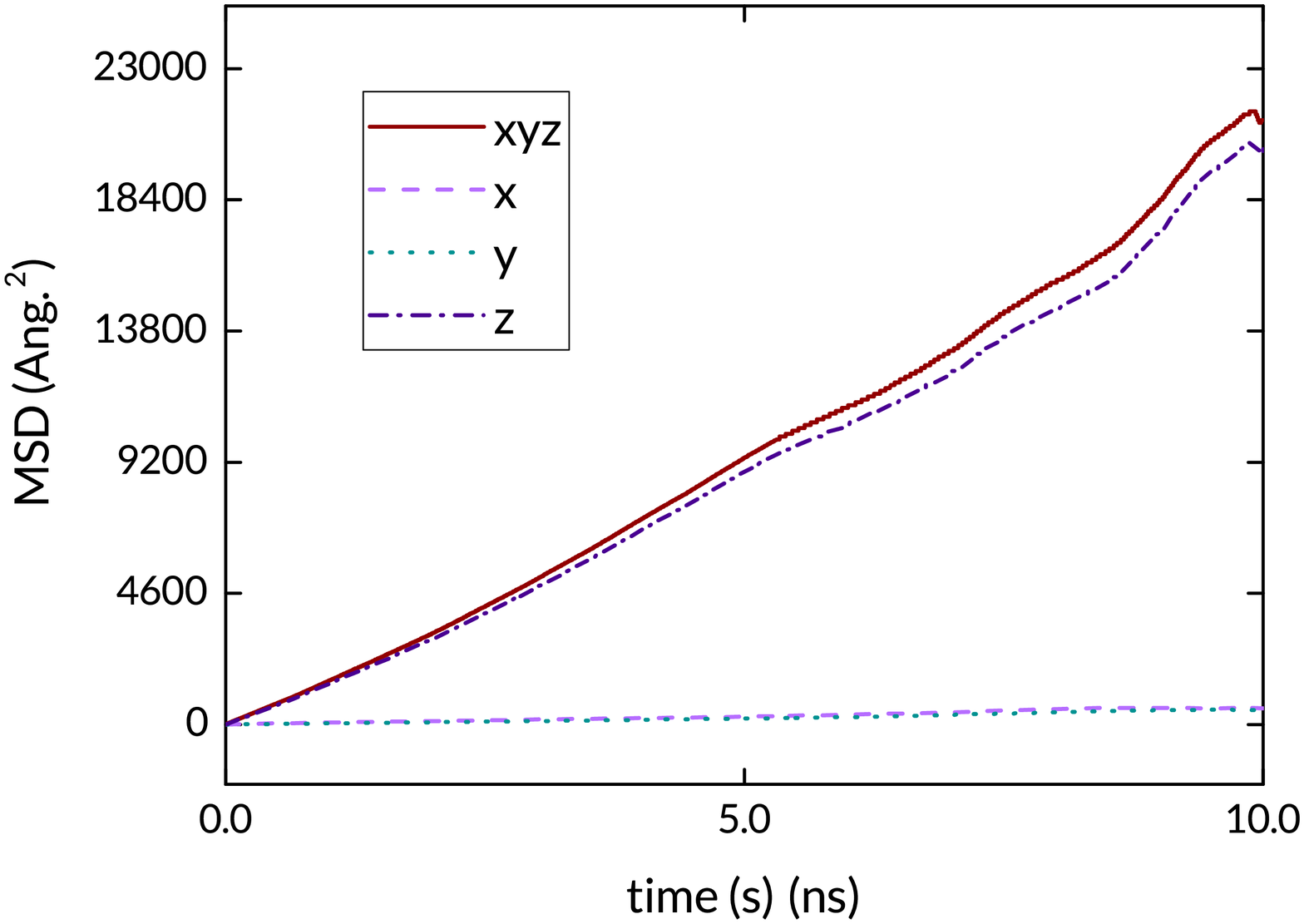}
\caption{900K}
\end{subfigure}
    \caption{MSD plots for NVT simulations of H in Ru tilt GB.}
    \label{fig:app_tilt}
\end{figure}
\FloatBarrier
For the twist GB, at the lower temperature of 500K, the square root of the MSD is less than half the dimension of the simulation box. This heuristic implies that longer simulations are necessary to properly sample diffusion at this temperature.

\begin{figure}[!h]
\captionsetup{justification=centering}
\centering
\begin{subfigure}[b]{0.45\textwidth}
\includegraphics[width=\textwidth]{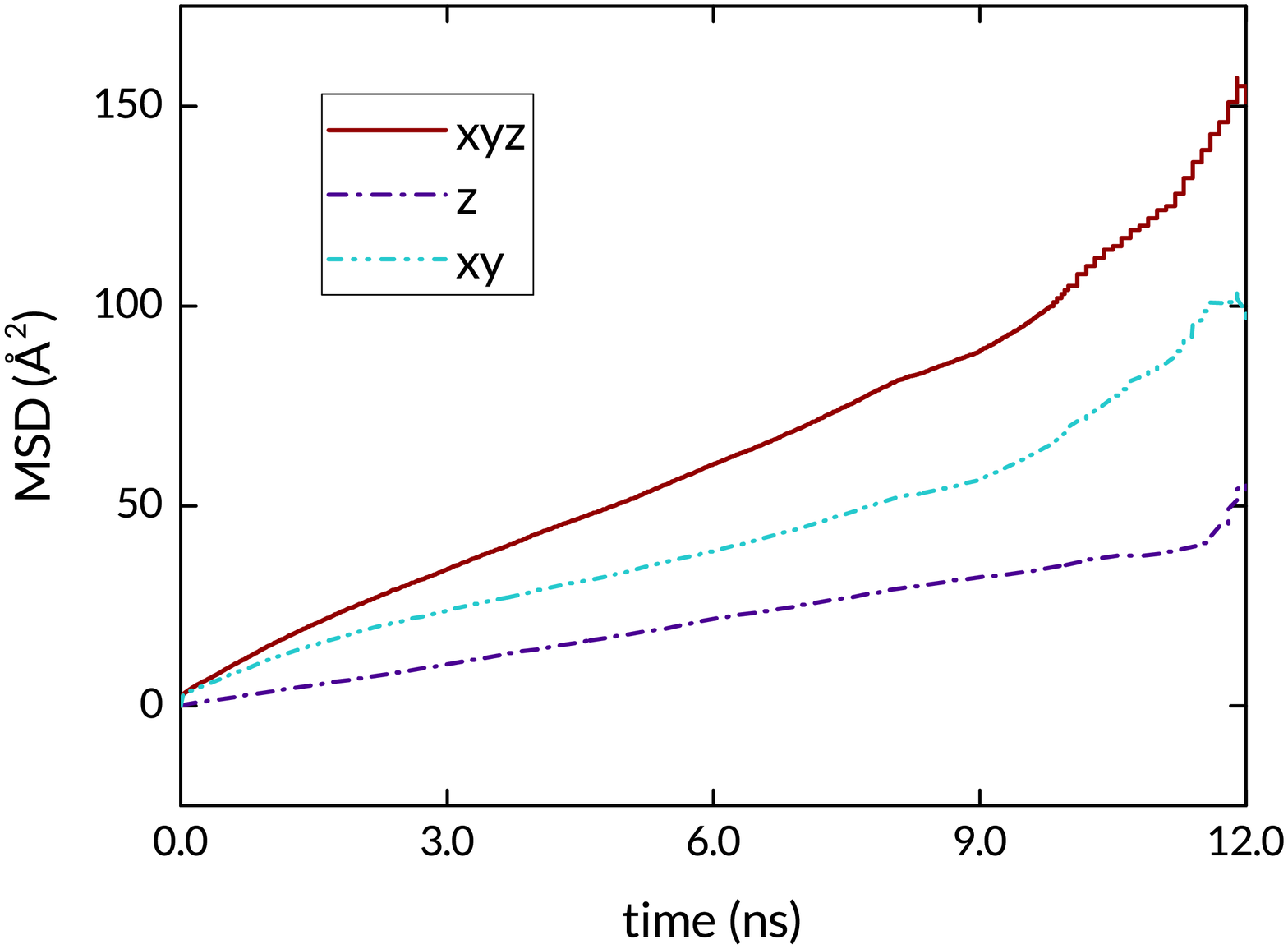}
\caption{500K}
\end{subfigure}
\begin{subfigure}[b]{0.45\textwidth}
\includegraphics[width=\textwidth]{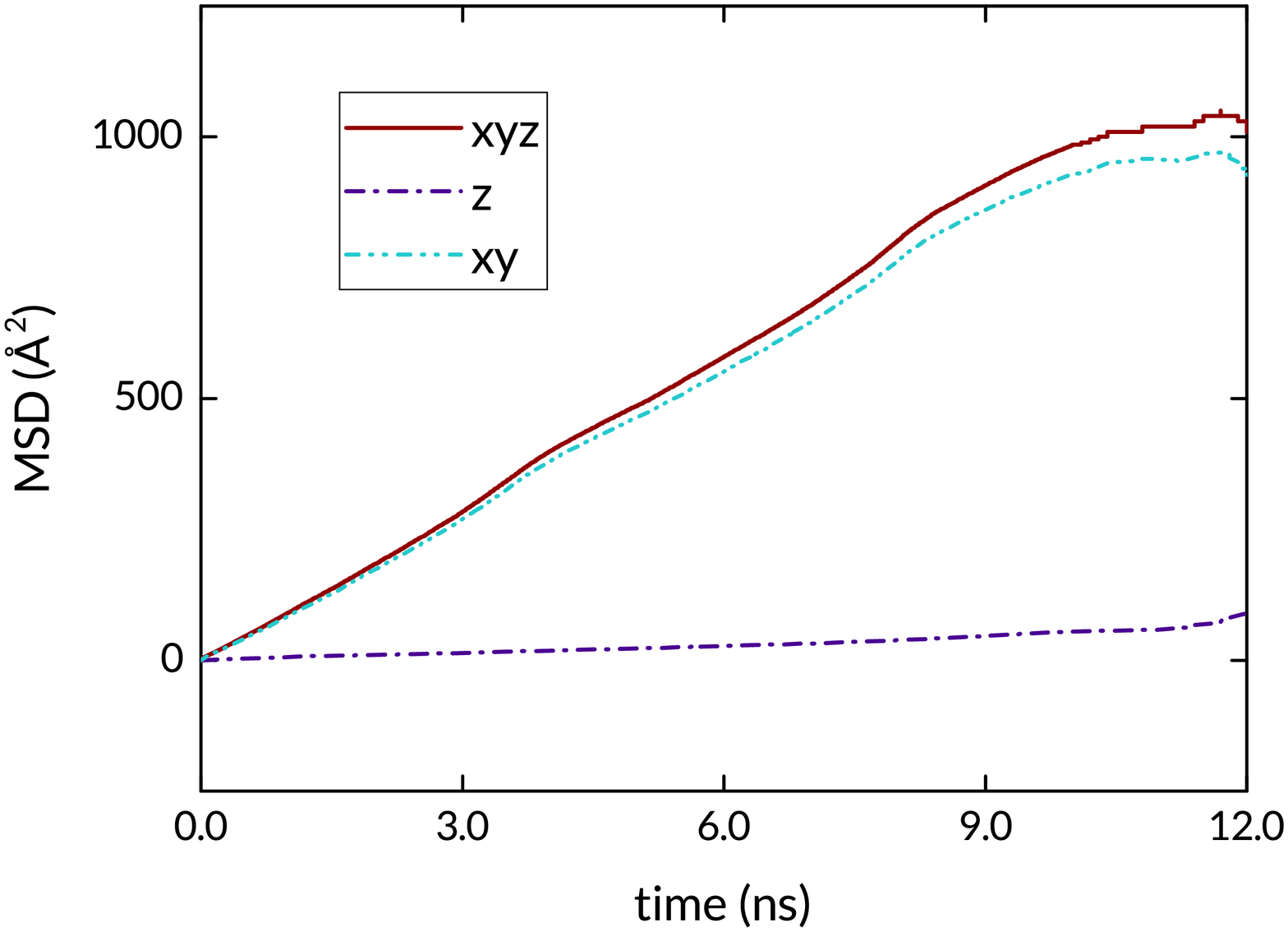}
\caption{600K}
\end{subfigure} \\
~\\
\begin{subfigure}[b]{0.45\textwidth}
\includegraphics[width=\textwidth]{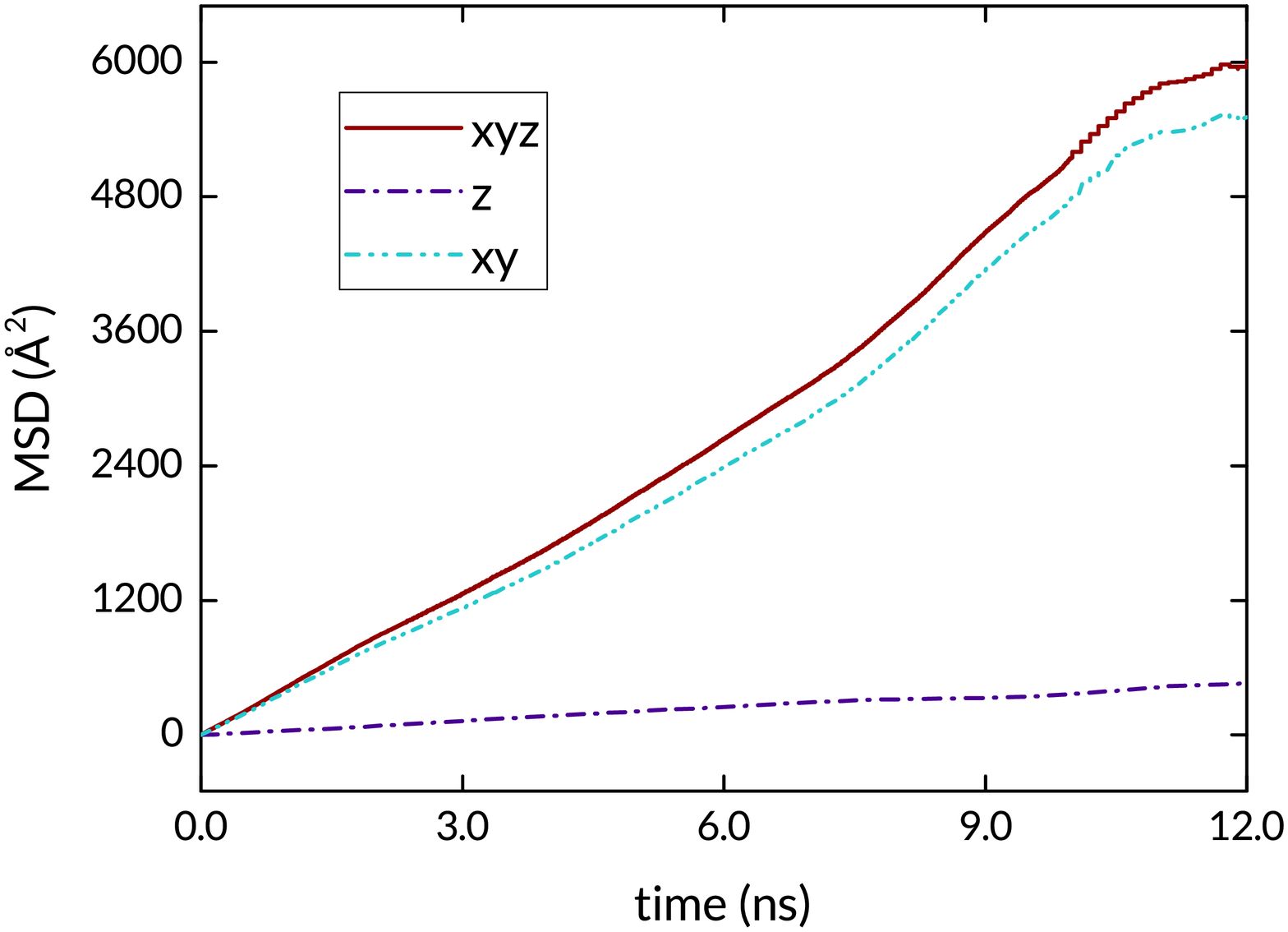}
\caption{800K}
\end{subfigure}
\begin{subfigure}[b]{0.45\textwidth}
\includegraphics[width=\textwidth]{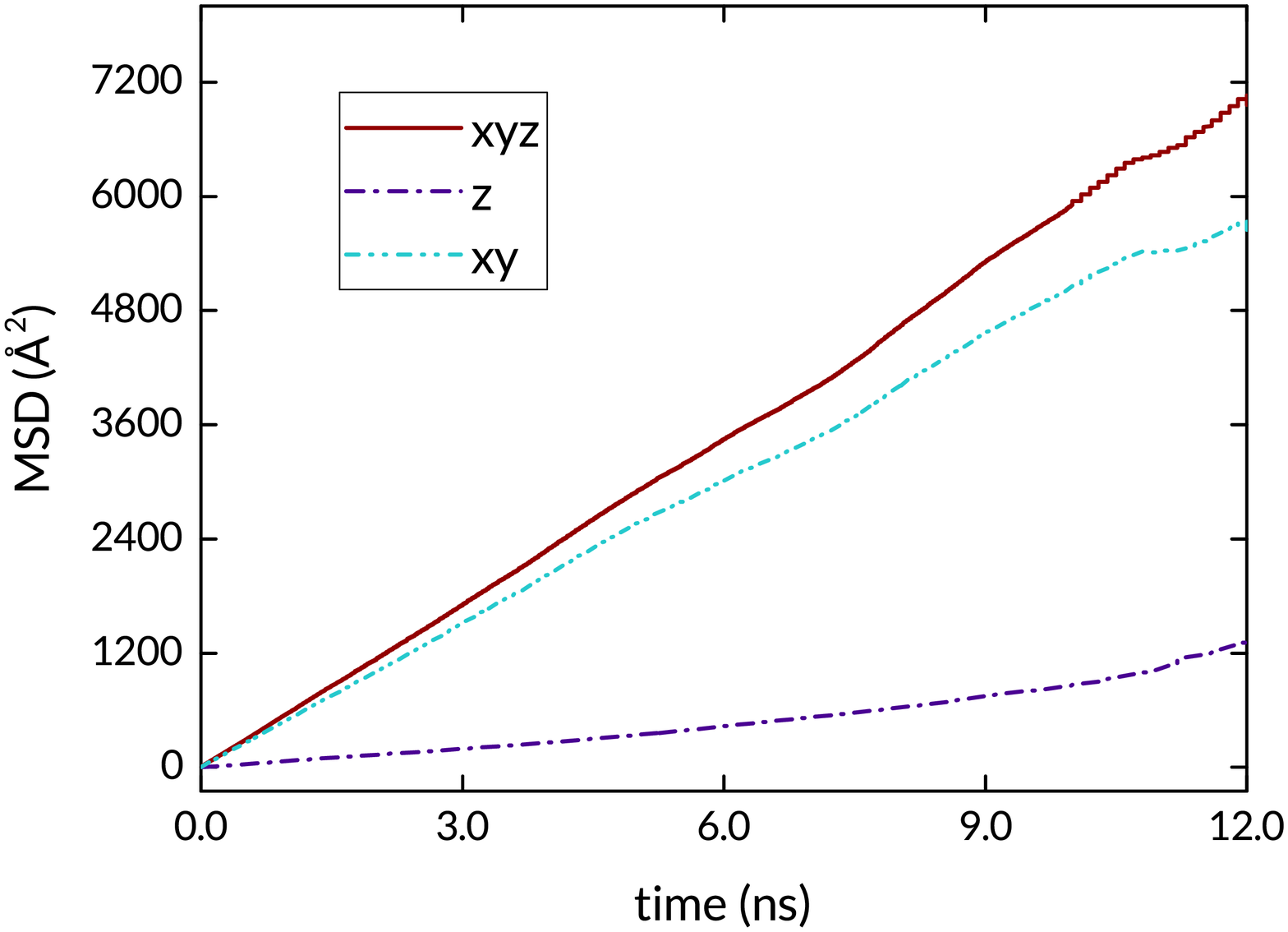}
\caption{900K}
\end{subfigure}
    \caption{MSD plots for NVT simulations of H in Ru twist GB.}
    \label{fig:app_twist}
\end{figure}

\end{suppinfo}

\end{document}